# Structures, Functions, and Mechanisms of Filament Forming Enzymes: A Renaissance of Enzyme Filamentation

**A Review**
**By**

**Chad K. Park**
**&**
**Nancy C. Horton**
**Department of Molecular and Cellular Biology**
**University of Arizona**
**Tucson, AZ 85721**

N. C. Horton (nchorton@email.arizona.edu, ORCID: 0000-0003-2710-8284)
C. K. Park (ckpark@email.arizona.edu, ORCID: 0000-0003-1089-9091)

## Abstract

Filament formation by non-cytoskeletal enzymes has been known for decades, yet only relatively recently has its wide-spread role in enzyme regulation and biology come to be appreciated. This comprehensive review summarizes what is known for each enzyme confirmed to form filamentous structures *in vitro*, and for the many that are known only to form large self-assemblies within cells. For some enzymes, studies describing both the *in vitro* filamentous structures and cellular self-assembly formation are also known and described. Special attention is paid to the detailed structures of each type of enzyme filament, as well as the roles the structures play in enzyme regulation and in biology. Where it is known or hypothesized, the advantages conferred by enzyme filamentation are reviewed. Finally, the similarities, differences, and comparison to the SgrAI system are also highlighted.

## Contents

## Introduction

For over 50 years it has been known that many enzymes form filamentous structures *in vitro* as assessed by various biophysical assays, including in some cases imaging by electron microscopy (EM)[1-16]. However, it was not generally known how filamentation affected enzyme activity. As protein structure determination by x-ray crystallography came to dominate enzyme structure and function studies, enzymes studied tended to be those which produced well-ordered crystals, and filament formation by enzymes seemed nearly forgotten. However, a handful of laboratories continued to work on this phenomenon and its role in regulation of their particular enzyme systems[17-22]. Then, an explosion of interest occurred with the discovery of widespread enzyme self-assembly in cells when viewed by confocal microscopy and with enzymes labeled with fluorescent proteins or antibodies[23-30]. These screens surprisingly found that many enzymes, not previously appreciated as filamentous, formed large scale self-assembled structures in cells, including foci, rods, and rings, which are sometimes referred to as cytoophidia. These membraneless, reversible subcellular structures were often seen in response to cellular stress (nutrient starvation, hypoxia) but in many cases they were also seen under normal physiological conditions[27, 31]. Controls with alternative tags, and the use of orthogonal techniques such as mass spectrometry, confirmed that these observations were not merely artifacts of fluorescent labeling such as GFP[25, 32-33]. In addition, several studies investigated the reversibility of the assemblies so as to distinguish from aggregates of misfolded proteins unlikely to represent regulatory states of the enzymes[25, 28].

Enzymes now shown to form nanoscale filaments *in vitro* and/or self-assemblies in cells come from a diverse array of biochemical and biological pathways, and from diverse cell types including bacteria, yeast, and metazoans (worms, flies, mice, humans). As such, many have medical significance, such as in metabolic diseases, cancer, neurodegenerative disorders, autoimmune disease, and infectious disease. Some have biotechnological or industrial applications, such as in the capture of $CO_2$ ($CO_2$ reductase) and production of specialized chemicals and bioremediation[34].

In this review, we attempt to comprehensively collate studies of enzymes found to either form large assemblies in cells (with unknown molecular structures) as well as those with filamentous structures known in atomic or near-atomic detail. For several enzymes both the molecular structure of the filament is known, at least to low resolutions via electron microscopy, and the cellular self-assemblies have been characterized. Our particular interest in this phenomenon originated with our studies of SgrAI, a type II restriction endonuclease with unusual allosteric behavior, where binding to one type of DNA sequence results in activation of the enzyme to cleave 14 additional DNA sequences[35]. Our investigation into the mechanism responsible for this behavior led to the discovery of filament formation by SgrAI when bound to the activating DNA (which is also a substrate for cleavage of SgrAI known as primary site sequences)[17, 36-37]. The filamentous form recruits additional copies of SgrAI bound to the second type of DNA sequence (secondary sites)[17, 38]. The filamentous state preferentially stabilizes the activated conformation of the enzyme, hence SgrAI in the filament is activated for DNA cleavage[39]. In reviewing the literature for precedence of this type of behavior, we discovered that such phenomenon were also under investigation in the regulation of IRE1 (the unfolded protein response kinase/ribonuclease)[18], CTP synthase[19, 24, 27, 40-41], and acetyl CoA carboxylase (ACC)[20]. Concurrently, the publication of proteome-wide screens for self-assemblies in cells, as well as older literature showing filamentation by a number of metabolic enzymes, became relevant to our studies and are also covered in this review.

Several other reviews have been published recently regarding enzyme filamentation[26-27, 42-49]. We focus here in this review on structure-function studies of filament forming enzymes, while also attempting to provide an up-to-date, comprehensive listing of enzymes known to form cytoophidia or intracellular self-assemblies. We attempt to find similarities and differences in the structures and mechanisms, and are particularly interested in why filament formation is necessary in addition to more "traditional" enzyme regulatory mechanisms. Our own studies with SgrAI indicate that filament formation provides for a much faster activation of the enzyme[50]. We also find that due to a particularity of the enzyme kinetic pathway, namely a relatively slow, rate limiting second order association rate constant for filament assembly, that filament formation can provide a means to sequester enzyme activity on only particular substrates of interest, those with high local concentrations[50-52]. This could be a general phenomenon for enzymes that have more than one class of substrate and require regulation of when and where that secondary substrate activity will occur. However, it remains to be seen for the vast majority of enzymes known to filament, what advantage filamentation has towards enzyme function and/or regulation. What is known in general, considering all enzymes reviewed herein, is that filamentation can occur through linear polymer assembly or more commonly helical assembly (left or right-handed). The filament form may either be the more active form, or the less active form, or may have altered activity (substrate preference or even a completely different type of activity). The purpose of filamentation may be to perform a structural function, such as determining cell shape, or may form a scaffold for the binding of other proteins. As such, it can sometimes perform functions in signaling. Filament formation can be responsive to cellular conditions, thereby regulating enzyme activity and any other activity such as signaling or chaperone function. All in all, enzyme filamentation has been found to perform many different functions in cells, and we are likely to continue to discover new roles and functions for this interesting phenomenon.

## Structurally Characterized Enzyme Filaments

### *Acetyl CoA Carboxylase (ACC)*

Acetyl CoA carboxylase (ACC) has a central role in primary metabolism, and its upregulation is linked to obesity related diseases[53-55] and tumor growth[56-58]. This enzyme catalyzes the carboxylation of acetyl-CoA to malonyl-CoA, the first and a rate-limiting step of fatty acid biosynthesis. The ACC enzyme uses biotin as a carboxyl carrier, requires adenosine triphosphate (ATP), and uses bicarbonate as a carbon donor[59-61] (**Fig. A1A**). Mammals contain two isoforms, ACC1 and ACC2[59]. While both isoforms produce malonyl-CoA, ACC1 is found in lipogenic tissues, is predominantly cytosolic, and generates malonyl-CoA that is used for fatty acid synthesis. ACC2 is mitochondrial and is found mostly in oxidative tissues, such as heart and skeletal muscle where it regulates fatty acid beta-oxidation because its product, malonyl-CoA, is a negative regulator. Hence, ACC2 has been identified as a potential target for treating metabolic syndromes because knockout of ACC2 reduces fat content and increases the resistance to high fat/high carbohydrate-induced obesity and diabetes[62-64].

Mammalian ACCs are typically homodimeric, but can also assemble into polymers with increased activity (~60 fold)[2-3]. ACCs are regulated by a complex interplay of phosphorylation, binding of allosteric regulators and protein-protein interactions, which is further linked to filament formation[1, 53, 65-68] (**Fig. 1b**). These filaments were discovered 50 years ago[1-2, 69] but only more recently were they shown to be relevant to *in vivo* activity and to be characterized structurally[20]. Polymerization of ACC1 increases enzymatic activity and is induced *in vitro* by supraphysiological concentrations of citrate (> 5 mM)[2-3]. Citrate is produced in the mitochondria and is used as a substrate by ATP citrate lyase, which converts citrate to acetyl-CoA, the substrate of ACC.

Filament formation is also modulated by other proteins. For the mitochondrial isoform of ACC (ACC2), midline-1-G12-like-interacting protein (MIG12) binding alone is sufficient to drive polymerization[20]. The presence of MIG12 further reduces the concentration of citrate needed for polymerization fivefold for the cytoplasmic isoform ACC1 (to <1 mM), and also increases the activity by ~60-fold[20]. However, Spot14, a paralog of MIG12, forms assemblies with MIG12 which attenuates activation and filament formation by both ACC1 and ACC2[70].

A recent structural investigation showed the formation of three different types of filaments of human ACC1[71] (see **Fig. 1c**). The authors used cryo-electron microscopy and single particle reconstruction to reveal filaments at 4-6 Å resolution, allowing the fitting of previously determined crystal structures into the cryo-EM envelopes. The basis of activation or inactivation in the filaments resulted from the locking in of either an active or inactive conformation, compared to nonfilamentous ACC1 that was shown to populate both active and inactive conformations. One type of filament is activated, due to the dimerization of its two biotin carboxylase (BC) domains (which is required for activity), while the other two show these domains are not in close proximity, and are inactive.

The active filament (ACC-citrate) is formed in the presence of citrate (at 10 mM), and only with a form of ACC1 that is free of phosphorylation. The filament is a run-on oligomer, as it is composed of a linear polymer of ACC1 dimers which appear to add singly at either end. The filament is a left-handed helix, with three ACC1 dimers per turn and a rise (distance between adjacent ACC1 dimers in the filament) of 154 Å. In the active filament, not only are the BC domains dimerized, but also the BCCP (biotin carrier protein domain) is in a position where it may rotate to carry the biotin from the BC to the CT (carboxyl transferase) domain (**Fig. 1d**). These filaments can be quite long at 0.5-1.0 microns in length.

A second filamentous form of ACC1 is induced by the addition of a tenfold molar excess of the feedback inhibitor palmitoyl-CoA to the filament form of ACC1 induced by citrate. This form, ACC-citrate$^{palm}$ is presumably inactive, as the BC domains are no longer dimerized. The helical backbone is thinner from which the BC domains now protrude, although the filament still shows 3 copies of the ACC1 dimer per helical turn (left-handed)(see **Fig. 1b**).

A third type of filament results from binding of the BRCT domains of BRCA1 to a phosphorylated form of ACC1 (ACC-BRCT). Phosphorylation at Ser78, Ser80, Ser1201, and Ser1216 are known to inactivate human ACC1, with Ser80 and Ser1201 phosphorylation having the greatest effect[65]. Mutations in the BRCT domains eliminating BRCA1 binding to ACC result in elevated lipogenesis, which is a prerequisite for cancer cell growth[72-73]. The C-terminal tandem BRCT domains of BRCA1 bind ACC1 by recognizing phosphorylated Ser1263, a residue that is phosphorylated in a cell cycle dependent manner, and this binding inhibits dephosphorylation of Ser80 of ACC1[72-74]. This filament is distinct in that it has a two-stranded appearance, with a rise of 190 Å between adjacent ACC1 dimers. However, it is also a left-handed filament with 3 ACC1 dimers per turn. The BRCT domains laterally "decorate" the filament, and interlink adjacent ACC1 dimers. Like the ACC-citrate$^{palm}$ filament, the BC domains are monomeric and the BCCP domains appear to be sterically unable to reach any of the active sites, hence the ACC1 in this form is predicted to be enzymatically inactive.

Large assemblies (*i.e.* cytoophidia) of ACC are seen in screens, including that of Shen *et al* (2016)[26] under conditions of nutrient starvation (in stationary phase)(**Fig. 1e**). These are much larger than the filaments viewed by EM (**Fig. 1c**), but may be made of bundles of likely inactivated enzymes.

### *Phosphofructokinase (PFK)*

Phosphofructokinase (PFK) is an important metabolic enzyme that acts as a gatekeeper since it unidirectionally catalyzes the committed step of glycolysis, namely the conversion of fructose 6-phosphate (F6P) to fructose 1,6-phosphate (F16P)(**Fig. 2a)**. Glycolysis is the core of central carbon metabolism; its intermediates provide precursors important for generating ATP through glucose oxidation, serine for one-carbon metabolism, sugars for protein glycosylation, and building blocks for nucleotide synthesis through the pentose phosphate shunt. Due its importance in cellular metabolism, it is highly regulated with allosteric regulation by over 10 metabolites to allow cells to fine tune their energy needs. PFK from yeast, *C. elegans*, and human liver have been shown to form filaments, fibers, and/or punctate foci[26, 75-76]. The *in vitro* self-association of purified rabbit liver PFK has been known since at least 1971, using size exclusion chromatography[4]. Follow-up studies using sedimentation, analytical ultracentrifugation, and fluorescence anisotropy also showed PFKL self-assemblies[5-6]. Finally, negative stain EM performed in 1980 showed these to be composed of filaments[77], which was further investigated recently (**Fig. 2b**)[76]. Filament formation occurs with starvation in yeast[26], stress from increased activity and/or hypoxia in neurons of *C. elegans*[75], and high concentrations of citrate or its substrate F6P in human PFKL (**Fig. 2c**)[76], suggesting a role in enzyme regulation. However, the effect of filamentation on enzyme function are not currently known. Dysregulation of PFK activity has implications for cancer [78-81], neuronal disfunction [75], type II diabetes[82-83] and autoimmune disease[84]. A low-resolution structure of the filament formed by human PFKL has been reported using single particle reconstruction and negative stain TEM (**Fig. 2d**)[76]. No high-resolution structures or structures of PFK filaments are known, though crystal structures of non-filamentous PFK have been reported[78, 85-88]. Fitting of the non-filamentous structure into the low-resolution EM envelope of the PFKL filament shows a right-handed helix with stacked PFKL tetramers related by a rotation of 221° and translation of 83 Å between subunits (**Fig. 2e**)[76]. The average size of filaments observed by EM is 6 tetramers, although filaments of 11 or more were also observed[76]. Filaments were observed to be kinked at random locations (**Fig. 2b)**, and structural analysis revealed that for every addition of PFKL tetramer, two possible interfaces are possible. One interface leads to an unkinked junction, the other to a kinked junction of ~130° (**Fig. 2f-h)**[76]. *In vivo*, TIRF was used to show punctate foci of PFKL in cells, and indicated that filaments must be smaller than 15 tetramers in length, consistent with the EM results (**Fig. 2b)**[76].

No detailed kinetic analyses of the role of filament formation in PFK function have been reported. Therefore, the following questions remain unanswered: how does filament formation affect PFKL enzyme activity? It's stimulation by its substrate suggests activation, however it is also stimulated by the inhibitory effector citrate. Also, how do its 10 allosteric effectors affect filament formation and PFKL activity?

Allosteric effectors include: ATP (a required reactant but also an allosteric inhibitor at higher concentrations), ADP (a product of the reaction, but also an allosteric activator), fructose 2,6-phosphate (activator) and citrate (inhibitor), among many others. Further, what advantages does filament formation confer to PFKL function?

### *Glucokinase (Glk1)*

Glucokinase (Glk1) is found in fungi, and one of three kinases in S. cerevisiae that initiate glycolysis by phosphorylating glucose in an ATP-dependent reaction (**Fig. 3a**), and also possesses an actin fold. The enzyme is induced in the absence of glucose, and expressed in high glucose[89]. A fusion protein of Glk1 and GFP is found to be diffused throughout the cytoplasm in the absence of glucose (**Fig. 3b**, left), but formed fiber-like self-assemblies in the presence of glucose (**Fig. 3b**, right)[90]. These bundles rapidly disassemble when glucose was removed from the growth media[90]. Purified Glk1 at 7.5 µM is nonfilamentous in the absence of ligands (**Fig. 3c**, left), but was found to form helical filaments *in vitro* in the presence of its substrates (ATP and glucose, mannose, glucosamine) or products (ADP and sugar-6-phosphate)(**Fig. 3c**, right). No filaments were formed in response to pH, fructose, or galactose (fructose and galactose result in fiber formation *in vivo*)[90]. Some polymerization is found with N-acetylglucosamine-6-phospahte and N-acetylglucosamine, inhibitiros of Glk1[90].

Polymerization of Glk1 in vitro was found to occur with a critical concentration of 2 µM, below which the enzyme remains unpolymerized and active, but above this concentration the enzyme forms polymers with little to no activity[90]. Therefore, below 2 µM, product formation increases with enzyme concentration, however above 2 µM, the rate of product formation is independent of enzyme concentration. Above 2 µM, the concentration of non-polymerized enzyme remains constant while the concentration of enzyme in the filaments increases with increasing overall enzyme concentration. This in effect, results in an upper limit to the enzyme activity, which may play a role in protecting the cell against excess toxic enzyme activity. Assembly of filaments was also found to be rapid, reaching the steady-state equilibrium in a matter of seconds[90].

The filament structure was determined using cryo EM (**Fig. 3d**) and found to be an anti-parallel double stranded, right-handed helix[90]. The enzyme is confirmed to be in the closed state, and evidence of ATP binding in the active site is found[90]. Each enzyme copy makes extensive contacts to enzyme copies before and after in the same filament, as well as the copy opposite it in the other strand[90]. The contacts between enzymes are distinct from those found in actin, and evolutionary considerations indicate that polymerization by Glk1 evolved independently from that of actin[90]. Being an anti-parallel, double helical filament, it has no polarity and one end is equivalent chemically to the other. No larger bundles of the dimensions formed in vivo were found in vitro, indicating that those larger self-assemblies in cells are the result of filaments assembling due to crowding, filament binding proteins, or even possibly due to the GPF tag.

Polymerization was found to be important for cell fitness, as mutants that eliminate filamentation without affecting the enzymatic activity of non-polymerized Glk1 show reduced survival compared to wild type[90]. Protecting the cell from toxic excess sugar kinase activity is important, and Glk1, unlike the two other similar kinases Hxk1 and Hxk2, which do not polymerize, is not inhibited by trehalose-6-phosphate (a metabolic intermediate that transiently accumulates as a result of elevated glucose 6-phosphate)[90]. Stoddard, *et al* (2019) argue that Glk1 polymerization evolved as a strategy to allow cells to adapt to environmental transitions occurring faster than the cell can change its protein levels[90].

### *Cytosine Triphosphate Synthetase (CTPS)*

CTP synthase (CTPS) is a universally conserved enzyme that forms filaments *in vitro* and rod-shaped self-assemblies in cells, both the bacterial and eukaryotic cells. This enzyme is a focal point for regulating pyrimidine levels through its formation of CTP from UTP (**Fig. 4a)**, and is a target for cancer, immunosuppressive, and antiparasitic drugs[91-95]. Cytoophidia, or large rod-like self-assemblies, were seen with GFP labeled CTPS in yeast under conditions of nutrient starvation[24] (**Fig. 4b**). Such structures were also discovered in the bacteria *Caulobacter crescentus*[19] and *E. coli*[96]. Filament formation by the bacterial enzyme inhibits its enzymatic activity, is required for nucleotide homeostasis, and its disruption significantly affects cell growth and metabolism[96-97]. In bacteria, the product (CTP) drives filamentation, which may also have a role in cell shape[19]. The filaments are also sensitive to the balance of substrate and product and can rapidly dissociate into active tetramers in response to changes in substrate concentration[98-99]. These dynamics may act to buffer the catalytic activity of CTPS[97]. A negative stain EM image of filaments formed by a bacterial CTPS is shown in **Figure 4c**.

Human CTPS, in contrast to the bacterial, forms filaments with increased enzymatic activity, and an image of negative stain EM is shown in **Figure 4d**[98]. The bacterial and human CTPS filaments also differ considerably in overall filament architecture and conformation of the enzyme[98]. Filament formation by human CPS is favored in the presence of substrates UTP and ATP, but not the products, CTP and ADP[98]. Filaments assembled with substrate are disassembled upon addition of glutamine presumably due to accumulation of CTP, but using nonhydrolyzable AMPPNP stabilizes filaments against disassembly[98]. In eukaryotes, cytoophidia and filament formation is part of a stress response and varies with developmental state in some organisms[32, 40, 97, 99-101]. CTPS in *Drosophila* germ cells form cytoophidia at developmental stages with a high demand for CTP[100]. CTPS activity is regulated by phosphorylation[102-104], but it is not known if phosphorylation has an effect on polymerization. CTPS cytoophidia colocalize with other metabolic and signaling enzymes too, raising the possibility that other enzymes may regulate CTPS filament formation and/or be incorporated into the filaments[31].

The homotetrameric structure of CTPS is conserved between eukaryotes and bacteria[105-108]. Each monomer is glutamine amidotransferase doman (GAT) fused to kinase-like ammonia ligase domain (AL) via an alpha helix linker. Ammonia from GAT is transferred to AL, and ligated to UTP forming CTP in ATP hydrolysis dependent reaction[105]. The 4.6 Å Cryo-EM reconstruction of the bacterial (*E. coli*) CTPS (ecCTPS) filament (**Fig. 4e**, right) with bound inhibitor CTP shows that it is formed from stacked tetramers that interact through the GAT & linker domains[98]. These authors also solved a crystal structure of ecCTPS in a non-filamentous form with bound CTP, and found the same conformation of the enzyme and CTP binding. This result is significant in that it supports an idea that the filament locks in an inhibitory conformation of the enzyme, and that CTP binding allosterically controls ecCTPS and filament formation by inducing a filament competent conformation in the enzyme. Notably, a mutation associated with drug resistance in CTPS, E155K, eliminates the ability to form filaments[96, 109]. Significantly, the mutant CTPS lacks the large reduction in activity found with wild type upon filament formation, and binds CTP less tightly. These results indicate a role for filament formation in providing an added layer of regulation in the form of rapid and efficient inhibition.

To further investigate the mechanism of inhibition of ecCTPS via filament formation, disulfides were engineered to create a filament that could be formed without the addition of product nucleotides (*i.e.* CTP, ADP)(**Fig. 4c**), and it was found to have a five-fold reduction in activity, consistent with the filament directly inhibiting activity. Products soaked into these filaments bind, as observed by cryo-EM, hence the filament structure is not occluding CTP or ADP binding sites. However, when repeated with substrates UTP and ATP, no binding was seen. It was concluded that the filament allosterically regulates ecCTPS activity by stabilizing an intrinsically lower-activity state upon incorporation into filaments, independent of CTP binding[98].

The structure of the filament of human CTPS has been determined to 6.1 Å by cryo-EM (**Fig. 4e**). These form in the presence of substrates UTP and ATP (and the allosteric effector GTP), but not CTP and ADP products, exactly the opposite of ecCTPS[98]. The structure also shows stacked tetramers, however with different assembly contacts, tetramer interfaces, and protomer conformations than the bacterial CTPS filament[98]. Filament assembly is mediated primarily via an insertion in the GAT domain that appeared early in eukaryotic evolution. But like ecCTPS, filament formation is driven by binding to allosteric effectors, however with the opposite result on activity, enzyme conformation, and with different filament architectures (**Fig. 4e**). A mutation in the GAT domain insert that mediates filament interactions between tetramers, H355A, forms the native tetramer but not the filament, and shows a 6-fold reduction in activity. This is consistent with the filament having the function of holding the enzyme in an active conformation. The active conformation appears to be the result of a 10° rotation between the GAT and AL domains, relative to all previously reported structures, which results in a tunnel that could be used for ammonia transfer between the two active sites. The rotation connected to filament assembly since it is also necessary for positioning the GAT domains for interactions between CTPS tetramer in the filament.

Both types of filament appear to have ~3-4 CTPS tetramers per turn **Fig. 4e)**. The ecCTPS associates via interdigitation of stacked Xs (each X is a CTPS tetramer), while hCTPS associates by Xs interacting at their “tips”. The “X” is a little different in the two as well, as it is somewhat elongated in one direction (that which coincides with the filament axis) and compressed in the other (by 5-6 Å each). Both polymerize via their GAT domains. The short helical insertion in human CTPS provides the primary assembly interface and also prevents the formation of ecCTPS-like contacts[98].

What remains unclear is the relative advantage of stabilizing hCTPS in an active conformation in the filaments. It could be to keep the bulk of the enzyme under conditions of cellular stress in a state primed for maximal activity immediately on return to normal growth conditions.

***Inositol Monophosphate Dehydrogenase (IMPDH)***

The precise regulation of nucleotide biosynthesis is crucial for normal cell metabolism[110]. Like CTPS, which forms the rate limiting metabolite in the *de novo* synthesis of CTP, IMPDH catalyzes the rate limiting step for the synthesis of GTP (**Fig. 5a)**. The substrate of IMPDH, inositol monophosphate (IMP), is generated by a large complex of related enzymes called the purinosome[111-112]. However, IMPDH is not part of the purinisome, perhaps why it forms super-structures of its own. IMPDH is an important drug target (antiviral, antiparasitic, antileukemic, antibacterial, and immunosuppressive[113-117]) because inhibition causes not only a reduction of the guanine nucleotides pools, but also more importantly an imbalance between adenine and guanine nucleotides, leading to wide-ranging repercussions[117]. Humans code for two differentially expressed IMPDH genes (IMPDH1 and IMPDH2)[118], and both have been shown to assemble into the large rod and ring shaped assemblies visible in cells as cytoophidia (**Fig. 5b**)[119-120]. In *Drosophila*, IMPDH has also been shown to bind to DNA and repress transcription of histone and E2F genes[121]. E2F is critical for the G1/S transition and DNA replication, hence this activity of IMPDH would result in slowing the cell cycle. In contrast, the enzymatic activity of IMPDH is under higher demand in proliferating cells.

IMPDH has been shown to form cytoophidia *in vivo* in mammalian cells as well as in *Drosophila*[31, 40, 122]. CTPS also forms these types of structures, and cytoophidia of CTPS and IMPDH are often are observed at the same time, often colocalizing but other times forming separate structures[40, 123-124]. Keppekke *et al.* (2015)[125] showed that CTPS and IMPDH cytoophidia were independent but related structures[124]. The sizes and locations can also be very different for the two types of assemblies, even in the same cell[125]. There is also currently no evidence that these two enzymes form mixed filaments.

IMPDH cytoophidia are found to occur spontaneously in a number of cell types, but can also be induced[46, 126]. IMPDH cytoophidia assembly is promoted by its substrate IMP[124]. They form spontaneously in mouse pancreatic islet cells in response to nutrient uptake[126]. Formation of the IMPDH cytoophidia in mouse embryonic as well as induced pluripotent stem cells is correlated with rapid cell proliferation[40, 124]. When cells launch division, they must increase nucleotide production before or during S phase to meet the needs of protein expression and DNA replication. IMPDH cytoophidia can also be induced in cells by the addition of inhibitors that impede GTP biosynthesis, such as MPA[122], ribavirin[46], or by depleting essential purine precursors[99]. Disassembly of cytoophidia can be induced by the addition of the downstream product guanosine or GTP[40, 120, 122]. Cellular IMPDH cytoophidia can also be reduced by disrupting the cell cycle through the PI3K-AKT-mTOR pathway using a PI3K inhibitor[126]. These results again support the hypothesis that upregulation of purine nucleotide synthesis promotes IMPDH cytoophidia assembly. IMPDH form abundant cytoophidia in mouse pancreatic islet cells, which correlate with insulin secretion (occurring in the fed, but not starved state)[126], and they are also developmentally regulated[97]. In summary, cytoophidia by IMPDH1 and IMPDH2 appear to be required for producing sufficient amounts of guanine nucleotides to maintain normal cell proliferation when the intracellular IMPDH level is not adequate[124].

Cellular cytoophidia are large structures, which are likely composed of bundles of filaments. Human IMPDH2 filaments have been structurally characterized by electron microscopy[127-128] (**Fig. 5c-d**). EM revealed two types of filament structures, one appearing under conditions giving rise to activated IMPDH (the octamers have an “open” configuration and they form with $NAD^+$), and the other with inactive (the octamers have a “closed” conformation and form with GTP)[128] (**Fig. 5d**). Both types of filaments are induced by ATP, and are composed of stacked octamers with 30° (open conformation) or 35.5° (closed conformation) between them[128](**Fig. 5d-e**). A domain of IMPDH, the CB domain, binds MgATP and is responsible for allosteric activation of *Pseudomonas aeruginosa* IMPDH[127], and human IMPDH without CB domain do not form cytoophidia[124].

Although the cellular superstructure known as ctyoophidia (or "rods and rings") appear to form under conditions of active IMPDH, the EM study finds that filaments may be composed of either active or alternatively of inactive enzymes. To investigate the relationship between enzyme activity and filamentation, a study was performed with mutations that either impair filamentation or result in constitutive formation of filaments[128]. It was found that wild type IMPDH and mutants that either promote or inhibit polymerization have comparable catalytic activity, substrate affinity, and GTP sensitivity. This is also true when expressed in living cells (HEK 293)[128]. Unlike other metabolic filaments (*i.e.* CTPS), which selectively stabilize active or inactive conformations, recombinant IMPDH filaments accommodate multiple states. These conformational states are finely tuned by substrate availability and purine balance[128]. Anthony, *et al.* (2017)[128] suggest that polymerization may allow cooperative transitions between states, where effectors can rapidly modulate the activity of the entire IMPDH population perhaps using a switch-like transition[128]. They also postulate that the filaments could also be a platform for other regulatory proteins, the conformational state providing a physical signal for the level of guanine biosynthetic activity[128]. Finally, given the transcriptional activity observed in *Drosophila* IMPDH, the filaments could serve to sequester IMPDH it from its nuclear functions[121, 128].

***Glutamine synthetase (Gln1)***

Glutamine Synthetase is an essential enzyme that synthesizes glutamine from glutamate and ammonium in an ATP dependent manner (**Fig. 6a**). It is found in all forms of life, and for plants, it is a critical factor in nitrogen fixation. In bacteria, it functions as a regulator of glutamine level. In higher organisms, it has critical functions in maintaining proper hormonal levels in neurons, particularly astrocytes. The enzyme catalyzes the activation of glutamate to phosphoglutamate with energy from ATP. Upon activation with ATP, the structure of the active site around the phosphoglutamate undergoes a rearrangement allowing binding of ammonium[129]. After enzymatic ammonium deprotonation, the ammonia attacks the intermediate, yielding products which may leave by opposite sides of the enzyme ring structure[129].

The various isoforms of glutamine synthase are regulated in many ways[129]. Many free amino acids inhibit the enzyme by binding the substrate site. Glycine, alanine, and serine have backbones that bind in the same orientation as the substrate glutamate. The activating nucleotide site can be inhibited by ADP, AMP, and GDP. In prokaryotes, glutamine synthase is regulated by adenylylating a critical active site tyrosine. However, all 12 protomers of the dodecamer need to be modified in order to stop the enzyme's activity. This can be an energy intensive process because the adenylylation requires ATP. In *E. coli* it is known that regulation is also carried out by specific proteins at the genetic level. Presumably genetic regulation is found in all branches of life. In eukaryotic non-brain tissue, GS is inhibited by glutamine metabolism end products. However, brain tissue does not seem inhibited this way. Filamentation may be another form of controlling activity, likely by inhibition, at the structural level and we are just beginning to investigate this.

Gln1 is an isoform of glutamine synthetase that was found in metabolic screens of yeast cells under starvation as well as heat stress conditions where fluorescent versions of the enzyme were seen microscopically in foci, rods, rings, fibrils, and related large self-assemblies (often referred to as cytoophidia)[130]. Petrovska *et al.* (2014)[101] have directly tested the effect of various fluorescent tags on Gln1 in yeast and found cytoophidia persist under starvation conditions. Although there is a propensity for GFP to induce aggregation, cytoophidia were found when Gln1 was fused to mCherry, and tetracysteine-binding fluorophores (**Fig. 6b**). The conclusion from Petrovska *et al.* (2014)[101] is that Gln1 in yeast is a cytoophidia-forming enzyme that conserves energy involved in protein synthesis by storing this important metabolic catalyst in until starvation conditions are reversed. Interestingly, the cytoophidia formation started nearly one hour (50 minutes) after advanced starvation conditions. Upon reintroduction of glucose, cytoophidia dissolved within 18 minutes and cells were available to reenter the cell cycle.

Glutamine synthetase cytoophidia were investigated separately by Narayanaswamy *et al.* (2009)[25] and Petrovska *et al.* (2014)[101]. The fluorescently labeled structures in yeast resemble punctate foci or short rods. Narayanaswamy *et al.* (2009)[25] showed that the foci formation by Gln1 was stimulated by absence of adenine in growth media, and could be reversed upon the addition of glucose. Gln1 punctate body formation was not inhibited by the addition of the protein translation inhibitor cycloheximide, while those of Gln1 were not. These foci did not colocalize with P-body marker proteins and only in a limited way with actin bodies. Pelleted purified Gln1 punctate foci could not be dissolved into soluble protein by the addition of rich media.

He, *et al.* (2009)[131] solved the crystal structure of Gln1 from yeast and found pentameric rings that stacked in the asymmetric unit (a total of 4 rings)**(Fig. 6d)**. This suggests that stacking of the rings is the basis for forming the larger self-assemblies visible in cells and by EM. There is an asymmetry to the stacking. The decamer forms from two pentameric rings that sit next to each other (*i.e.* front-to-back) with a 30° rotation between the rings. The interaction between pentamers in the functional decamer seems to be mediated by a large amount of looping structures and driven primarily by hydrophobic interactions (1900 $Å^2$) and 10 hydrogen bonds. The back-to-back interaction (1700 $Å^2$) that allows decamers to self-associate into the stack of 4 pentameric rings seen in the asymmetric unit has a 10° rotation in the same direction as that between rings in decamers. The residues in the area defining the decamer-decamer interaction seem to be highly charged which seems to mediate binding presumably in pH or charge dependent manner.

Petrovska *et al.* (2014)[101] also formed an inducible Gln1 mutant yeast strain that could be controlled. By stopping further translation of Gln1 under starvation conditions, they found that the yeast was able to respond to new supplies of glucose. This indicates that the formed cytoophidia from Gln1 could form the basis of new active enzymes once the cytoophidia dissolved. Petrovska *et al.* (2014)[101] used correlative light electron microscopy to show the foci and punctae seen in fluorescent images are in fact clusters of aligned fiber segments approximately 1 micron long (**Fig. 6e**). Further, attempts to purify the fibers from mutants or starved cells, resulted in dispersions of decamers. However, fibers could be reconstituted *in vitro* by addition of 200 mg/mL Ficoll 70 as a crowding agent.

Enzymatic activity falls 60% for mutants in low pH media or starved cells relative to well-fed wild type enzyme. This activity can be recovered for the wild type or in some mutants, by reversing cellular or *in vitro* conditions. The loss in activity seems to correlate with the extent of cytoophidia/fiber formation[101]. Gln1 has a theoretical pI of about 6. Changes in cytosolic pH would then reduce its net charge indicating a possible electrostatically governed interaction. Alternatively, protons could act as allosteric effectors changing the structure and/or interfaces. Charge flipping by the mutation E186K was seen to be important in keeping pentameric rings dispersed[101]. But then charge reversal for R23E stabilizes cytoophidia and filaments made of the same rings[101]. Thus, there may be an overall electrostatic effect, for example, increasing the negative charge stabilizes filaments and filaments, while increasing the positive charge stabilizes pentameric rings.

Petrovska *et al.* (2014)[101] tested the hypothesis that these residues are responsible for cytoophidia formation in a pH dependent manner. Interestingly, as the cells become acidic due to starvation conditions[132-134], the cytoophidia formation seems to be enhanced. In glucose containing media with well-fed yeast, the pH was dropped by addition of the chemical reagent 2,4 dinitrophenol and cytoophidia were observed. Mutants tested were in several classes. E186K, P83R, T49E were capable of preventing cytoophidia formation. Y81A had enhanced starvation-induced cytoophidia and R23E formed cytoophidia in both starvation and well-fed conditions. In fact, they found that the Y181A mutant had decreased rate of return to normal cell cycle after starvation induced cytoophidia. They concluded that the return to normalcy is limited by the rate of fiber dissolution. These cytoophidia-stabilizing mutants grew poorly in regular media but could grow normally when the media was supplemented with glutamine. Interestingly, the mixed effect T49E, R23E double mutant shows a strong resistance to cytoophidia formation indicating the relative strength of these two mutants. Imaging of the self-assemblies of purified R23E Gln1 from yeast (a mutant form that constituitively formed cytoophidia in cells[101]) using negative stain EM shows filaments that appeared to be composed of stacked rings, as was predicted by the x-ray crystal structure (**Fig. 6c**).

There are a number of advantages thought to be important for Gln1 self-association. There is likely a resistance to bulk autophagy for these enzymes. Also, having functional, oligomerized enzymes ready to react as cellular conditions revert from starvation is an advantage over having to synthesize new enzymes from nucleic acid sequences. This further conserves cellular energy as has been proposed for mammalian cells containing ADF/cofilin filaments[135-136]. In fact, it may be that there is a small amount of inactive cytoophidia that provides buffering from metabolic fluctuations (enzyme inactivation by filament formation may provide a buffer against metabolic fluctuations, thus preventing accidental re-entry into the cell cycle).

Activity in the fibrillar (cytoophidic) state seems to be reduced. It’s not clear if this is due to blocking of the active site or constraint of conformational changes necessary for activity or if there’s simply a charge interaction within the active site itself. Fibrillar disassembly presumably happens as rings dissociate. In the microscope images of Petrovska *et al.* (2014)[101] the fibers are seen to "disappear" on the micron length scale with the reestablishment of well-fed conditions or increased pH. This suggests that the unbinding of fibers from each other as well as disassembly of the stacked ring filament structure happens in concert.

Gln1 cytoophidia in yeast, then, may be a storage depot for enzymes. This is seen in other enzymes like Ura7/Ura8, fibers of Gcd2, Gcd6, Gcd7, Gcn3 and Sui2[24]. Other researchers have noticed that quiescence-associated subcellular structures, such as proteasome storage granules and actin bodies, also form in a pH-dependent manner and has concluded that the drop in intracellular pH is important for their formation[137]. The establishment of pH as an important messenger is furthered by work of Isome, *et al.*(2013)[138] who show that the Gα subunit of G-proteins acts as a pH sensor. In these systems, signaling becomes more quiescent as the pH is dropped. This strongly reinforces the notion that intracellular pH changes serve as a global messenger to signal the depletion of energy during starvation and cellular quiescence[133, 139].

***β-Glucosidase***

A plastidal β-glucosidase from *Avena sativa* (oat) is activated during fungal infections. Antifungal precursors are stored in plant vacuoles and the enzyme activates these avenacosides by hydrolysis to form antifungal 26-desglucoavenacosides. The enzyme is a globular α/β barrel with two catalytic glutamic acid residues that act as nucleophiles or as an acid/base catalyst. β-glucosidase from oat (*Avena sativa*) hydrolyzes the β-glucoside avenacoside to C26-desgluco-avenacosides. This enzyme is nominally a homohexamer that hydrolyzes β1->4 glucose bonds; it also cleaves avenacosides as an antifungal defense. Antifungal precursors are stored in plant vacuoles and the enzyme activates these avenacosides by hydrolysis to form antifungal 26-desglucoavenacosides.

It has been known since 1965 that the oat β-glucosidase forms *in vivo* fibrils named "stromacenters"[9]. There are two isomers of oat β-glucosidase, As-Glu-1 and As-Glu-2. As-Glu-1 is sufficient to form filaments but can also induce fibrillar or filamentous formation with As-Glu-2 which normally forms a homodimer[140]. This filament resists pH and non-ionic detergents but can be disassembled with $CaCl_2$[141]. Interestingly, *Hevea* β-glucosidase also forms long filamentous assemblies, but the enzyme in this species has not been studied extensively by higher resolution techniques or kinetic modeling[9, 142].

Finite sized oligomers of oat β-glucosidase, trimers, hexamers and multimers, are seen by EM[140-141]. The protein is seen as a trimeric ring that dimerizes along the ring's axis into hexamers. Larger ordered oligomers also stack along the axis of the ring into a filament (**Fig. 7B-D**). Through detailed reconstructions there seems to be a difference in the twist between neighboring trimers forming the "tunnel" housing the active sites. In the multimers, the twist between trimers is 38°, however in the non-filamented hexamers this is 60°. The smaller twist seems to result in smaller side fenestrations presumably changing access to the active site. This may explain both the reduced $K_M$ and $k_{cat}$, since a caging effect may increase binding affinity to both substrates and products[141]. In addition, the $k_{cat}/K_M$ of the multimers (*i.e.* filament) is twice that of the hexamer[141]. Hence, the caging effect may result in increased substrate discrimination, such as in preference for avenacoside substrates over other kinds of β-glucosides. Alternatively, or in addition, the physiological role of filamentation may be to create a strongly localized concentration gradient of antifungal product near the formation of the fibers of assembled filaments.

***Hydrogen-Dependent $CO_2$ Reductase (HDCR)***

The interconversion of $CO_2$ and formic acid is an important reaction in bacteria, and enzymes catalyzing the oxidation of formic acid to $CO_2$, and concomitant reduction of protons to molecular hydrogen have been known for decades[143-144]. However, only relatively recent has an enzyme capable of the reverse been characterized. This enzyme utilizes molecular hydrogen as electron donor for the reversible reduction of $CO_2$ (the first reaction of the acetyl-CoA or Wood-Ljungdahl pathway) and has been named the hydrogen-dependent $CO_2$ reductase (HDCR)(**Fig. 8A**). It derives from an acetogenic bacterium, *Acetobacterium woodii*[145]**.** This pathway is utilized for carbon fixation as well as energy conservation[145-147]**.** Reduction of $CO_2$ is also of interest in biotechnological applications such as removal of $CO_2$ from the atmosphere, hydrogen storage, and hydrogen production.

HDCR is composed of four subunits in equal stoichiometry and forms filaments with increased enzymatic activity at concentrations of 0.05 mg/ml and higher, and in the presence of divalent cations[148].

The lengths of the filaments are varied and can be more than 0.1 micron in length but with apparent bending indicating flexibility (**Fig. 8B**)[148]. Their formation is reversible and dependent on only the presence of divalent cations (*e.g.* $Mg^{2+}$, $Mn^{2+}$, and $Ca^{2+}$) at concentrations of 2-20 mM[148]. The width of the filaments is 10-15 nm, with an apparent helical structure[148], however no three-dimensional analyses or higher resolution structures of the filament, or enzyme protomers has been reported.

Light scattering studies showed that the formation of polymers/filaments follows a first order process complete after 500 s, but with a fast initial jump[148]. Enzyme activity assays indicate that the polymeric form is more active by approximately 2-fold than the non-polymerized form[148]. Hence authors speculate that filamentation may be a regulatory mechanism to activate HDCR. Alternatively, it could stabilize the enzyme, regulate its degradation, or act as a scaffold for other enzymes, making use of close proximity for electron tunneling between iron-sulfur clusters of adjacent enzymes. Finally, it may serve to protect the iron-sulfur clusters from oxidative damage[148].

***Nitrilase***

Nitrilases are a family of enzymes that convert nitriles to the corresponding acid and ammonia (**Fig. 9A**), and occasionally release an amide product (Fernandes *et al.* (2006)[149], Pace & Brenner (2001)[150]). The majority of known enzymes are from bacteria, fungi, and plants. The enzyme composition is homodimeric with a αββα- αββα sandwich fold and an active site containing four positionally conserved residues: Cys, Glu, Glu and Lys. Nitrilases are attractive biocatalysts for the production of fine chemicals and pharmaceutical intermediates and also are used in the treatment of toxic industrial effluent and cyanide remediation. Nitrilases have been visualized as dimers, tetramers, hexamers, octamers, tetradecamers, octadecamers, and variable length helices, all with the same basic dimer interface, and was known since 1977 to form large aggregates upon activation[10-11]. Self-assemblies are also visible in cells. Cutler & Somerville (2005)[21] found GFP labeled Nitrilase I in *Arabidopsis* to form "aggregates" in cells following wounding and just prior to cell death (**Fig. 9B**). One function of nitrilases *in vivo* may be in detoxification, hence why these self-assemblies (which likely contain activated nitrilases) form upon wounding with herbicides. Negative stain electron microscopy shows the fine structure of activated, self-associated nitrilase enzymes as short spirals consisting of 8 to 22 subunits or long filaments (**Fig. 9C**)[34, 151-154]. These show a left-handed spiral with about 4-5 copies of the dimer per turn **(Fig. 9D-F)**. Nagasawa *et al* (2000)[155] show that the activity of a nitrilase increased with formation of large aggregates from inactive homodimers (presumably the helical filaments). This aggregation could be induced by addition of benzonitrile, as well as the presence of salts and organic solvents (the best conditions were 10% saturated ammonium sulfate and 50% (v/v) glycerol, and by preincubation at increased temperatures or enzyme concentrations). Thuku *et al* (2007)[152] found that the helical filamentous form could be formed by truncation of the C-terminal 39 amino acid residues, perhaps indicating their involvement in regulating activation of the enzyme via helical filament formation. Interestingly, Woodward *et al* (2008)[34] found that the substrate specificity of plant nitrilase complexes is affected by the helical twist of its filamentous state. Previously, attempts to swap specificities between nitrilase enzymes by swapping active site residues failed. However, helical twist and substrate size were found to be correlated among different nitrilase enzymes, and when binding pocket residues are exchanged between two nitrilases that show the same twist but different specificities, their specificities could be swapped. Modifying nitrilase helical twist, by either exchanging an interface residue or by imposing a different twist without altering any binding pocket residues, changes substrate preference. This is significant, as it shows that one function of filament formation is in determining substrate specificity among evolutionarily related enzymes (**Fig. 9G**). In addition, Thuku *et al.* (2009)[156] speculate that helix could forms docking sites for other associated enzymes and that the central hole channels substrates and products.

***CoA-Dependent acetaldehyde and alcohol dehydrogenase (AdhE)***

The AdhE enzyme is type of enzyme is found only in bacteria and photosynthesizing unicellular organisms. AdhE catalyzes the conversion of an acyl-coenzyme A to an alcohol via an aldehyde intermediate, which is coupled to the oxidation of two NADH molecules to maintain the $NAD^+$ pool during fermentative metabolism (**Fig. 10A**). AdhE from *Escherichia coli* is a homopolymer of 96-kDa subunits harboring three $Fe^{2+}$-dependent catalytic functions: acetaldehyde-CoA dehydrogenase, alcohol dehydrogenase, and pyruvate formatelyase (PFL) deactivase. AdhE forms helical structures in cells (as well as *in vitro*) call "spiralosomes". Spiralosomes were first identified by Kawata, *et al.* (1975)[157] in *Lactobacillus fermti F-4*, but have been identified in several bacterial species since (see Laurenceau *et al.* (2015)[158])(**Fig. 10B**). Having confirmed the identity of spiralosomes as composed of the enzyme AdhE, Kessler *et al.* (1992)[22] describe their structure as a left-handed helical assembly of 20-60 subunits in rods of 45-120 nm in length (see **Fig. 10C-D**, from Laurenceau *et al.* (2015)[158]). They also show that subunit packing is widened along the helix axis when $Fe^{2+}$ and NAD are present, which is accompanied by a change in width and pitch consistent with opening and stretching of the helix. Extance *et al.* (2013)[159] purified recombinant *Geobacillus* AdhE and found it to form large assemblies refractory to crystallization. Crystallization of the ADH domain however was successful. By combining this structure with a homology model of AldDH they were able to propose a molecular model for the AdhE spiralosome (**Fig. 10E**). This model has 7 copies per turn with right handed helical geometry. Though it is not known what effect on enzyme activity this structure has, or if its structure is modulated in the cell, Extance *et al.* (2013)[159] propose that it may enhance catalytic efficiency through substrate channeling of the reactive acetaldehyde intermediate, or alternatively, provide some stabilization of the protein. Kessler *et al.* (1992)[22] hypothesize that the two forms visualized in their study represent the inactive, closed helical form, and the open, active form.

***Glutamate dehydrogenase***

Glutamate dehydrogenase (GDH, EC 1.4.1.2) is found in all living organisms and catalyzes the reversible oxidative deamination of L-glutamate to 2-oxoglutarate using $NAD(P)^+$ as a coenzyme[160] (**Fig. 11A**). In bacteria, the ammonia produced by GDH is assimilated to amino acids via glutamate and aminotransferases[161]. In plants, the enzyme can work in either direction depending on environment and stress[162-163]. Transgenic plants expressing microbial GDHs are improved in tolerance to herbicide, water deficit, and pathogen infections[164]. GDH from animals, but not other kingdoms, is allosterically regulated by a wide array of ligands[165-171]. Mammals encode two important GDHs, GLUD1 and GLUD2. GLUD1 is a mitochondrial matrix enzyme and plays a key role in nitrogen and glutamate metabolism and energy homeostasis. It is allosterically activated by ADP and inhibited by GTP and ATP. This dehydrogenase is expressed at high levels in liver, brain, pancreas and kidney, but not in muscle. In the pancreatic cells, GLUD1 is thought to be involved in insulin secretion mechanisms. In nervous tissue, where glutamate is present in concentrations higher than in the other tissues, GLUD1 appears to function in both the synthesis and the catabolism of glutamate and perhaps in ammonia detoxification. Two clinical conditions are associated with mutations in GLUD1: familial hyperinsulinism, characterized by hypoglycemia that ranges from a severe neonatal-onset, difficult to manage disease to a childhood onset disease with mild symptoms and difficult todiagnose hypoglycemia, and hyperammonemia/ hyperinsulinism, associated with mild to moderate hyperammonemia and with relatively mild, late-onset hypoglycemia[172-173]. These mutations affect regulation by GTP, causing failure to be inhibited by GTP. It's link to insulin has made it a target for drug discovery for activators to increase insulin secretion in patients with diabetes[174-176]. GLUD2 is localized to the mitochondrion and acts as a homohexamer to recycle glutamate during neurotransmission[177].

Over 60 years ago, GDH purified from bovine liver was shown to form polymers *in vitro*. Olson & Anfinsen (1952)[178] used sedimentation equilibrium to show it behaved anomalously, sedimenting at much larger molecular weights than expected. Eisenberg & Reisler (1971)[12] used light scattering to show evidence of polymerization. Huang & Frieden (1972)[13] showed how ligands affect polymerization; GTP, a known inhibitor, causes depolymerization, and ADP, a known activator, can mitigate this effect[13]. Josephs & Borisy (1972)[14] investigated the structure of assemblies of glutamic dehydrogenase and found that they formed linear polymers that further assembled into a helical tube, with 4 linear chains lining the tube (**Fig. 11B**). Each linear polymer chain was found to be inclined to the tube axis at an angle of 28.5°±1.5°, with a pitch of ~800Å, with 9-9.3 molecules per turn. However, at least 3 variants of this geometry were also found. Zeiri & Reisler (1978)[15] studied the catalytic activity of the polymers, although the relationship between them remained unclear. It is now known that the catalytically active form of GDH is a homohexamer of composed of six 54 kD subunits[179-181], which can also associate further into filaments[180-182]. Binding of leucine, ADP, succinyl-CoA, or BCH to the allosteric sites increases GDH enzyme activity and polymerization of its polypeptide chain, while binding of GTP or palmitoyl-CoA to these sites decreases GDH enzyme activity and causes dissociation of the polypeptide chains from one another[180-181, 183]. The activator and inhibitor sites are overlapping[180-181], consequently, for example, leucine can displace GTP from the allosteric sites and activate the enzyme[184]. Despite all these studies, the role of GDH polymerization on enzyme regulation remains elusive. Finally, Shen *et al.* (2016)[26] (re)discovered this phenomenon (although on a much larger scale and *in vivo*) using GFP labeled GDH, which occurred upon nutrient starvation (**Fig. 11C**), sparking interest in this phenomenon once again.

***Glutaminase***

Glutaminase (EC 3.5.1.2), also known as L-glutamine amido hydrolase, catalyzes the cleavage of glutamine to ammonia and glutamate (**Fig. 12A**). Olsen and coworkers[16] purified this enzyme from pig renal cortex and found it to consist of three forms: T, P and PB forms. After purification from pig kidney, the enzyme was in the P-B form. It was converted to the T form by dialysis overnight against 10 mM Tris-HCl (pH 8.0) containing 3 mM EDTA, and recovered in the supernatant fraction following centrifugation at 80,000 g for 1 hour. The T form was converted to the P form by dialysis overnight against phosphate buffer, 50 mM (pH 8) containing 3 mM EDTA. The P form was further converted to the P-B form by addition of sodium borate, 15 mM (pH 8) to the enzyme in phosphate buffer. The T form (with a molecular weight estimated at 140 kD by sedimentation velocity, likely a dimer) is nonpolymerized and has low to no activity. The P-B form was found to be composed of long, double-stranded left-handed helical structures of various lengths, with a diameter of ~12 nm. The diameter of each of the two strands composing the helix is about 3 nm, and the length of each half-period along the helical axis is about 22 nm. Helices with 2-3 turns, as well as 25-30 turns were observed. The P form also displayed helical structures in negative stain EM, but were much shorter than the PB form (produced by adding phosphate-borate buffer. Most were 4-5 turns.

Since this early work many x-ray crystal structures of glutaminase have been revealed. DeLaBarre *et al.* (2011)[185] published the first full length structure of human glutaminase (**Fig. 12B**). The structure shows a homotetramer with C4 symmetry. There is now considerable interest in glutaminase as a drug target to treat cancer because glutamine from blood can be converted to glutamate, which can be oxidized to α-ketoglutarate to feed the TCA cycle or be used for the biosynthesis of several other amino acids and lipids[186-187]. In humans, there are two distinct genes encoding glutaminase enzymes: GLS1 and GLS2. The GLS1 enzyme is expressed at high levels in kidney and brain tissue. In the kidney, GLS1 is believed to maintain acid−base balance during metabolic acidosis[187-188].

GLS2 is predominantly expressed in the liver where it provides nitrogen for the urea cycle. In the central nervous system, GLS1 activity is proposed to generate a significant amount of the total neuronal glutamate pool that in turn acts as an excitatory neurotransmitter[189-191]. GLS1 activity may therefore be essential for the optimal function of multiple central nervous system glutamate receptors and their downstream roles in both the normal and pathological brain. In the peripheral nervous system, GLS1 expression and activity in the dorsal root ganglia have been proposed to generate glutamate pools involved in inflammatory pain, suggesting a role for glutaminase inhibitors in the modulation of nociceptor function[192]. Finally, HIV associated dementia has been linked to upregulation of glutaminase activity and subsequent glutamate excitoxicity derived from HIV-infected macrophages[193-194].

In 2013, Ferreira and coworkers[195] published a detailed work describing the filamentous form of human glutaminase C (**Fig. 12C-E**). They propose it to be the most active form of the enzyme, composed of a right handed helical double-stranded helix with seven tetramer copies per turn per strand and interacting via the N-terminal domains (53±2 nm rise per turn, strand inclination of 25°, and an average width for a single strand of 6.6±0.7 nm, estimated resolution of model is 35 Å). They identify a lysine which becomes acetylated *in vivo* that down-regulates the enzyme by disrupting the filament. Previous models of enzyme activation involved promotion of inactive dimers into active tetramers by binding to phosphate, however, the inhibitor BPTES stabilizes the tetrameric form with a rigidly open gating loop at the active site. Ferreira *et al.* (2013) found that BPTES stabilizes the tetramer but disrupts the filamentous form, thereby explain the mechanism of inhibition as disruption of the filament. They also found that a mutation found in cancer cells (K325A) leads to enzyme much more prone to form filaments and therefore is hyper-activated.

***β-lactamase-like protein (LACTB)***

LACTB is a 55 kDa homolog of bacterial penicillin-binding proteins found in the intermembrane space of mitochondria of mammals[196-197]. Though its enzymatic activity is not known, homologues are serine proteases and all important motifs of such enzymes are preserved in LACTB [196]. Studies have found a causal relationship between LACTB and obesity[198]. LACTB has been shown to polymerize into stable filaments both in vitro and in the mitochondrial intermembrane space (**Fig. 13A**). These may be important for submitochondrial organization and metabolon organization (Plianskyte (2009)[199]). Analysis of the filaments suggests that LACTB forms tetramers that further oligomerize into the observed polymers. In mitochondria, the LACTB filaments are found tethered to the inner membrane but spanning intracristal regions of the intermembrane space (**Fig. 12B-C**). The purpose of the filaments, or function of LACTB, are not currently known.

***2-Cys Peroxiredoxins (2CPrxs)***

2-Cys Peroxiredoxins (2CPrxs) are thioredoxin-dependent peroxidases that catalyze the reduction of hydrogen peroxide using conserved active-site cysteine residues (**Fig. 14A**). These enzymes are found in archea, bacteria, and eukaryotes. Hydrogen peroxide produced in cells can be toxic although is also an important second messenger, hence its levels must be tightly controlled[200-201]. A functional switch from the detoxification/peroxiredoxin activity to ATP-independent molecular chaperone activity can be triggered by oxidative stress, site-specific phosphorylation, or high temperature in many 2CPrxs[202](**Fig. 14A**). Both states have a role in human disease, from resistance to neurodegenerative disorders to host-pathogen interactions, up to the onset of cancer[203-204]. Foci of 2CPrx form in mouse epithelial cells under oxidative stress[201] (**Fig. 14B)**.

Reduced 2CPrx forms homodimers that can associate into toroidal decameric rings during the peroxiredoxin catalytic cycle (**Fig. 14C**)[200, 205]. Hyper-oxidation triggers stacking of decamers into filaments, which stabilizes the decameric structure (**Fig. 14D**), simulated by substitution of active site cysteines with serine (**Fig. 14E-F**)[20082,188]. The filaments appear as tubes with a central cavity that binds unfolded proteins[206-207]. **Fig. 14G** shows a theoretical model of filamentous 2CPrx based on x-ray crystal structures of the decameric ring forms of the enzyme[207]. Hence, 2CPrxs are also redox sensitive holdases, responding to redox stress by filament formation which then sequesters unfolded proteins and thereby prevents protein aggregation[208].

***Ribonucleotide reductase (RNR)***

Ribonucleotide reductases are responsible for all *de novo* biosynthesis of DNA precursors in nature by catalyzing the conversion of ribonucleotides to deoxyribonucleotides (**Fig. 15A**). Hence human RNRs are targets for anti-cancer drugs. Both class 1a (from human) and class 1b (from *Bacillus subtilis*) RNRs have been shown to form various oligomeric forms, including monomers, dimers, hexamers, and filaments[209-211](**Fig. 15B**). Punctate foci of RNR (both subunits, Rnr2 and Rnr4) were also discovered in yeast upon nutrient starvation[25]. The *B. subtilis* RNR is composed of NrdE and NrdF (α and β subunits, respectively), both being required for activity. The structures of various oligomeric states have been investigated for the class 1b RNR from *B. subtilis* by SAXS, x-ray crystallography, and cryo-EM[210-211]. Three distinct dimers (NrdE S type and I type dimers, and NrdF dimers) were characterized by x-ray crystallography (**Fig. 15C**)[210-211]. Two filamentous forms were investigated using cryo-EM to 4.7-4.8 Å resolution (**Fig. 15D-E**)[210-211]. Both occur under conditions of enzyme inhibition, namely high (100 µM) concentrations of dATP[210-211]. Considering only the NrdE subunit of RNR, a right-handed double helical filament is found which contains alternating dimeric interfaces (**Fig. 15D**, closeups of I and S type dimers shown in **Fig. 15Ci, iii**). The presence of the NrdF dimers results in a filament with a single NrdE filament bound by NrdF dimers (**Fig. 14E**). A model for the allosteric regulation of RNR is shown in **Figure 16.**

***Recombinase A (RecA)***

RecA (and its eukaryotic homologue Rad51) are responsible for strand exchange in homologous recombination, and process important for DNA repair, viral integration, and recombination of DNA segments between homologous DNAs as in meiosis[212]. These functions have roles in generating genetic diversity and preventing DNA damage that could otherwise result in diseases such as cancer. RecA forms filaments upon binding to single stranded DNA and ATP, which then binds double stranded DNA, and through stretching the bound DNA, allows for homology searching by facilitating base pairing between the single and melted double stranded DNA duplex (**Fig. 17A**)[213-214]. Filament formation by RecA is highly cooperative. Hydrolysis of ATP releases the bound DNA and results in filament dissociation. RecA filaments have been characterized by electron microscopy (**Fig. 17B**) and x-ray crystallography (**Fig. 17C**). The x-ray crystal structure of short filaments of RecA (limited to 4-6 RecA protomers) bound to DNA (single stranded in one structure, double stranded in another) and a non-hydrolyzable ATP analogue have been solved to 2.8 and 3.15 Å[214]. These represent active filament structures, but the filament can also assume a separate, inactive structure, which was also characterized to 4.3 Å resolution[214]. In all cases, the filaments are right-handed with straight helical axes, ~6 RecA per turn and 94 Å pitch. The DNA is bound inside the helix formed by the RecA proteins, closer to the helical axis. The ATP analog binds at the interface between adjacent RecA protomers, explaining the ATP dependency of DNA binding and the dissocation from DNA upon ATP hydrolysis. The DNA is stretched globally, but adopts B-DNA-like conformation in the three nucleotides (or base pairs) bound to each RecA protomer, with stretching between. This stretching is thought to destabilize the donor duplex allowing local melting and sampling of base pairing by the original single stranded DNA[214].

### *Chromosomal replicator initiator protein (DnaA)*

DnaA and its homologues in bacteria are required for the initiation of replication[215-217]. DnaA possesses a sequence specific DNA binding domain, which recognizes specific sequences in the bacterial origin, and is also a AAA+ ATPase which binds and hydrolyzes ATP. DnaA binds to the sequences at the origin and is responsible for inducing strand opening (**Fig. 18A**) at a nearby sequence, which then allows for loading of the replicative helicases DnaB (**Fig. 18B**). The crystal structure of DnaA bound to single stranded DNA shows a right handed helical filament that binds DNA on the interior of the spiral, near the helical axis[218] (**Fig. 18C**), reminiscent of the structure formed by RecA (**Fig. 18D**). Similar to the mechanism of RecA, the DNA is stretched from B-DNA helical parameters which is thought to be important for destabilization of the double helix. The ATPase domain is responsible for protein-protein interactions forming the spiral, as well as making non-sequence specific contacts to the DNA (**Fig. 18B**). The analogue of ATP found in the crystal structure binds at the interface between neighboring DnaA protomers, similar to ATP binding to RecA filaments. The helical filament contains 8 copies of the single chain DnaA protomer per turn (45° per protomer), with a pitch of 178 Å. Also like RecA, this filament is polar, meaning the two ends of the filament are not equivalent[218].

### *SgrAI restriction enzyme (SgrAI)*

SgrAI is a type II restriction endonuclease from the bacterium *Streptomyces griseus*[219]. Restriction endonucleases protect their host cells from invading DNA, such as the bacterial viruses known as phage[220]. They are also considered to be coded by “selfish DNA” as their genes, as well as those for their cognate methyltransferases and possibly other related proteins have been shown to evolve through horizontal gene transfer from organism to organism[221]. They not only impart this benefit of innate immunity, but also other functions including genetic recombination, nutrition, and genome evolution among others[222]. Like most type II restriction endonucleases, SgrAI cleave particular sequences in double stranded DNA in a $Mg^{2+}$ dependent reaction (**Fig. 19A**)[219]. The primary DNA recognition sequence of SgrAI is shown in **Fig. 19B**, CRCCGGYG, where R represents A or G and Y represents C or T[219]. SgrAI also cleaves a second type of site, secondary sites, with sequences CRCCGGGG or CRCCGGY(A or C or T)[35]. However, these are only cleaved if the primary site is also present, either on the same contiguous molecule of DNA, or at high concentrations[35, 223-226]. SgrAI also cleaves primary site DNA faster when there is more than one copy on the same contiguous DNA (or in high concentrations).

SgrAI forms filaments when bound to primary site DNA sequences[17, 36-37], and its structure has been characterized to 3.5 Å resolution by helical reconstruction and cryo-electron microscopy[39]. The filament is a left-handed helix with approximately 4 copies of the SgrAI/DNA complex per turn (**Fig. 19C)**. It is stabilized by protein-protein and protein-DNA interactions between neighboring SgrAI/DNA complexes. Compared to the structure of non-filamentous SgrAI, a ~10° rotation is found between the two chains of the SgrAI dimer[39]. In the structure of the non-filamentous state, determined by x-ray crystallography, one of the two $Mg^{2+}$ is displaced from its expected position, providing an explanation for the low DNA cleavage activity[227-228]. To accommodate the rotation between subunits in the filamentous conformation, many residues at the dimeric interface shift position which is hypothesized to stabilize a more optimal position of the misplaced $Mg^{2+}$ (ref.[39]).

The working model for filament formation and activation of SgrAI is shown in **Figure 19D**[39]. SgrAI bound to DNA is in equilibrium between two conformations, a low activity T state and a high activity R state. Only the R state forms filaments. Filament formation in turn stabilizes the R state. In the absence of filament formation, the T state is favored, however the R state is populated sufficiently to self-assemble into filaments under certain conditions. When bound to the primary site, the R state is sampled sufficiently to form filaments if concentrations are also sufficient (**Fig. 19E, top**). When bound to the secondary site, the R state is less favored, and therefore filamentation is far less favored and secondary sites will not be cleaved (**Fig. 19E, middle**). Yet in the presence of filaments of SgrAI bound to the primary site, SgrAI bound to secondary sites will join, upon occasional sampling of the R state (**Fig. 19E, lower**). Structural studies suggest that the reason the R and T states differ energetically with the two sequences is due to sequence dependent DNA structural energetics. Base stacking is different in the R state conformation (in the filament), which is predicted to favor primary site sequences. Hence the origin of the differential DNA cleavage activity on the two types of sequences, primary and secondary, may originate in DNA structure and energy with filamentation serving as the means to detect this energy. In the filament, SgrAI cleaves both types of sites rapidly resulting in a 200 fold acceleration in the case of primary sites, 1000 fold in the case of secondary[38]. The expansion of DNA cleavage activity from primary to primary and secondary increases the number of cleavage sequences from 3 to 17.

Global kinetic modeling of filament association, dissociation, and DNA cleavage FRET reaction data indicate that the rate limiting step of the activated DNA cleavage reaction is filament association[51-52]. The second order rate constant is several orders of magnitude lower than diffusion limited (*i.e.* $10^5$ $M^{-1}s^{-1}$). Hence to overcome this slow association rate constant, high concentrations (100 nM or greater) of SgrAI/DNA are needed (and in the R state conformation). In the cell, this limitation is overcome by the local concentration effect of recognition sequences on the same contiguous DNA[50]. Hence sites on the same DNA appear to be at a higher concentration, relative to each other, than they actually are in the cell and are more likely to form the filament, stabilize the activated conformation of SgrAI, and become cleaved. SgrAI bound to sites on separate DNA molecules are far less likely to associate into a filament, since concentrations of DNA in the cell are low (~3 nM). Primary sites in the host DNA are protected by methylation, but secondary sites are not. SgrAI will not bind to methylated primary sites, but does bind to the secondary sites tightly (nanomolar affinity). Invading DNA will contain both types of sites, unmethylated, and SgrAI in the cell binds both types. SgrAI bound to the primary sites induces the filament, which then draws in SgrAI bound to secondary sites, but only those on the same DNA molecule, owing to the local concentration effect[50]. This effect then is useful to focus or sequester enzyme activity on only substrates of interest, in this case the invading phage DNA and not the host DNA.

The global kinetic model also allowed simulations to be performed to predict SgrAI behavior under *in vivo* conditions[50]. Compared to a non-filament mechanism (Binary, with all the same rate constants but limited to an assembly of 2 enzymes, **Fig. 20**), the Filament-forming mechanism is superior in both speed (*e.g.* the rate of DNA cleavage), as well as the ability to sequester its damaging DNA cleavage activity on invading phage DNA (and away from host DNA, **Figs. 19-20**). **Fig. 19A** shows why the Filament mechanism is superior in speed over the Binary. The Filament mechanism has multiple ways for enzymes to assemble (**Fig. 20A, left**), whereas the Binary mechanism has only a single association configuration (**Fig. 20A, center**). **Fig. 20B** shows that the filament mechanism is superior to the Binary mechanism in cleavage of secondary sites (golden boxes, **Fig. 20B**), because no competition occurs between enzymes bound to the two types of sites. SgrAI bound to primary sites (blue boxes, **Fig. 20B**) induce filament formation, and SgrAI bound to secondary sites (gold boxes, **Fig. 20B**) join, leading to rapid DNA cleavage of all sites (**Fig. 20B, upper**). In the case of the Binary mechanism, association of enzyme bound to primary sites is preferred (**Fig. 20B, lower**), and outcompete association with secondary site bound enzymes, and are therefore cleaved far slower than in the Filament mechanism (compare red to green line, **Fig. 20B, right**).

The Filament mechanism is also superior in sequestering enzyme activity (**Fig. 21**). In this figure, the *in vivo* role of SgrAI is depicted (**Fig. 21A-C**). SgrAI binds to primary and secondary sites on the phage DNA (**Fig. 21A**) and form the filament with activated DNA cleavage (**Fig. 21B**). However, due to the slow association rate constant of enzymes into the filament (*i.e.* $10^5$ $M^{-1}s^{-1}$), and the intrinsically low concentration of DNA in the cell (estimated at 3 nM), but high local concentration of SgrAI on the same contiguous DNA (estimated >300 times higher[50]), SgrAI bound to secondary sites on the host DNA (blue arrow, **Fig. 21B**) do not assemble into the filaments (rate of association = $10^5$ $M^{-1}s^{-1}$ * 3 nM * 3 nM = 0.0009 nM/s), while those on the same contiguous DNA as the primary site (*i.e.* the phage DNA, red arrow, **Fig. 21B**) do (rate of association = $10^5$ $M^{-1}s^{-1}$ * 1 μM * 1μM = 100 nM/s). Hence, secondary sites on the host DNA, which are not methylated by the SgrAI methyltransferase, are largely protected from damaging DNA cleavage. Concurrently, secondary sites on the phage are cleaved, which is expected to increase the effectiveness against phage transcription and replication (**Fig. 21C**). Simulations performed with the two mechanisms, having identical rate constants, show this effect (**Fig. 21D-E**). Cleavage on the host is simulated with DNA concentrations of 3 nM, while that on the phage is estimated at 1 μM, the estimated local concentration between primary and secondary sites on the same contiguous DNA, based on average expected distances between sites[50]. Both mechanisms effectively sequester their cleavage activity on the phage DNA, because of the slow association rate constants of enzymes into the assembly (Filaments or Binary assemblies) and the low concentrations of DNA in the cell (see "Cleavage on host DNA" compared to "Cleavage of phage DNA", **Fig. 21D**). However, the Filament mechanism is faster (compare red to green line, "Cleavage of phage DNA", **Fig. 21D**). Yet if speed is required, a faster rate of Binary assembly could instead be evolved, rather than a complicated Filament forming mechanism with a slow association step. **Fig. 21E** shows why the Filament mechanism is yet superior; when the association rate constant for the assembly of the Binary complex is increased so that its DNA cleavage rate matches that of the Filament mechanism (a 4.5-fold increase is necessary)(green solid line for "Cleavage of phage DNA", **Fig. 21E**), much more cleavage of secondary sites on the host DNA is predicted (blue arrow, **Fig. 21E**). The reason for this loss of sequestration is due to the much higher association rate constant necessary so that the DNA cleavage rate of the Binary matches the speed of the Filament mechanism[50].

***The unfolded protein response nuclease kinase (IRE1)***

In eukaryotes, the vast majority of secretory and transmembrane proteins are folded in the endoplasmic reticulum (ER). Inositol requiring protein 1 or IRE1 (also known as Ire1) is a transmembrane ER protein that senses unfolded proteins in the ER[229], and has both kinase and RNase (and splicing) enzymatic activities (**Fig. 22A-B**). The luminal domain of IRE1 senses the unfolded proteins by binding directly to them, as well as through the release of BiP (a protein that also binds unfolded proteins). This results in the assembly of IRE1 into dimers and higher order assemblies[18, 230-234], observed as foci in live imaging (**Fig. 22C**)[18, 234-236]. The assembly of IRE1 also allows for auto-phosphorylation via its cytosolic kinase domain, and consequential activation of its RNase activity in the assemblies[18] (**Fig. 22D-F)**. The RNase activity has two consequences: nonconventional cytosolic mRNA splicing and regulated IRE1-dependent decay (RIDD)[237]. The splicing involves two cleavages followed by ligation (by a separate RNA ligase) at conserved recognition sequences in the pre-mRNA for the transcription factor XBP1 (in metazoans) or HAC1 (in *S. cerevisiae*)[237]. Cleaved exons are then ligated by tRNA ligase in the case of HAC1[238] and RtcB for XBP1[239] generating the spliced form of HAC1 or XBP1 mRNA[240]. RIDD leads to decay of mRNAs[241-243]. XBP1/HAC1 drives a large transcriptional program to adjust the ER's protein-folding capacity according to the protein folding load in the ER lumen[238, 244-247]. The UPR is activated in cancers, viral infections, as well as neurodegenerative, metabolic, and inflammatory diseases[229, 248-249].

IRE1 engaged in RIDD must be coupled with mRNA degradation enzymes in order to prevent either translating or ligating cleavage products[240]. One report suggested that the two RNase activities, RIDD and splicing, were controlled by different protein oligomeric states, with splicing a result of the filamentous form, and RIDD the dimeric form[240]. Another suggests that the different outcomes following cleavage by IRE1 (ligation or degradation), have more to do with RNA sequence and structure, and perhaps cellular location of mRNAs[250]. mRNAs that are degraded via RIDD encode proteins to lessen the ER stress burden and in extreme cases initiate apoptosis. During the adaptive response, IRE1 conducts RIDD on mRNAs encoding ER translocating proteins to prevent further increases in protein-folding demand in the ER[251]. In metazoans, RIDD reduces the load of ER client proteins through mRNA degradation and cleavage of 28S rRNA[252], particularly of mRNA that is associated with the ER membrane[241-243, 251, 253], Also in metazoans it's been shown that 21% of RIDD substrates encode proteins associated with gene expression regulation that might also participate in the global decrease in protein production[254], and that RIDD reduces the total protein influx into the ER by 15%[243].

Mammals have two IRE1 paralogs, IRE1α and IRE1β. IRE1α is expressed ubiquitously and knockout mice exhibit embryonic lethality. IRE1β expression is restricted, and knockout mice are viable, therefore the α is thought to be more important[237]. The two IRE1 modalities (splicing and RIDD) co-exist in metazoan cells[241-242, 255], yet are evolutionarily separated in the two yeast species, *S. cerevisiae* and *S. pombe*. *S. cerevisiae* IRE1 has splicing, but no RIDD[243, 256], while *S. pombe* IRE1 has RIDD but no splicing[243]. Mammalian IRE1α is more efficient in XBP1 mRNA splicing, while IRE1β prefers to cleave (leading to degradation) ribosomal RNA[252, 257-258].

If attempts to restore ER homeostasis fail, IRE1 activates apoptosis through RIDD[251, 259], degrading mRNA for UPR target genes and miRNAs that are anti-Casp2[237]. The latter results in upregulation of Casp2, a proapoptotic protease essential for the execution of apoptosis[259]. Therefore, IRE1α is a molecular switch and apoptosis executioner during ER stress[237].

X-ray crystal structures of the luminal and cytosolic IRE1 have been reported. Those of the luminal domain show the basis for their signal dependent (*i.e.* unfolded protein binding) dimerization/oligomerization for *S. cerevisiae*[233] and human[260]. Structures of the cytosolic domain have been solved in dimeric forms (yeast[261] and human[262]), as well as a filamentous form of yeast IRE1[18]. The yeast dimeric structure shows the kinase domain dimerizing back-to-back, which places the kinase active sites too far apart for auto-phosphorylation. However, a dimeric structure of the unphosphorylated human IRE1 shows face-to-face orientation of kinase domains, well positioned to autophosphorylate[262]. The authors interpret this structure as an early state, prior to phosphorylation, afterwhich the dimer assumes the back-to-back orientation that activates the RNase activity. The filament structure shows the *S. cerevisiae* IRE1 cytosolic domains in complex with a kinase inhibitor, which is a potent activator for the IRE1 RNase[18]. The oligomeric structure positions the kinase domains perfectly for autophosphorylation and orders the RNAse domain of each IRE1 protomer. The filament creates surface for RNA binding and positioning of the splice sites. The filament is a right-handed helical filament made by the addition of the back-to-back dimers, hence a run-on oligomer (**Fig. 22F**). Each dimer is separated by a 54.2° rotation, giving 7 dimers per turn. It was found in the structure that 17 residues of the IRE1 cytosolic domain were phosphorylated, which must have occurred via autophosphorylation during expression in *E. coli.* The three phosphorylated residues important for IRE1 activation *in vivo*, Ser 840p, Ser 841p, Thr844p, are well resolved in the crystal structure[18] and are ideally placed to help IRE1 oligomerization[18].

The trans-autophosphorylation as well as binding to ADP contribute to activation of the RNase of IRE1[18]. In the oligomeric structure, the RNase domains are ordered, unlike in the dimeric structures, and the ordering appears to be due to intermolecular interactions only found in the oligomer. The ordering of the RNase domain structure is expected to activate its enzymatic activity[18]. Binding of a cofactor occurs in the open state of IRE1 kinase, shifts the equilibrium from monomers towards multimers and provides an additional, phosphorylation independent level of positive modulation for the activating transition[18]. Oligomerization of the unphosphorylated IRE1 opens the kinase domain and positions it for trans autophosphorylation. ATP enters the opened kinase and phosphorylates the activation loop in trans to lock it in the oligomerization compatible open state and to introduce a phosphate-mediated salt bridge at the interface IF3c[18]. These events provide positive feedback for oligomer assembly[18]. Cofactor binding and phosphorylation enhance the self-association properties of IRE1, but neither is strictly required[18].

***Casein kinase 2 (CK2)***

CK2 or casein kinase 2 is a highly conserved serine/threonine kinase (**Fig. 23A**) that is crucial for cell viability and is involved in cellular processes such as cell cycle control, cellular differentiation, and proliferation, circadian rhythm, apoptosis, and gene expression[263-264]. Human CK2 is a heterotetrameric holoenzyme (α2/β2) composed of two catalytic alpha subunits attached to a central, regulatory dimer of β subunits. CK2α is constitutively active, and upon interaction with CK2β is remains active but its thermostability, substrate specificity, and ability to attach and penetrate cell membranes is altered.

Conventional kinase regulatory mechanisms such as phosphorylation, dephosphorylation, or second messenger binding are not observed for CK2[265]. Instead, regulation of enzyme activity is proposed to occur by controlling the oligomeric state of the enzyme[266]. A mixture of oligomeric species including heterotetramers, ring like oligomers, and linear polymers have been observed using electron microscopy, analytical ultracentrifugation and native mass spectrometry (**Fig. 23B-C**) and found to vary as a function of ionic strength, polycations, $Mg^{2+}$ concentration, pH, and temperature[266-269]. Enzyme activity is maximal under conditions favoring the ring-like structures, and decreases with dissociation into the heterotetramers or association into filaments[266]. CK2α without CK2β is constitutively active and does not form higher assemblies[266]. Crystal structures of CK2 show a circular trimeric oligomer (1JWH[270], and 4DGL[271]) or linear polymers (4MD7-4MD9 and 4NH1[272]). Native mass spectrometry using ion-mobility mass spec and H/D exchange confirmed the mixture of oligomeric states including heterotetrameric protomers, tetrameric ring like structures, and linear polymers which are sensitive to ionic strength (with high ionic strength causing disruption of the higher order forms)[269]. The stoichiometry of the complex also changes with ionic strength $(\alpha_1,\beta_2)_1$ to $(\alpha_2,\beta_2)_1$ ratio increases with decreasing ammonium acetate concentration[269]. Formation of oligomers and aggregates was shown in cells using BRET (bioluminescent resonance energy transfer) and found to be non-static[273]. The CK2 binding protein alpha subunit of the heterotrimeric G-protein that stimulates adenylyl cyclase (Gas), as well as the polycationic compound polylysine. Gas, but not the CK2 substrate β-arrestin2, reduced the BRET signal associated with aggregation of CK2[273].

***Adenylate kinase (AK)***

Adenylate kinase performs the reaction shown in **Figure 24A**, and is found in plants and bacteria. Wild and coworkers[274] determined the crystal structure of AK from maize with an inhibitor (adenosine-(5')pentaphospho(5')adenosine (Ap5A) and found that the 24.9 kDa enzyme crystallized in a manner that produced infinite, linear "proto-rods" composed of hexamers (**Fig. 24B-D**). Based on contact surface area arguments, the authors propose that this is more than merely crystallographic packing. The authors also argue that the enzyme must be inactive in this linear assembly due to closing of a "lid" domain that cuts off access to the active site, and this "lid" domain participates in close contacts within the rod. The function of the rod, proposed by Wild *et al.* (1997)[274], is to store AK during nighttime, when photosynthesis is low and AK activity is not required. Water would also be released from the rod, which is required during the night for plants. Once daytime arrives, the AK enzymes may quickly disassemble and become active again, avoiding the need for synthesis of new enzymes. However, no further enzymological studies of the formation of rods *in vivo*, or the effect of rod formation on enzyme activity have been reported.

***Receptor-Interacting serine/threonine-protein kinases (RIP1/RIP3 kinases)***

RIP1 and RIP3 are intracellular signaling Ser/Thr kinases (**Fig. 25A**) that play important roles in immune defense, cancer, and neurodegenerative diseases. RIP1 controls whether the pleiotropic cytokine TNFα induces NF-kB activation, apoptosis, or programmed necrosis[275]. Programmed necrosis involves formation of an assembly, the necrosome, with RIP3 with activated kinase activity[276-278](**Fig. 25B**). The necrosome recruits MLKL (Mixed Kinase Domain Like protein), which is phosphorylated by RIPK3 and immediately translocates to lipid rafts inside the plasma membrane[276, 279-280]. This leads to the formation of pores in the membrane, allowing the sodium influx to increase and consequently also the osmotic pressure, which eventually causes cell membrane rupture[276]. Phosphorylation stabilizes the assembly, which is mainly formed via interactions between a short segment (5-6 residues) of each protein known as the RIP homotypic interaction motif (RHIM)[281](brown rectangle in **Fig. 25B**). This motif is surrounded by unstructured regions but forms a cross-beta amyloid structure through self-association or association of RIP1 and RIP3 with each other[281] (**Fig. 25C**). This association is critical for function of RIP1 and RIP3 kinases in programmed necrosis[281]. Amyloid is distinct in its structure and binding properties, with this amyloid shown to be composed of alternately stacked RIP1 and RIP3 RHIMS to form heterotypic beta sheets. Two such beta sheets bind together along a compact hydrophobic interface featuring an unusual ladder of alternating Ser (from RIP1) and Cys (from RIP3) residues[282]. EM shows filaments of RIP1/RIP3 with its globular domains (kinase on both and death domain on RIP1) extending outward from a core protease resistant amyloid filament, and these structures appear to also interact to form larger aggregates[281] (**Fig. 25D** shows a filament of RIP1 and the RHIM of RIP3). These are also seen from RIP1/RIP3 complexes purified from cells undergoing programmed necrosis, but not controls, where foci are seen (**Fig. 25E**). The amyloid is remarkably resistant to denaturation, requiring 150 mM NaOH to disrupt[281]. The RHIM, which mediates amyloid formation, is found in a growing number of signaling adaptors with crucial functions in cell death and innate immunity[275]. Amyloids are fibrous protein aggregates composed of cross-beta structures and associated with many neurodegenerative[283] and infective prion diseases[284]. Amyloids can perform normal cellular functions, such as host interaction, hazard protection and memory storage[283]. The purpose of amyloid/fibril formation may be for feed-forward gain of function in which kinase activation and RIP1/RIP3 necrosome formation are mutually reinforcing[281]. Upon amyloid formation, transphosphorylation occurs which activates RIP1/RIP3[281].

### *Death Domain containing proteins that form filaments*

Filament forming proteins are increasingly recognized as critical intermediaries in signaling pathways including those of innate immunity, necrosis, and apoptosis[285-286]. These pathways contain proteins which upon initial stimulation (such as binding to a pathogen associated molecule in the case of innate immunity) leads to filament nucleation, propagation, and recruitment of downstream effectors such as kinases, transcription factors, caspases, and interleukins. The subcellular structures involved in innate immunity are known as inflammasomes. Filaments are formed similarly, via the Death Domain (CARD, PYD, DED). Example proteins include RIG-I, MDA5, and MAVS)[287-288], NOD1/NOD2 and RIP2[289], TLRs, MyD88, IRAK2, and IRAK4[290], NLRPs, ASC, and Caspase-1[291-293], and Fas-FADD[294]. Not all are enzymes, but many contain protein kinase domains. **Figure 26A** shows foci formed by GFP labeled RIP2, a CARD domain containing kinase that functions in innate immunity. **Figure 26B** shows a negative stain EM image of RIP2-CARD. **Figure 26C** shows a model from a three-dimensional reconstruction of a filament composed of the PYD domain of ASC. Though the different filaments formed by death domain containing proteins share many common features, studies show that they can have different geometries, including different handedness[291, 295]. **Figure 26D** gives models for inflammasome assembly of two different innate immune pathways that involve filament formation by DD containing proteins.

### *Target of Rapamycin Complex 1 (TORC1 )*

TOR is a eukaryotic serine/threonine kinase found in TORC1 and TORC2, and TORC1 has been found to form foci in cells (**Fig. 27A**) composed of hollow, helical, cylindrical assemblies (**Fig. 27B-C**) upon glucose starvation[296]. The kinase is inactive in this assembly, which forms at vacuoles, and is dependent on Rag GTPases for its assembly through some unknown process (**Fig. 27D**)[296]. Prouteau and colleagues suggest that this assembly, they call the TOROID, may have a function to inactivate an expensive multiprotein complex in a cost-effective way[296]. TORC1 functions as a nutrient/energy/redox sensor and controls protein synthesis.

## Enzymes shown to form foci, rods, or cytoophidia in cells, but assembly architecture not known

### *Glutamate Synthase (Glt1 in yeast)*

Glutamate synthase forms L-glutamate from L-glutamine and 2-oxoglutarate which is an alternative pathway to L-glutamate dehydrogenase for the biosynthesis of L-glutamate (**Fig. 28A**), and participates with glutamine synthetase in ammonia assimilation processes. The enzyme is specific for NADH, L-glutamine and 2-oxoglutarate. Both Noree *et al.* (2010)[24] (**Fig. 28B**) and Shen *et al.* (2016)[26] found that this enzyme forms foci and large rod-like structures (*i.e.* cytoophidia) in yeast cells when attached to GFP (or an HA tag, Noree *et al.* (2010)[24]). Noree *et al.* (2010)[24] performed a number of additional tests investigating the control of self-assembly formation, such as measuring the self-assemblies at different phases of cell growth (log-phase, diauxic, and saturation) and found similar self-assembly in all. Shen *et al.* (2016) found that self-assembly formation increased in diauxic and saturation in both quantity and length. Noree *et al.* (2010)[24] found that the Glt1p self-assemblies did not colocalize with other self-assembly forming enzymes (Ura7p or Ura8p (both CTPS), Psaip (GDP-mannose pyrophosphorylase), Gcd2p (eIF2b-delta), Gcd6p (eIF2b-ε, Gcd7pg (eIF2B-β), Gcn3p(eIF2b-α) or Sui2p(eIF2-α)). Gltp1 self-assembly formation was also unaffected by known regulators of prion biogenesis (RNQ1 and HSP104), nutrient deprivation, carbon source depletion, energy status of the cell (by addition of sodium azide), protein synthesis (by addition of cycloheximide), or low temperature[24].

Shen *et al.* (2016)[26] measured dynamics of Glt1p self-assembly formation and found that these structures are motile and located preferentially at the cell cortex. Some are highly dynamic, while others are relatively slow. Analysis of the dynamics indicates that the self-assemblies undergo sub-diffusion, *i.e.* random diffusion within confined areas that is not driven by motor proteins.

***Asparagine Synthetase***

Asparagine Synthetase (paralogous yeast genes Asn1 and Asn2) catalyzes the synthesis of L-asparagine from L-aspartate and L-glutamine in an ATP-dependent reaction and is part of the asparagine biosynthetic pathway)(**Fig. 28C**). Shen *et al.* (2016)[26] uncovered rod shaped self-assemblies formed by this enzyme (**Fig. 28D**), and that these self-assemblies grew on going from log-phase growth to diauxic and stationary. These enzymes can form self-assemblies structures in both the cytoplasm and nucleus[26].

Zhang *et al.* (2018)[297] also studied the self-assembly formation of asparagine synthetase in yeast, and found an increase in self-assembly formation with glucose deprivation. They also investigation colocalization of the two paralogs and found that they colocalize, and that self-assembly formation of Asn2 is dependent on Asn1. Although forming distinct structures (*i.e.* cytoophidia), both asparagine synthetase and CTPS cytoophidia are found adjacent to one another in cells, suggesting a connection, possibly metabolic.

***Eukaryotic translational initiation factor (eIF2/2B)***

Noree *et al.* (2010)[24] found proteins of the eIF2/eIF2B genes (Gcd2 (eIF2B-δ), Gcd7 (eIF2B-β), Gcd6 (eIF2B-ε), and Sui2 (eIF2-α), Gcn3 (eIF2B-α)) formed cellular self-assemblies resembling rods (with either GFP or HA tags). These authors also tested for colocalization, using representative subunits of eIF2 and eIF2B (Gcd2 or eIF2B-δ and Gcd7 or eIF2B-β) and found that these do not colocalize with other self-assembly forming proteins (CTP synthase, glutamate synthase, GDP-mannose pyrophosphorylase) but do colocalize with each other (and Gcd6/eIF2B-ε)(**Fig. 28E**). They found that eIF2/2B self-assemblies are not affected by known regulators of prion biogenesis (RNQ1 or HSP104), declined somewhat in saturated cultures (although varied by type of subunit), were not affected by carbon source depletion, or sodium azide (affecting energy status of the cell), were affected by an inhibitor of protein synthesis (cyclohexamide)(see also Campbell *et al.* (2005)[298]), and were not affected by lowering the temperature to 0°C or addition of the kinase inhibitor staurasporine. Shen *et al.* (2016)[26] confirmed the findings of Noree *et al.* (2010)[24], and discovered self-assembly formation by an additional eIF2B subunit, Gcd2 (eIF2B-δ). The eukaryotic translation initiation factor eIF2 mediates the first step of translation, by binding to the initiator met-tRNA and GTP and recognizing the start codon in an mRNA (Dever *et al.* (1995)[299]). GTP is hydrolyzed by eIF2 to GDP after association with the ribosome, which in turn initiates translation. Regeneration of the active, GTP bound form of eIF2 is performed by the nucleotide exchange factor (GEF) eIF2B. Therefore, self-assembly formation by eIF2/eIF2B prevents it from doing its job, no doubt, which is initiation of translation. Confirming this, Nuske, *et al.* (2018)[300] found that starvation induced acidification of the cytosol causes self-assembly of eIF2B (Gcn3/eIF2B-α) and downregulation of translation that is independent of the kinase Gcn2.

***GDP-mannose pyrophosphorylase (Psa1)***

The yeast gene Psa1, or GDP-mannose pyrophosphroylase (reaction catalyzed by this enzyme shown in **Fig. 28F**) was found in Noree *et al.*, (2010)[24] and Shen, *et al.* (2016) to form self-assemblies in yeast cells during stationary phase (*i.e.* nutrient starvation)(**Fig. 28G**). Noree, *et al.* (2010)[24] performed several additional tests and found that these self-assemblies were distinct from the others (Ura7p, Glt1p, and eIF2/2B) tested. Like Glt1p self-assemblies, the Psa1p self-assemblies were no affected by knockouts of prion formation requiring proteins RNQ1 and HSP104, nor were they affected by overexpression of HSP104. Self-assemblies of Psa1p were significantly increased as nutrient starvation increased. Carbon source depletion had no effect on Psa1p self-assemblies. They were induced by treatment with sodium azide (affecting energy status of the cell). Cyclohexamide, or reduction in temperature to 0°C had no effect, but addition of the kinase inhibitor staurosporine greatly increased the number of cells with self-assemblies.

### *Xanthene oxidase*

Xanthene oxidase from mammals is important in purine catabolism and nitrogen metabolism (**Fig. 29A**). A report in 1987 (Angermuller *et al.* (1987)[301]) indicated that self-assemblies from of homotetrameric rings of this enzyme form *in vitro*. No studies have been reported of the effect on enzyme function, or formation of such structures *in vivo*.

### *Purinosome*

Enzymes comprising the "purinosome" are responsible for *de novo* synthesis of purines (**Fig. 29B**) and include: phosphoribosyl-pyrophosphate amidotransferase (PRPPAT, step 1, **Fig. 29B**), Trifunctional GART enzyme (GAR Tfase, steps 2-3 and step 5, **Fig. 29B**), FGAMS/PFAS enzyme (step 4, **Fig. 29B)**, Bifunctional PAICS enzyme (steps 6-7, **Fig. 29B)**, ADSL enzyme (steps 8, **Fig. 29B)**, and Bifunctional ATIC enzyme (steps 9-10, **Fig. 29B**). These have been shown to form reversible cytoplasmic foci under purine starvation (HeLa cells[111], **Fig. 29C**) and general nutrient starvation (yeast[25, 112]) as well as in the female germ line in Drosophila[29]. Narayanaswamy *et al.* (2009)[25] showed that the foci formation by Ade4 (*i.e.* (PRPPAT)) was stimulated by absence of adenine in growth media, and could be reversed upon the addition of adenine. Ade4 punctate body formation was inhibited by the addition of the protein translation inhibitor cycloheximide. These Ade34 foci did not colocalize with P-body marker proteins and only in a limited way with actin bodies.

### *Pyruvate kinase*

Pyruvate kinase in yeast (gene Cdc19) is involved in carbohydrate metabolism (**Fig. 30A**) and found to form reversible punctate foci in yeast under starvation conditions[25], as well as glucose starvation, heat shock, and hypoxia[33, 302]. Pyruvate kinase also colocalizes with phosphofructokinase-1 (PFK) and fructose-bisphosphate aldolase in the self-assemblies also called G-bodies[33]. Formation of the large assemblies is postulated to depend upon a low compositional complexity region (LCR) within Cdc19, as they are prevented by tetramer formation, which sequesters the LCR[302]. The tetrameric form is favored by binding to fructose 1,6-bisphosphate. Cdc19 aggregation appears to trigger its localization to stress granules and modulates their formation and dissolution. Reversible aggregation is thought to protect Cdc19 from stress-induced degradation, thereby allowing the cell cycle restart after stress[302]. Addition of glucose causes rapid dissolution of the aggregates, and did not depend on protein synthesis (*i.e.* cycloheximide had no effect). The assemblies were also unaffected by inhibitors of the proteasome or vacuolar protein degradation.

### *G-bodies*

Enolase, (yeast-gene Eno2)(**Fig. 30C, E**) and fructose bisphosphate aldolase (yeast gene Fba1)(**Fig. 30D-E**), as well as phosphofructokinase (yeast genes Pfk1 and Pfk2) and pyruvate kinase (see above), all part of carbohydrate metabolism, form assemblies known as G-bodies under hypoxic conditions[33]. G-bodies have also been shown to be composed of other glycolytic enzymes including phosphoglucose isomerase(yeast gene Pgi1), glyceraldehyde-3-phosphate dehydrogenase (GAPDH) (yeast gene Tdh3), tetrameric phosphoglycerate mutase (yeast gene Gpm1), enolase I (yeast gene Eno1), synthase subunit of trehalose-6-P synthase/phosphatase complex (yeast gene Tps1), beta subunit of fatty acid synthetase (yeast gene Fas1), alpha subunit of fatty acid synthetase (yeast gene Fas2), and heat shock proteins (yeast genes HSP70, Ssa1, and Ssa2)[33]. Proteosomal subunits were also found to localize near G-bodies[33]. G-bodies were found to be required for increased glycolytic demand, and are distinct from stress granules or P-bodies[33]. G-bodies form specifically from hypoxia, and not glucose deprivation (in fact their formation required glucose), nitrogen deprivation, or after being grown on a nonfermentable carbon source[33]. The intrinsically disordered region (IDR) of PFK1 was found to be important for its colocalization to G-bodies, suggestive of a similarity to factors contributing to other membraneless compartments such as stress granules and P-bodies[33]. The AMP activated protein kinase Snf1 is also required for G-body formation[33, 303]. G-bodies also appear to depend on RNA, as treatment of cells with ribonucleases decreases their formation[303], hence reminiscent again of stress granules and P-bodies[33, 304].

***Mps1/Polo kinases***

MPS1 protein kinases are found widely, but not ubiquitously, in eukaryotes. This family of potentially dual-specific (threonine/tyrosine) protein kinases is among several that regulate a number of steps of mitosis. The most widely conserved MPS1 kinase functions involve activities at the kinetochore in both the chromosome attachment and the spindle checkpoint. MPS1 kinases also function at centrosomes. Beyond mitosis, MPS1 kinases have been implicated in development, cytokinesis, and several different signaling pathways. Polo kinases are serine/threonine kinases that participate in 3 metabolic pathways: cell cycle, cell cycle - yeast, and progesterone-mediated oocyte maturation.

Gilliland *et al.* (2009)[305] report that Mps1 and Polo kinase localize to numerous self-assemblies in the ooplasm of prometaphase *Drosophila* oocytes. They first appear throughout the oocyte at the end of prophase and are disassembled after egg activation. The self-assemblies form reversibly in response to hypoxia (**Fig. 30F**). The return to identical positions suggests an unseen filamentous structure in the cell, yet the kinases did not colocalize with tubulin, actin, anillin, septin, or lamin[306]. The addition of collagenase prevents filament formation by the kinases, suggesting that the underlying filament is sensitive to this protease. Gilliland *et al.* (2009)[305] hypothesize that the self-assemblies formed by Mps1 and Polo could serve in signaling, that absence of the kinases results in a signal throughout the oocyte (a surrogate kinetochore).

***D-Amino acid oxidase***

DAO catalyzes the oxidation of D-amino acids to their corresponding imino acid (**Fig. 30G**). The enzyme is most active toward neutral D-amino acids, and not active toward acidic D-amino acids. One of its most important targets in mammals is D-serine in the central nervous system. By targeting this and other D-amino acids in vertebrates, DAO is important in detoxification. The role in microorganisms is slightly different, breaking down D-amino acids to generate energy[307]. In 1966, Antonini *et al* (1966)[308] used light scattering to demonstrate polymerization of the protein *in vitro*. Lower pH and temperature, and higher chloride concentration, decreased this polymerization. Similarly, infinite self-association was also shown to best fit sedimentation and light-scattering data of DAO purified from pig kidney[309-310].

***Additional self-assembly forming enzymes discovered in large-scale screens***

18 additional enzymes, in addition to several discussed above, were uncovered in Narayanaswamy, *et al.* (2009)[25] as forming punctate foci in yeast under nutrient starvation conditions (stationary growth phase). 90% of the proteins identified as forming foci in the GFP screen were also found in pelleted fractions of cell lysate (without GFP tags), showing the behavior in the absence of GFP. All proteins listed in **Tables 1-2** as identified in the Narayanaswamy screen were found to reversibly associate into these foci and dissociate upon addition of nutrients.

Enzymes identified as forming cellular self-assemblies by Narayanaswamy, *et al.* (2009)[25] and not discussed above include alanyl-tRNA synthetase, Leucyl tRNA synthetase, Cystathionine β-synthase, Trifunctional CAD enzyme, adenylosuccinate synthetase, alcohol dehydrogenase, UDP-Glucose-pyrophosphorylase, sterol 3-β-glucosyltransferase, peptidyl-prolyl cis-trans isomerase, glutamine tRNA synthetase, histidine tRNA synthetase, Isoleucine tRNA synthetase, Protein kinase of the PAK/Ste20 family, ribosome-associated molecular chaperones, ATPases, threonyl tRNA synthetase, trehalose-6-P synthase/phosphatase complex, γ-aminobutyrate (GABA) transaminase, valyl-tRNA synthetase, and serine hydroxymethyltransferase.

In addition to acetyl-CoA carboxylase, CTP synthase, GDP-mannose pyrophosphorylase, and glutamine synthetase described in sections above, 13 additional enzymes were also found by Noree, *et al.* (2010)[25] to form self-assemblies and foci in cells upon nutrient starvation. These include 5-phospho-ribosyl-1(alpha)-pyrophosphate synthetase, glutamine amidotransferase (GATase II), threonine aldolase, glycogen synthase , glycogen phosphorylase, aminolevulinate dehydratase, multifunctional enzyme containing phosphoribosyl-ATP pyrophosphatase; phosphoribosyl-AMP cyclohydrolase, and histidinol dehydrogenase activities, S-adenosylmethionine synthetase, disaggregase (ATPase), ATPase involved in protein folding and NLS-directed nuclear transport, HSP70 family ATP-binding protein, non-ATPase regulatory subunit of the 26S proteasome, homoserine kinase (yeast gene Thr1, also Thayer, *et al.* (2014)), and glycogen debranching enzyme. Werner, *et al.* (2009)[23] identified UDP-N-acetylmuramate-alanine ligase as a cellular self-assembly-former. Shen, *et al.* (2016)[26] confirmed many of those previously discovered, and also identified thioredoxin peroxidase as forming self-assemblies. Suresh, *et al.* (2016) discovered that the fatty acid synthase complex (yeast genes Fas1 and Fas2) form distinct foci under starvation conditions, and that these contain active enzymes and are not substrates of quality control proteins such as Hsp104 and did not co-localize with self-assemblies of Ura7, Pre6, or Ade4. They were also reversible, dispersing in the presence of added nutrients. Finally, a protein of unknown function with glutathione-S-transferase domain (*Drosophila* gene Fax, for Failed axon connections (CPTI-002774)) was found to form foci in the developing oocyte (Lowe *et al.* (2014)[29]).

# Discussion

### *Filaments vary in structure*

Over twenty different enzymes are now confirmed to form linear self-assemblies, polymers, or helical filaments that have been structurally characterized by electron microscopy or x-ray crystallography **(Table 1**). Very likely, these structures are present in a biological context (*in vivo*) as well as *in vitro*. Another 50 more form self-assemblies in cells that resemble those formed by filament forming enzymes, but have yet to be structurally characterized in molecular detail and therefore are not yet confirmed to form filaments (**Table 2**). This review prioritized those self-assemblies that are known to be reversible - to distinguish from those assemblies formed by aggregates of misfolded proteins, though reversibility has not been tested in every case (see **Table 2**). Of those filament forming enzymes for which the structure is known at the molecular level, most form helices (about half of the known filaments are left-handed, half right-handed) and without polarity (either end is equivalent due to internal symmetry of the enzyme building block)(**Table 1**). Four form double helical structures (glutaminase, ribonucleotide reductase, glucokinase, and Filament 3 of acetyl-CoA carboxylase, **Table 1**). Three form an intermediate structure composed of a ring which then stacks linearly to create a filament (glutamine synthetase, β-glucosidase, and 2-cys peroxiredoxin). Adenylate kinase and casein kinase 2 form linear filaments, without any helical nature. The death domain (DD) containing proteins can form complicated filaments with different proteins assembling in a specific order. Finally, in some cases, the filaments have a large internal cavity and form hollow tubes (TORC1, β-glucosidase, nitrilase, glutamate dehydrogenase, and 2-cys peroxiredoxin). Hence, filaments form in a wide array of structures.

Although we have made an effort to collate a comprehensive list of proteins found to form filaments or at least cellular self-assemblies, we have not included the additional proteins uncovered in screens of protein aggregation in response to heat stress or arsenic, because our interest is in enzyme regulation and these screens likely identify proteins undergoing irreversible aggregation due to misfolding. Although it should also be noted that many of those proteins found to form cellular self-assemblies under nutrient stress also are found in the heat shock (or arsenic) induced cellular aggregates as well[30, 130, 311]. We have also not covered the proteins found in a screen for prion-like assembly formation[312], as these protein assemblies have several distinct biophysical features affecting their solubility, reversibility, and heritability. Further, many other proteins and enzymes can be found in RNA-rich self-assemblies such as P-bodies, stress granules, and other membraneless compartments that are not given in **Table 2**, with filament formation yet to be investigated[304, 313-314].

### *Effect of filamentation on enzyme activity*

Perhaps counterintuitively, most of the enzymes known to be modulated in activity by filament formation are activated in the filamentous state (13 total including filaments 1 and 2 of ACC, human CTPS, IMPDH, β-glucosidase, $CO_2$-reductase, nitrilase, glutaminase, SgrAI, IRE1, RecA, DnaA, RIP1/3 kinases, and death domain containing, **Table 1**). Six enzymes are known to be inhibited (ACC filament 3, glucokinase, bacterial CTPS, glutamine synthetase, ribonucleotide reductase, and TORC1, **Table 1**). IMPDH is also suspected of having an inactive filamentous state. The enzyme 2-cys peroxiredoxin shifts from peroxireductase activity in the non-filamentous state, to a chaperone in the filament form which is inactive in peroxireductase activity. Hence it is inactive in the dimeric state, but gains an entirely new activity in the filamentous state. In the case of LACTB, which has sequence homology with β-lactamases, it is not actually known if it possesses any enzymatic activity, much less how such activity is modulated by filamentation. Bacterial CTPS is thought to possibly act in cell shape determination when filamented[43]. It is not yet known if PFK, AdhE, and glutamate dehydrogenase are activated in the filament form, but other considerations suggest that they may be activated because the filaments form under conditions of high demand in the cell or high concentrations of substrate (PFK[75-76]), filaments form under conditions where the enzyme is more active (glutamate dehydrogenase[180-181, 183]), or have active sites more accessible to substrates (AdhE[22])(**Table 1**). CK2 and adenylate kinase may be inhibited in the filamentous state (**Table 1**). For enzymes of **Table 2**, it is not known how their activity is affected by filament formation, but many form filaments under nutrient starvation conditions suggesting that they filament in order to store inactivated enzymes until needed. Alternatively, filamentation may activate enzymes under conditions of stress due to high demand (such as G-bodies composed of enolase, PFK, and/or fructose bisphosphate aldolase)[33].

***Filamentation can change enzyme specificity***

In only a few enzymes does the substrate specificity change when in the filament form. SgrAI, a sequence specific endonuclease, cleaves 17 different 8 bp DNA sequences in the filamentous site, but only 3 in the non-filamentous state[17, 35, 38]. IRE1 may also have expanded substrate specificity when filamentous, cleaving other RNAs in addition to its canonical RNA substrate[240]. The enzyme 2-cys peroxiredoxin changes specificity and enzymatic activity when filamentous, changing from a peroxireductase to a chaperone[200, 206]. A change in specificity to smaller substrates, due to reduced access to active sites located within the central tunnel of a fenestrated tube, was suggested for β-glucosidase[141]. Nitrilases from different organisms assemble into filaments with different helical twists, which correlate with their exhibited substrate specificities[315]. Hence, although not changing within the same filament, the family of nitrilase enzymes show that filaments may evolve to fine-tune activity and specificity for particular cellular needs.

***Regulation of filament formation***

Regulation of filament formation also varies. Enzyme filaments stimulated by their substrates (and which become activated) include human CTPS, nitrilases, SgrAI (which is also stimulated by its product), RecA, and DnaA (**Table 1**). In the case of PFK, it is not known but likely to be activated in the filament, because filament formation is stimulated by its substrate[76]. Filaments of glucokinase form in response to substrates ATP and glucose[90], when the enzyme concentration is above a critical concentration (2 µM), but is inactivated in the filamentous form. 2-Cys peroxiredoxin is stimulated by its substrate, $H_2O_2$, to form filaments with a new activity (binding unfolded proteins) but which is inactive in peroxiredoxin activity[202]. Filaments forming in response to their enzymatic product include SgrAI (cleavage products of their primary recognition sites which stimulate the enzyme activity), glucokinase (inhibited in the filament), and bacterial CTPS (also inhibited as a consequence of filamentation)(**Table 1**). Many enzyme filaments form in response to another signal, including binding to allosteric effectors (PFK, ACC Filaments 1-2, IMPDH, ribonucleotide reductase and glutamate dehydrogenase), or ligands ($CO_2$ reductase, IRE1 and many death domain containing proteins)(**Table 1**). At least two enzymes are known to form filaments as a result of phosphorylation (ACC filament 3 and 2-cys peroxiredoxin), one from protein binding (ACC Filament 3), and others in response to buffer components (nitrilase, glutaminase, glutamine synthetase, $CO_2$ reductase, 2-cys peroxiredoxin)(**Table 1**). Phosphorylation stabilizes IRE1 filaments, as does ADP binding[18]. Just as filaments are stimulated in response to environmental stimuli, they may also be induced to disassemble. Those disassembled by their products, direct or downstream, include glutamate dehydrogenase[13, 15] and human CTPS[98]. Filament formation by ACC1 and ACC2 are reportedly inhibited by binding to SPOT14 and MIG12 proteins[70] (although another report suggests MIG12 stimulates filament formation[20]). Others are known to be disassembled by buffer components (pH for Gln1, $CaCl_2$ for β-glucosidase[9, 142], and high ionic strength for casein kinase[266]).

***Direct transformation from one type of filament to another***

Some filamentous enzymes may be capable of responding to environmental conditions without dissociating and reassembling into a new filamentous structure. Instead, these remain filamentous but coordinately (or not) change conformation to active or inactive enzymatic activity. Three possible examples of this have been described. First, ACC forms the Filament 1 type filament in the presence of citrate and when the enzyme is not phosphorylated. The addition of a product of a reaction downstream of ACC, palmitoyl-CoA, results in what appears to be an altered Filament 1, but which now is inactive, and is named Filament 2[71]. However, it is not known for certain at this time if this transformation from the active Filament 1 form to the inactive Filament 2 form occurs without dissociating into ACC protomers. The second example is IMPDH. ATP stabilizes both types of filaments which are composed of stacked octameric IMPDH enzymes. The octameric IMPDH enzyme has the appearance of a hollow sphere with openings on the sides presumably for substrate and product diffusion. The open form is stabilized in the presence of a cofactor for the reaction, $NAD^+$, while the closed form is stabilized by a downstream product of the enzyme pathway, GTP. The closed form is likely inactive, due to lack of substrate access to the active site, while the open form is likely the active state. Both states, open and closed octamers, have been visualized within the same filament[128]. This may indicate less cooperativity in the filament, since both states coexist, but it also indicates that the filament can be sensitive and responsive to molecular signals without dissociating. The third example is AdhE. Upon addition of cofactors $Fe^{2+}$ and $NAD^+$, the filaments appear slimmer and more extended[22]. This indicates responsiveness of the filament to the environment, although transition from one filamentous form to the other was not strictly investigated, hence these two forms may have required dissociation and reassociation. Further investigations are needed.

***Enzymes that form more than one type of filament***

Five enzymes were shown to form more than one type of filament. Two were discussed above, ACC and AdhE, and in addition, ACC forms a third type of filament that requires phosphorylation of the enzyme and binding of the BRCT domains of BRCA1[71]. Further, ACC may also form yet additional filament types, as filamentation *in vitro* and *in vivo* has been shown to be stimulated by the protein MIG12, which appears to form a co-filament with ACC[20]. Ribonucleotide reductase also forms different filamentous forms, both inactive, but which differ in architecture (one is a double helix, the other is a single helix) and the presence or absence of a protein NrdF[210-211]. CTPS also forms two types of filaments, one active and one inactive. Although in this case it is not the same CTPS enzyme; instead it is CTPS from a bacterial source that forms an inactive enzyme filament[32], and CTPS from human that forms an active enzyme filament[98]. The filaments have similarities in structure, in that they are formed by stacking the tetrameric CTPS, but differences in the conformation of the tetramer and hence also the interfaces between tetramers within the filaments. These two filaments appear to be representatives of the type of activation or inhibition that derives from stabilization of either the active or inactive conformation of the enzyme by contacts between adjacent enzymes within the filament. Finally, nitrilases from different organisms have been examined and found to differ in the helical twist between the enzyme protomers, as mentioned above[315]. Interestingly, this difference in twist correlates with substrate preference, and hence evolution may have modified the filament structure to fine tune enzymatic specificity.

***Morpheins***

Some filament forming enzymes are also morpheins[48]. Morpheins are proteins which can form more than one type of oligomer, each having distinct enzymatic properties, and where conversion from one type to the other requires dissociation into a more fundamental building block under normal physiological conditions. The interconversion is also post-translational modification independent. The shifting between oligomeric states is controlled by binding to allosteric effectors. The classic example of a morpheein is porphobilinogen synthase[316-319]. As for filament forming enzymes, the filament being one oligomeric state, 2-cys peroxiredoxin and ribonucleotide reductase fit the strict requirements of a morpheein. Other enzymes, such as SgrAI, are related to the morpheein model in that a fundamental building block (the SgrAI dimer) is stimulated to form a different, more active oligomeric form (a filament). However, it is stimulated by is substrate and product, rather than a distinct allosteric effector. It also does not have a second oligomeric state beyond the fundamental building block.

***Biological roles and possible selective advantages of filament formation by enzymes***

It is clear now that many enzymes form filaments and large self-assemblies, and for some the role and advantage of those structures are yet to be uncovered. Many have stimulated enzymatic activity in these self-assemblies, yet others are reduced or inhibited, showing filament formation to be another layer of enzyme regulation. Why filaments and large self-assemblies are necessary rather than (or in addition to) other mechanisms is an important next question for researchers. It may be that formation of the filament enables more rapid, and more cooperative activation or inactivation of the enzyme, as has been suggested for SgrAI (rapid activation[50]) and *E. coli* CTPS (rapid inactivation[96]). Many enzymes regulated by filament formation control key steps in biochemical pathways, such as PFK, which may require the additional layer of regulation[76]. Why a filament and not a discrete oligomer, such as a dimer, tetramer, hexamer, etc. is formed may be due to the fact that evolving a filament requires the selection of changes that result in a single type of interface between protomers that can propagate without steric hindrance, vs. a discrete oligomer such as a dimer that would require evolving an interface with two-fold symmetry that occurs only once between protomers[47]. Evolving other discrete oligomers would require such interfaces to lead to a ring, to "close the ends". Interestingly, such closed ring oligomers can themselves form filaments by stacking, as discussed above.

Various strategies for controlling enzyme activity by filament formation are found. One strategy is to evolve an interface between protomers that preferentially stabilizes one conformation of the enzyme, either the active or the inactive state. Hence, the filamented enzyme is locked in one conformation, and cannot access the other state. Such is proposed for several enzymes, including SgrAI, glucokinase, and the CTPS enzymes. In the ribonucleotide reductase filament, binding of a second protein (NrdF) to the filament formed by NrdE leads to an inter-subunit gap that is too large to allow for radical transfer, thereby inactivating the enzyme[211]. In adenylate kinase, the active site is blocked by the filament structure[274]. Another strategy, for activating an enzyme by filamentation, is to create active sites from multiple enzyme protomers within the filament, such as Filament 1 of ACC[71], DnaA[218], and RecA[214]. Alternatively, formation of an internal tunnel within a hollow filament can aid in channeling substrates and products to different active sites as proposed in AdhE[159], and may be a central feature of G-bodies, purinosomes and other metabolically related structures. The internally located active sites in hollow filaments means that substrates and products must diffuse through gaps between protomers, which may allow for mechanisms of selectively of substrates, as suggested for nitrilases[156] and β-glucosidase[141]. In addition, forming a large assembly in the cell could localize products of the enzymatic reaction to particular locations. This was suggested for several enzymes activated by filament formation, but also β-glucosidase, where the filamentous structures produce anti-fungals and are activated upon infection[141]. Conversely, filament formation may sequester enzymes away from one another and their substrates, modulating metabolic flux.

The filaments may be necessary to bind large substrates, as in the case of DnaA, RecA, IRE1, nitrilases, and significantly, for altering substrate specificities, as in IRE1 which cleaves additional mRNAs using an activity known as RIDD upon filamentation, and SgrAI, which cleaves an additional 14 DNA sequences when in the filament form. Significantly, simulations of enzyme activity based on kinetic modeling indicate that the filamentation mechanism can act to sequester enzymatic activity on only a subset of available substrates[50-52]. For example, as a direct consequence of filament formation, the expansion of substrate specificity by SgrAI occurs only on DNA molecules containing the SgrAI primary recognition sequence. The filament takes advantage of the higher local concentration of different recognition sites present on the same molecule of DNA to induce filamentation and cleavage activity on the secondary site sequences on the same contiguous molecule of DNA. Though this could be accomplished by a finite, discrete oligomer, the filament has the advantage of also maximizing speed of enzyme activation as well[50]. Local concentration effects are easily possible on large multi-substrate molecules such as DNA, but such activity may also be possible wherever high concentrations of a preferred substrate occur, which can induce filament formation to act upon a second type of substrate. Hence the second type of substrate is acted upon only where the high local concentrations of the preferred substrate occurs (**Fig. 31**), which could result from the activity of nearby enzymes, or in particular compartments including phase separated assemblies (liquid separated droplets in cells often formed from RNA and proteins)[45].

Some filaments appear capable of switching between active and inactive conformations in the same filament, such as ACC, AdhE, and IMPDH. The purpose of the filament may be for more rapid communication of a signal for activation or alternatively, for inactivation. In this case, it's akin to a multimeric enzyme with a large degree of cooperativity.

The filaments and self-assemblies may be a way to store unneeded enzymes, to protect against degradation, and to be rapidly redeployed upon, for example, reintroduction of nutrients to the environment without the need for new transcription and/or translation. Such has been suggested for the cytoophidia and foci formed upon nutrient starvation[25, 101]. Filament formation may also be used to protect active enzymes from degradation or oxidation, as has been suggested for $CO_2$ reductase[148] and AdhE[159].

Buffering of enzymatic activity is another proposed role of enzyme filaments and self-assemblies. Active enzymes in the non-filamentous state can reach an equilibrium with the inactive, filamentous state and by modulation of the on and off rates of this equilibrium, different levels of active enzyme in the pool can be achieved, as proposed for bacterial CTPS[97]. Thus, the level of active enzyme in a cell can be easily maintained and adjusted as needed. In the case of glucokinase, active monomeric enzymes reach an equilibrium with those in the filaments (composed of inactivated glucokinase), thereby maintaining a constant concentration of active enzyme and creating an upper limit to the maximum enzyme activity. This protects the cells from sudden “surges” in glucokinase activity that could be toxic. Binding of proteins to the filament to stabilize it from dissociation, such as Filament 3 of ACC with the binding of BRCT1[71], could further regulate filament dissociation.

Another reason for filament formation by enzymes may be to control water activity. Formation of large complexes such as enzyme filaments reduces the available protein surface area, thereby significantly reducing the amount of water required for full hydration of the protein[320]. Trihalsoe-6-P forms self-assemblies in cells in response to desiccation[33]. Adenylate kinase, found in plants, is thought to form filaments at night when the enzyme is not needed, to free up available water[274].

Several enzymes gain additional functionality in the filament form. The enzyme 2-cys peroxiredoxin is a striking example which switches from a peroxiredoxin to a chaperone upon oxidative stress[205]. The enzyme functions to eliminate $H_2O_2$ at lower levels, but higher levels will induce greater oxidation and eventually the formation of a hollow filament capable of binding unfolded proteins in its interior. This type of chaperone activity is non-ATP dependent and termed “holdase” since it sequesters the unfolded proteins from one another to prevent their aggregation. CTPS from *C. crescentas* is another example, which regulates cell shape in its filamentous form[19]. Acting as a scaffold for the binding of other proteins and enzymes is another gain of function that has been suggested for many filament forming enzymes. Certainly the death domain (DD) containing proteins and RIP1/RIP3 kinases do this, allowing binding and activation of downstream effector enzymes such as kinases, proteases, and membrane pore forming proteins[276, 279-280, 293, 295]. Relatedly, the filaments could be a signal themselves to the cell of the levels of metabolites or other environmental stimuli. In addition to the proteins and enzymes of innate immunity (DD and RIP1/RIP3 kinases), this function has been proposed for IMPDH[128] and the kinases Mps1 and Polo[305]. In addition, it was suggested for IRE1 that filament formation could prolong the signal induced by unfolded proteins since self-assembly is self-propagating and stabilizing via auto-phosphoryation[18], similar to inflammasomes composed of RIP1/RIP3 kinases and also DD enzymes like RIP2[281, 289].

Finally, a role for filamentation in sequestering enzymes to particular compartments, such as cytoplasm, nucleus, or the mother cell in dividing cells. This was proposed for IMPDH, since in *Drosophila* it acts also as a transcription factor[121]. Thayer *et al.* (2014)[321] proposed sequestration upon cell division for homocysteine kinase.

***Unique features of SgrAI***

The most defining and unusual characteristic of the SgrAI endonuclease is its ability to use filament formation to modulate its DNA sequence specificity in a very targeted way. It does this by using the DNA sequence of its recognition sequence to modulate the equilibrium between two conformational states, an inactive, non-filament forming state, and an activated, filament forming state[39]. Binding to its primary recognition sequences shifts this equilibrium to the active, filament forming state more than does binding to its secondary site sequence. Hence it is SgrAI bound to primary sites that will drive filament formation, although SgrAI bound to secondary site sequences will join filaments under particular conditions. Joining the filament leads to activation of SgrAI and rapid cleavage of the bound DNA. However, joining is limited by a very slow association rate constant[51]. To overcome this limitation, high concentrations of filament forming states are necessary. In the cell, this is accomplished by SgrAI binding to primary recognition sequences on the same DNA molecule, and the high local concentration therefore drives the association of these SgrAI into a filament. Primary sites on the host are methylated and will not be bound by SgrAI, but invading DNA is not expected to have the same protection. Hence unmethylated primary sites on the phage DNA drive SgrAI to form filaments, which also recruit SgrAI bound to secondary site sequences on the same DNA. Secondary sites are not methylated on the host, and therefore are at risk of being cleaved by activated SgrAI. SgrAI binds to secondary sites but does not cleave them unless activated by recruitment into a filament. However, the slow association rate constant for their incorporation into the filament, and the low concentration of DNA in the cell, prevents their recruitment into the filament formed on the invading DNA. Only the high local concentration of SgrAI bound to secondary sites on the invading DNA can drive their incorporation into filaments formed by SgrAI bound to primary sites there[50].

The computational model developed from the kinetic study of DNA cleavage by SgrAI also found that the filament mechanism imparted greater speed in product production, and with a greater ability to sequester its activity (between sites on invading and host DNA)[50]. The sequestration of activity derives from the slow association rate constant for filament formation. Association and dissociation rates of enzymes into and out of filaments to our knowledge have not been measured, with the exception of the observation in some limited cases of cytoophidia dynamics. It remains then to be seen how general or specific this particular but distinguishing property of the SgrAI system is. As for the property of rapid activation, the opposite, rapid and efficient inactivation, has been suggested for the bacterial CTPS system. A mutation in CTPS, E155K, eliminates the ability to form filaments, but also does not show the significant reduction in activity shown by wild type CTPS[96, 109].

Some other unique features of the SgrAI system are also found, including filament formation being stimulated by both substrate and product (of the primary site sequence), participation of the substrate (and product) in stabilization of the filament via direct interactions with neighboring protomers, and the requirement to dissociate from the filament in order to release product[17, 37-38, 50-52]. We found that the dynamics of SgrAI enzymes into and out of the filament is sufficiently fast to prevent trapping of the product, despite this requirement. Many filamentous enzymes with smaller substrates and products would not have such barriers to product dissociation.

Only IRE1 may share the unique property of substrate specificity modulation via filament formation, since some evidence suggests that the filamentous form is more effective in RIDD, an activity which results in degradation of other mRNAs in addition to that which is spliced by IRE1[243]. SgrAI cleaves 14 additional (secondary site) DNA sequences in the filamentous state, but only 3 (primary sites) in the non-filamentous state. It does so by drawing SgrAI bound to secondary site DNA sequences into filaments formed by SgrAI bound to primary site, thereby activating SgrAI to cleave the secondary sites. Secondary sites will not induce SgrAI to form filaments. A somewhat similar situation is found with β-glucosidase, two isomeric forms of the enzyme are known, As-Glu1 and As-Glu2. As-Glu2 forms a homodimer, but will be drawn into filaments formed by As-Glu1[140]. However, both forms of the enzyme, as well as both states (filament and non filamentous) possess the same enzyme specificity.

As for the qualities of becoming activated in the filamentous state, thirteen or more enzymes share this property, however, quantification of exactly how much more active is lacking for most systems. For SgrAI, the observed rate constants for cleavage of primary DNA show a 200-fold increase under filament forming conditions compared to non-filament forming conditions, and for secondary site sequences, this number is 1000-fold[38]. However, kinetic modeling found the rate constant for activated DNA cleavage in the filamentous state is 0.8 $s^{-1}$, 480 times faster than the observed DNA cleavage rate constant of primary sites in the non-filamentous state, and nearly 4000 times faster than with secondary sites[38, 52].

Finally, the issue of length and time scales of filament forming enzymes requires addressing. **Table 1** summarizes enzymes where the molecular structure of the filamentous state is known, but in many cases, it is also known that these enzymes form large assemblies in cells that must be composed of either assemblies of these filaments, or perhaps some other structure. **Table 2** summarizes enzymes for which only the cellular assembly has been observed, but studies of the molecular details of the assemblies have not yet been performed, hence it is unknown if they form filaments at the nanoscale. In one case, electron microscopy of a cell showed the large assembly to be composed of layers of aligned filaments. One could imagine that this would be most efficient for the purposes of storing and protecting inactive enzymes. In other cases, the assemblies may have other geometries, perhaps less organized (networks) and more like gels[45] (PFK, known to form filaments, also forms liquid-like droplets in cells[322]). This may allow substrates and products to diffuse in and out of the large-scale assembly/droplet, perhaps making use of local concentration to alter enzyme specificity, as in the SgrAI case, but using phase separation to induce the high local concentrations. As for timescale, our studies show association/dissocation of SgrAI with rate constants of $3x10^5$ $M^{-1}s^{-1}$ and 0.02 $s^{-1}$, resulting in very fast (subsecond) association under optimal conditions and a half-life of the complex on the order of seconds[52]. The timescale of filaments and cellular assemblies is less well characterized, though studies of Mps1/Polo kinases show complete dissociation after 30 minutes, and complete re-filamentation upon reintroduction of filament inducing conditions after another 30 minutes[305].

## Conclusions

In conclusion, nearly 80 enzymes have been shown to form some sort of self-assembly either in cells or *in vitro*. Of these, 23 have been characterized for their assembly structure. Structures vary from the less common linear filament and ring stacks, to more common helical filaments, either right or left-handed. Enzyme activity may be enhanced in the filament, or inhibited, and less commonly, altered in some form. Some enzymes form more than one type of filament each with unique enzyme activities. Beyond controlling enzyme activity, other functions such as scaffolding, controlling cell shape, and signaling have also been shown. In the case of SgrAI, a type II restriction endonuclease, studies show that filament formation provides a clever means to control DNA cleavage behavior onto only invading DNA. The existence of filaments *in vivo*, their regulation and activity, and the wide range of diversity in structure and complexity is an emerging phenomenon in our biophysical understanding of cell biology. In the coming years, as more studies on filament forming enzymes are performed, no doubt new and interesting purposes and advantages will be uncovered.

**Acknowledgements**
This work was supported by the National Science Foundation under Grant No. MCB-1410355 (to N.H.).

# Figures

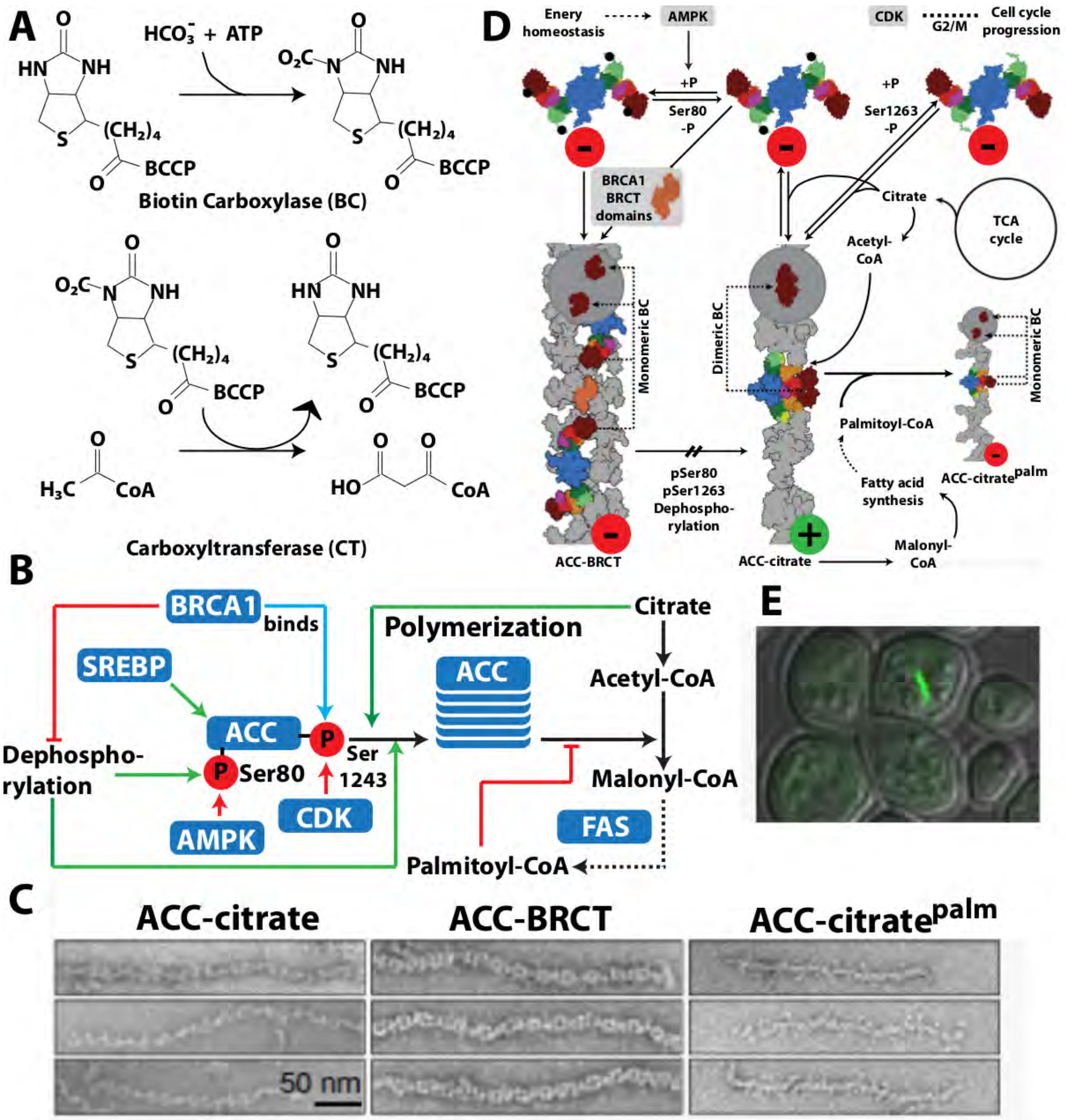

**Figure 1** Filamentation of acetyl-CoA carboxylase (ACC). **A.** Reactions catalyzed by ACC. **B.** Complex regulation of ACC function involving filamentation (*a.k.a.* polymerization), phosphorylation, BRCA1 binding, and allosteric inhibition (adapted with permission from Hunkeler *et al.* (2018)[71]). **C.** Negative stain electron micrographs of three different filament forms of ACC (adapted with permission from Hunkeler *et al.* (2018)[71]). **D.** Roles of the three filament forms in regulating ACC activity (indicated by red or green circles) (adapted with permission from Hunkeler *et al.* (2018)[71]). **E.** GFP labeled ACC form cytoophidia (large self-assemblies) in yeast cells under nutrient starvation (adapted with permission from Shen *et al.* (2016)[26])

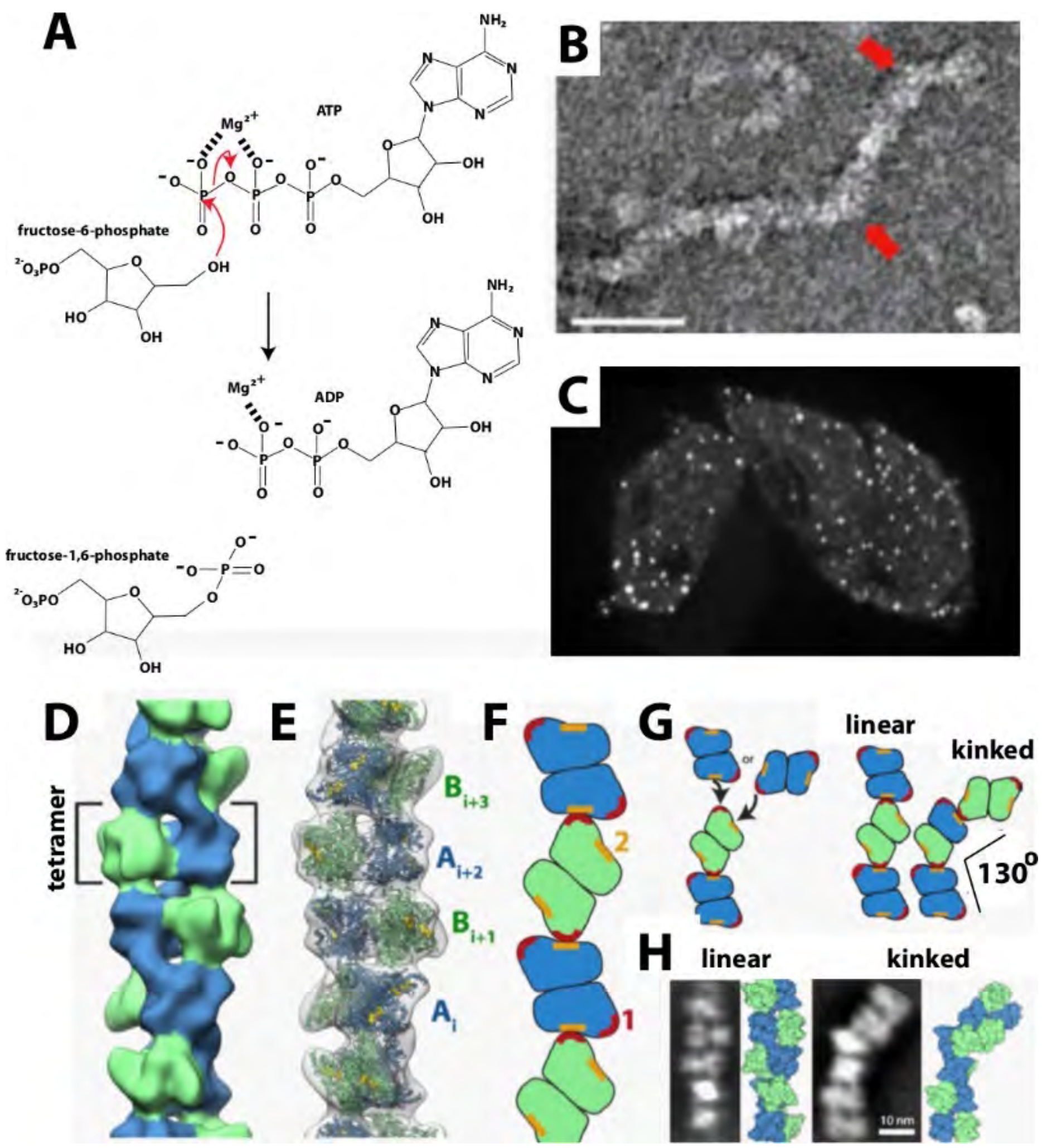

**Figure 2** Reactions and filament formation by phosphofructokinase-1 (PFK). **A.** Reaction catalyzed by PFK (EC 2.7.1.11 ). **B.** Negative stain EM image of PFKL (liver isoform of human PFK) filament. Bar represents 100 nm. Arrows identify "kinks" (adapted with permission from Webb *et al.* (2017)[76]). **C**. PFKL-EGFP filaments in cells MTln3 cells following addition of 10 mM citrate using confocal microscopy showing punctae. Not such puncta were apparent before the addition of citrate (not shown)(adapted with permission from Webb *et al.* (2017)[76]). **D.** Three-dimensional reconstruction of PFKL filament (adapted with permission from Webb *et al.* (2017)[76]). **E.** Coordinates of the tetramer of the PFKP crystal structure fit into the three dimensional reconstruction (adapted with permission from Webb *et al.* (2017)[76]). **F.** Schematic of PFKL tetramer associations in PFKL filament (adapted with permission from Webb *et al.* (2017)[76]). **G.** Left, schematic showing two different associations of PFKL tetramers leading to either linear or kinked filaments. Right, schematic of linear and kinked filaments (adapted with permission from Webb *et al.* (2017)[76]). **H.** Left images, 2-D class averages and right images, reconstructions, for linear and kinked filaments (adapted with permission from Webb *et al.* (2017)[76])

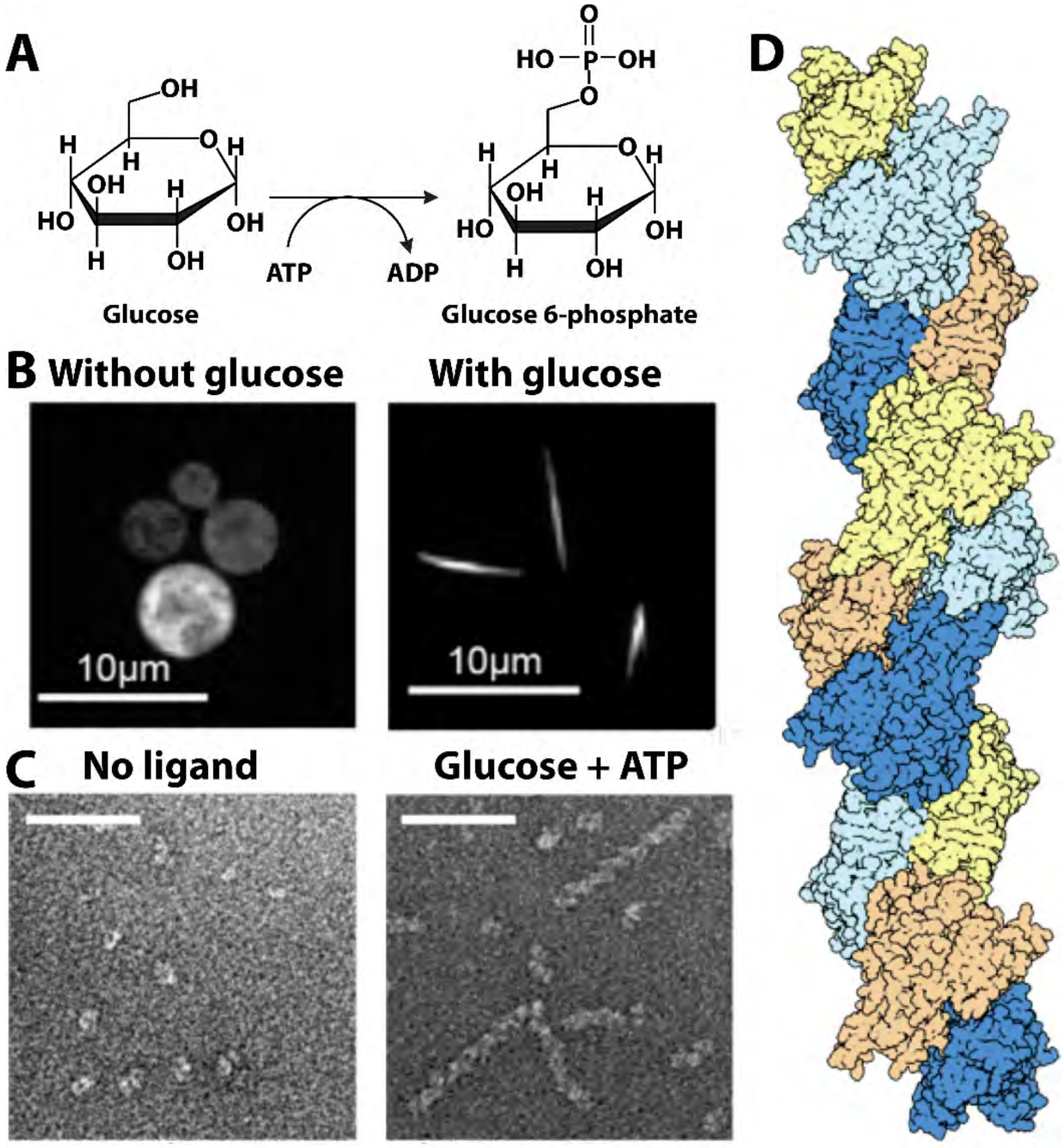

**Figure 3** Reaction catalyzed and structures formed by glucokinase (Glk1). **A.** Reaction catalyzed by glucokinase (Glk1), phosphorylation of glucose using ATP. **B.** Fluorescence images of stationary phase Glk1-GFP before (left) and after (right) the addition of glucose bars are10 µm. (Adapted from Stoddard *et al* (2019)[90]). **C.** Negative-stain electron micrographs of Glk1 in the absence of ligands (left) or in saturating glucose and ATP (right). Scale bars are 50 nm. (Adapted from Stoddard *et al* (2019)[90]) **D.** Surface representation of Glk1 filaments derived from cryoEM structure. Yellow/orange represent copies of the monomeric enzyme on one strand, light and dark blue on the other strand. The double-stranded filament is antiparallel, with close contacts between enzymes in each strand, as well as between strands. (Adapted from Stoddard *et al* (2019)[90])

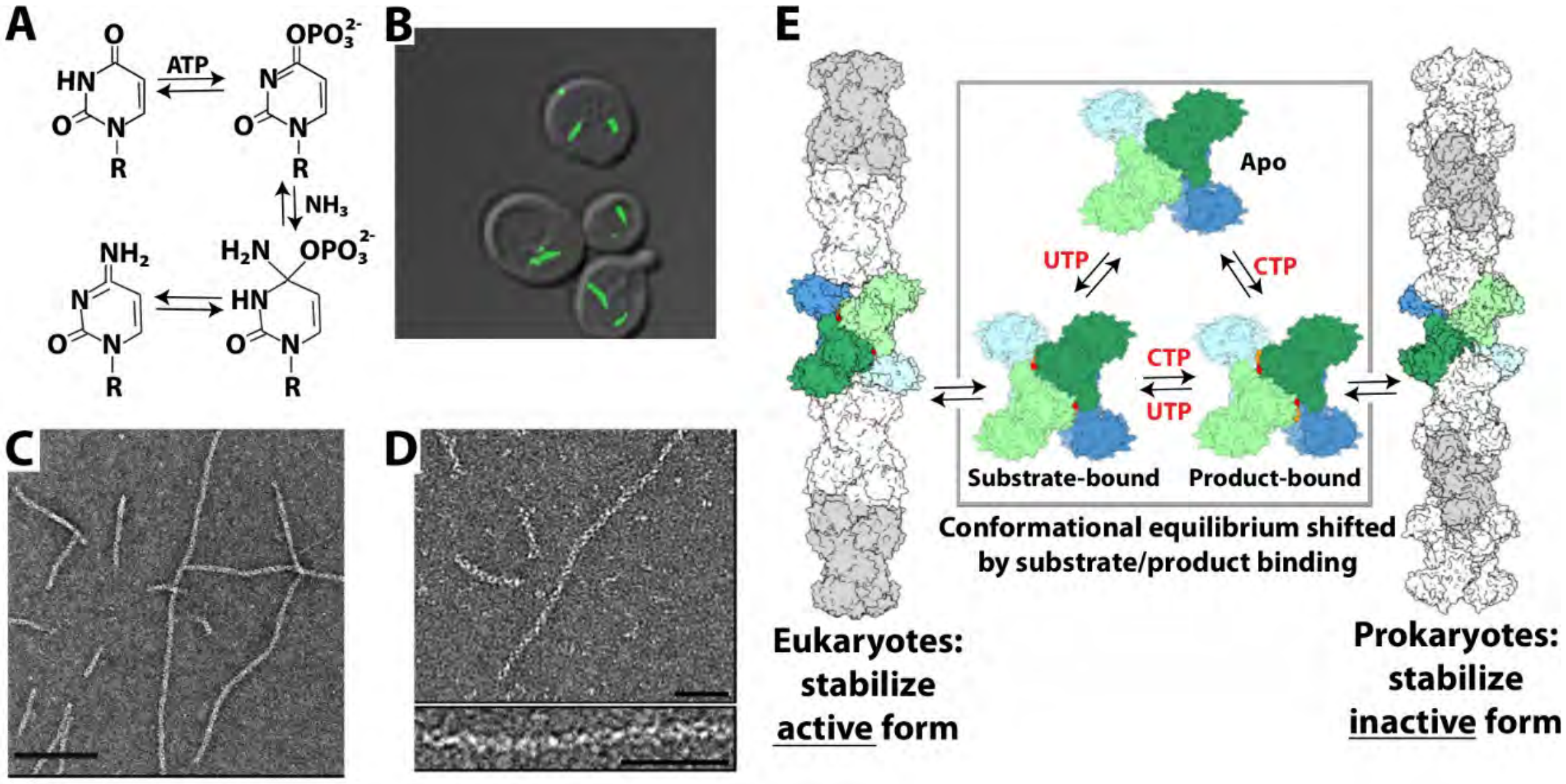

**Figure 4** Reaction catalyzed by, and filaments formed of, CTP Synthase (CTPS). **A.** Reaction catalyzed by CTPS (EC 6.3.4.2). **B.** Cytoophidia or large rod-like structures of GFP labeled self-assemblies observed in yeast under nutrient starvation conditions (adapted with permission from Noree *et al.* (2010)[24]). **C.** Filaments of ecCTPS formed by a mutant designed to for crosslinks to stabilize the filamentous structure viewed by negative stain EM. Line represents 200 nm (adapted with permission from Lynch *et al.* (2017)[98]). **D.** Filaments of human CTPS formed in the presence of UTP, ATP, and GTP. Bar represents 50 nm. (adapted with permission from Lynch *et al.* (2017)[98]). **e.** Models of CTPS filaments from human (left) and bacteria (right) based on single particle cryo-EM reconstructions and their relationship to the presence of substrates (UTP) and products (CTP)(adapted with permission from Lynch *et al.* (2017)[98])

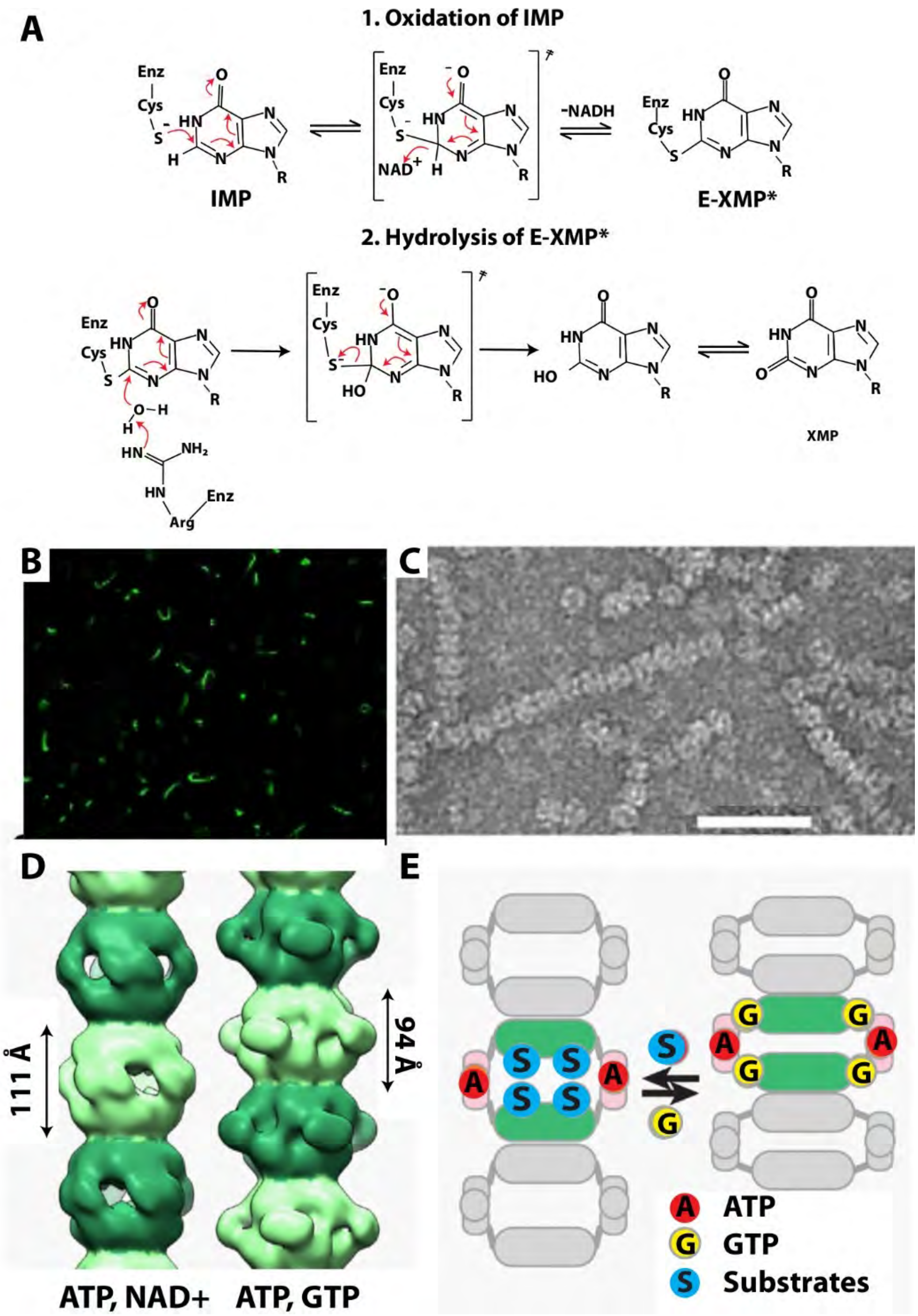

**Figure 5** Reaction catalyzed and filaments formed by IMPDH. **A.** Reaction catalyzed by IMPDH (EC 1.1.1.205)( adapted with permission from Hedstrom, (2009)[117]). **B.** Cytoophidia, or large rod and ring like structures, formed in HEK293 cells by GFP labeled IMPDH after introduction of the IMPDH inhibitor ribavirin (adapted with permission from Anthony *et al.* (2017)[128]). **C.** Negative stain EM micrograph of filamentous IMPDH induced with purified IMPDH in vitro with 5 mM $NAD^+$ and 1 mM ATP (adapted with permission from Anthony *et al.* (2017)[128]). The scale bar represents 50 nm. **D.** Single particle reconstructions using cryo-EM images of filaments of IMPDH formed with either ATP and $NAD^+$ (left) or ATP and GTP (right) (adapted with permission from Anthony *et al.* (2017)[128]). **e.** Model relating open (left) and collapsed (right) filaments of IMPDH (adapted with permission from Anthony *et al.* (2017)[128])

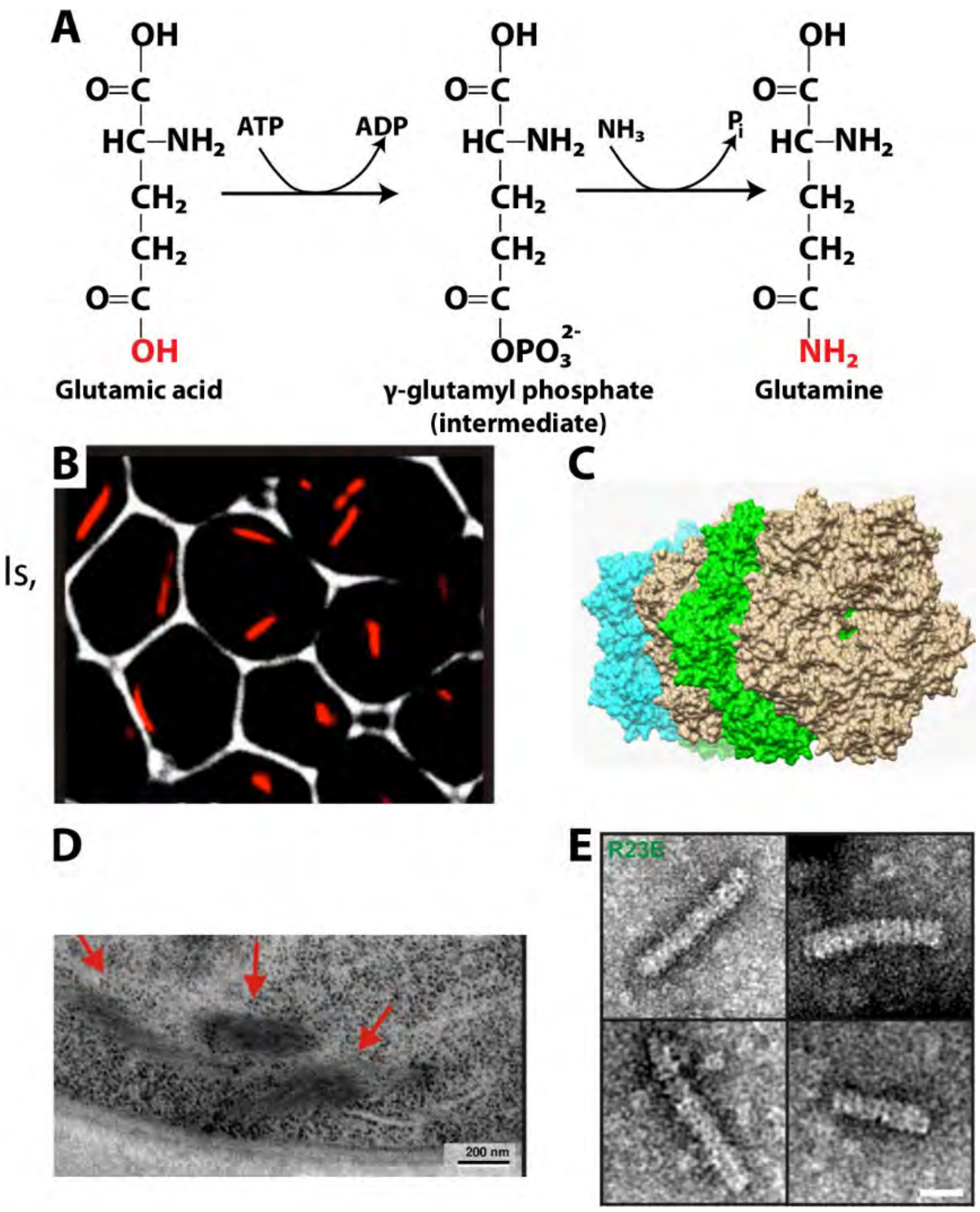

**Figure 6** Reactions catalyzed and structures formed by glutamine synthetase. **A.** Reaction catalyzed by glutamine synthetase (EC 3.5.1.2). **B.** mCherry labeled glutamine synthetase forms cytoophidia in yeast cells under nutrient starvation (adapted with permission from Petrovska *et al.* (2014)[101]). **C.** Asymmetric unit of x-ray crystal structure (PDB code 3FKY[131]) showing pentameric rings that stack head-to-tail to form a decamer, than further stacks with another decamer in the head-to-head fashion. **D.** Electron microscopy of glutamine synthetase cytoophidia in yeast cells showing stacked filamentous fine structure (adapted with permission from Petrovska *et al.* (2014)[101]). **E.** Negative stain EM of the constitutive cytoophidia forming mutant R23E of yeast glutamine synthetase showing filaments that appear to be composed of stacked rings. (adapted with permission from Petrovska *et al.* (2014)[101])

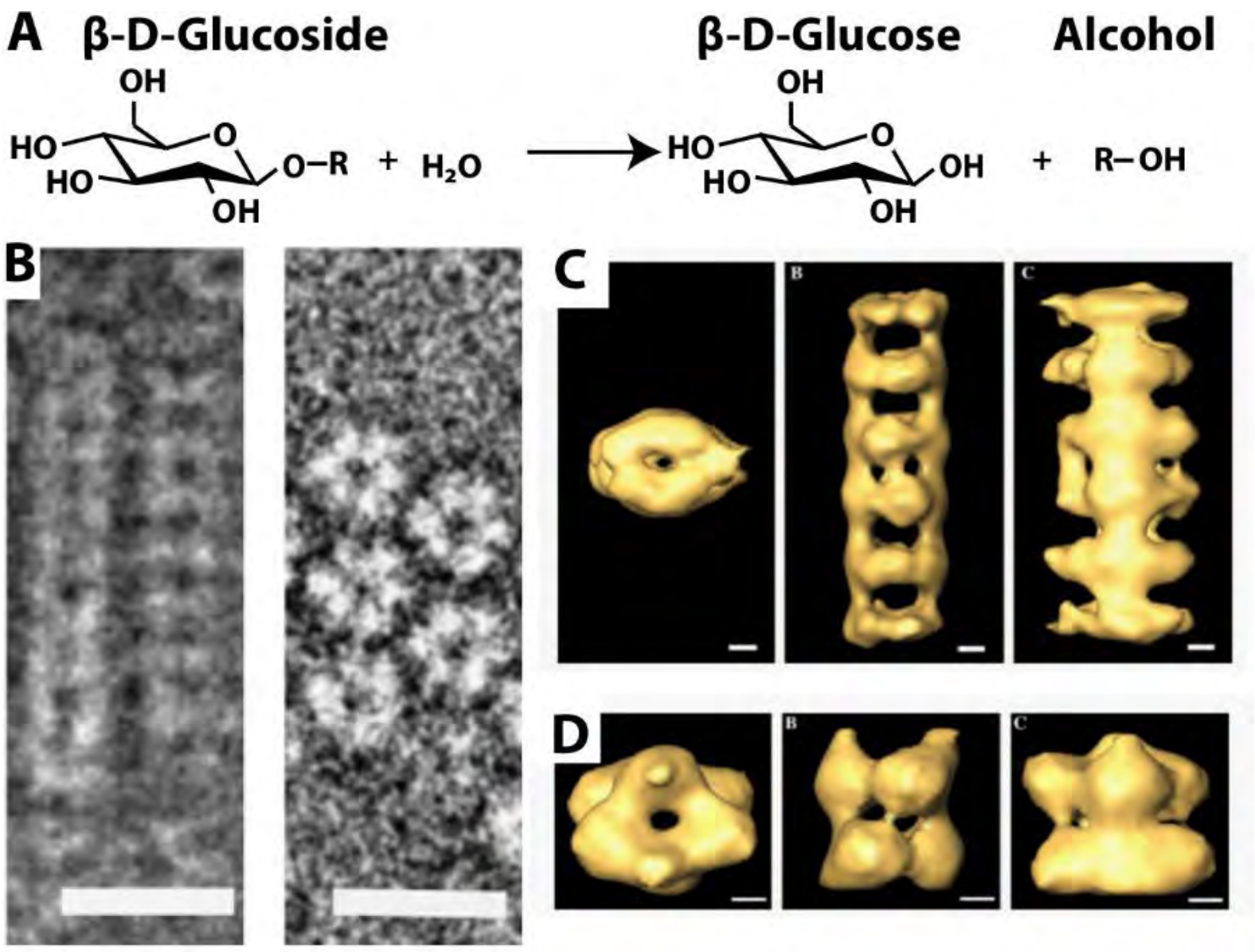

**Figure 7.** Reaction catalyzed and filamentous structures formed by oat β-glucosidase. **A.** Reaction catalyzed by oat β-glucosidase (EC 3.2.1.21). **B.** Negative stain EM micrograph of filamentous oat β-glucosidase (adapted with permission from Kim *et al.* (2005)[141]). **C.** Single particle reconstruction using cryo-EM images of oat β-glucosidase showing the filamentous form which forms from stacks of hexamers (themselves made of stacked trimeric rings). Scale bars are 20 nm (adapted with permission from Kim *et al.* (2005)[141]). Left, top view, middle and right, side views. **D.** Hexameric unit from filamentous form shown in C, with additional symmetry applied during refinement. Scale bars are 2 nm (adapted with permission from Kim *et al.* (2005)[141]). Left, top view, middle and right, side views

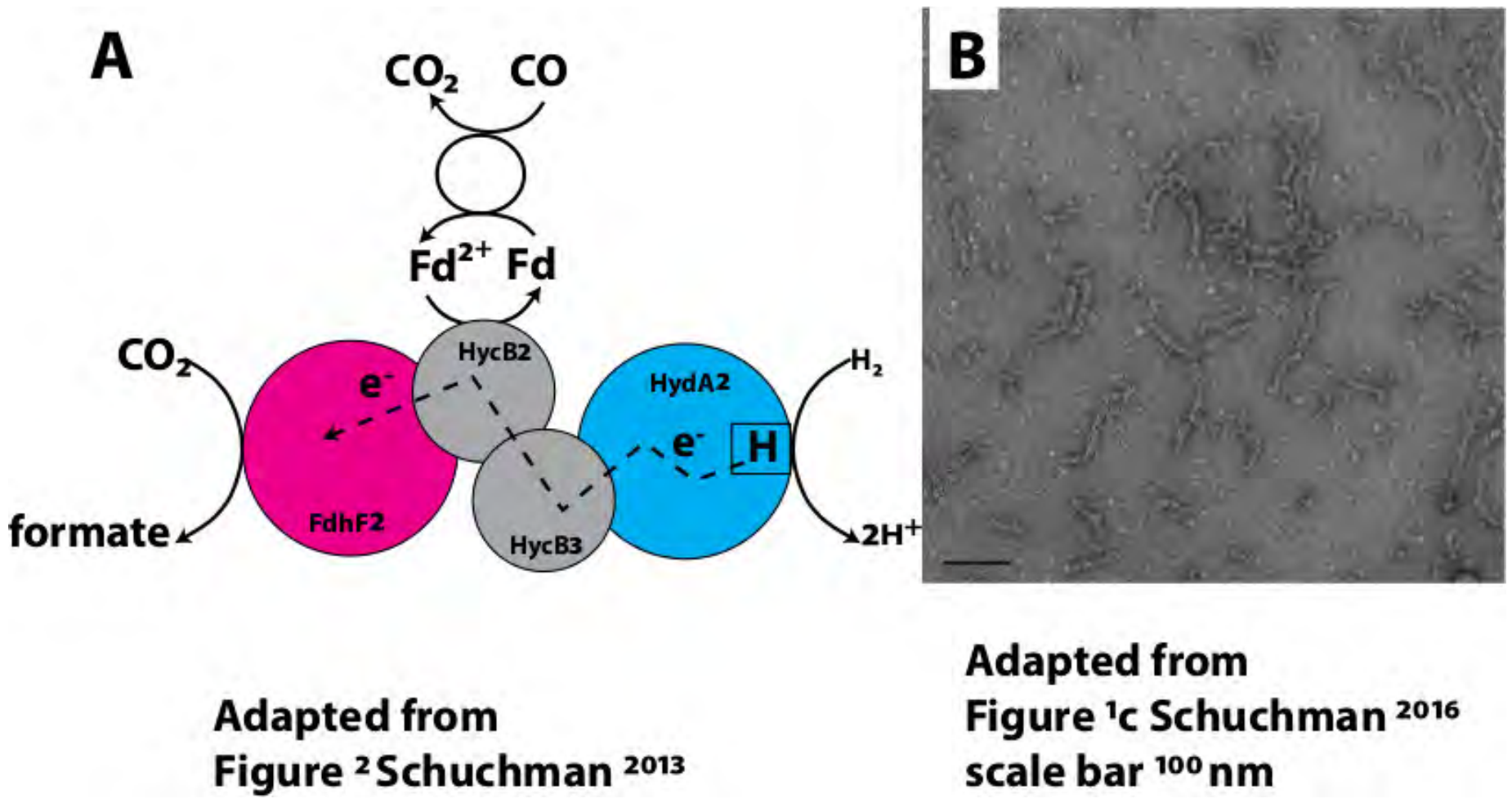

**Figure 8.** Reaction catalyzed and assemblies formed by *A. woodii* $CO_2$ reductase. **A.** Enzyme reaction carried out by hydrogen dependent $CO_2$ reductase (HDCR) from *A. woodii*. Electrons for $CO_2$ reduction are either provided by the hydrogenase subunit HydA2, where hydrogen oxidation takes place, or by reduced ferredoxin (Fd). Electrons are delivered to the active site for $CO_2$ reduction in FdhF2 via the electron-transferring subunits HycB2/3. Fdh, formate dehydrogenase, Hyd, hydrogenase, CODH, CO dehydrogenase (adapted with permission from Schuchman *et al*. (2013)[145]). **B.** Negative stain EM micrograph of $CO_2$ reductase showing filamentous assemblies (adapted with permission from Schuchman *et al.* (2016)[148]). Scale bar represents 100 nm.

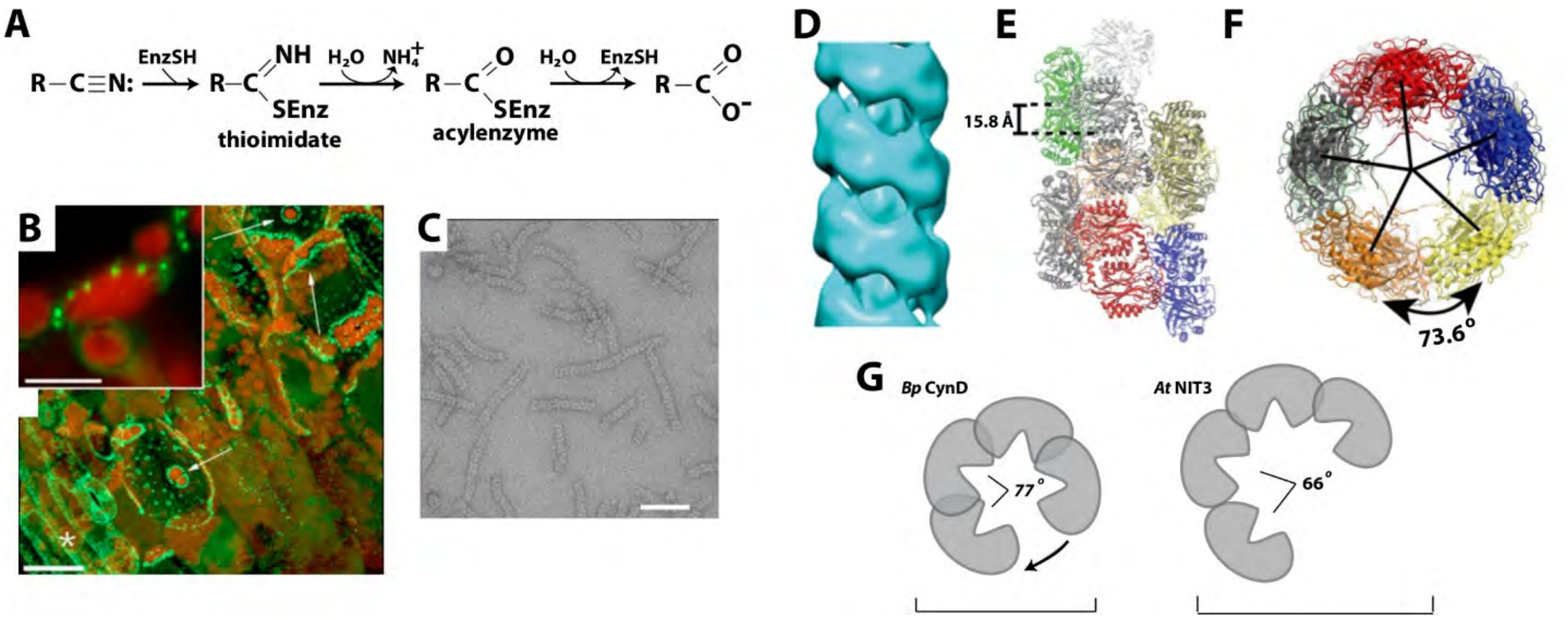

**Figure 9.** Reactions catalyzed and structures formed by nitrilases. **A.** An example of one of the many reactions carried out by nitrilase superfamily members (EC 3.5.5.1). **(**adapted with permission from Pace & Brenner (2001)[150].) **B.** In a plant cell, GFP labeled Nitrilase forms self-assemblies (green) around chloroplasts (red) following cell injury. Scale bars represent 10 microns (top) and 25 microns (bottom). (adapted with permission from Cutler *et al*. (2005)[21]). **C.** Negative stain EM micrograph of *Capsella rubella* nitrilase 2 filaments formed *in vitro*. Scale bar represents 50 nm (adapted with permission from Woodward *et al*. (2018)[315]). **D.** Single particle reconstruction of *Capsella rubella* nitrilase 2 from negative stain EM (adapted with permission from Woodward *et al.* (2018)[315]) **E-F.** Orthogonal views of model built from flexible fitting of the *Rhodococcus rhodochrous* nitrilase structure into the EM reconstruction (adapted with permission from Chan *et al*. (2011)[323]). **G.** Cartoons of two nitrilase structures (*Bacillus pumilus* CynD and *Arabidopsis thaliana* nitrilase 3)( adapted with permission from Woodward *et al.* (2018)[315]).

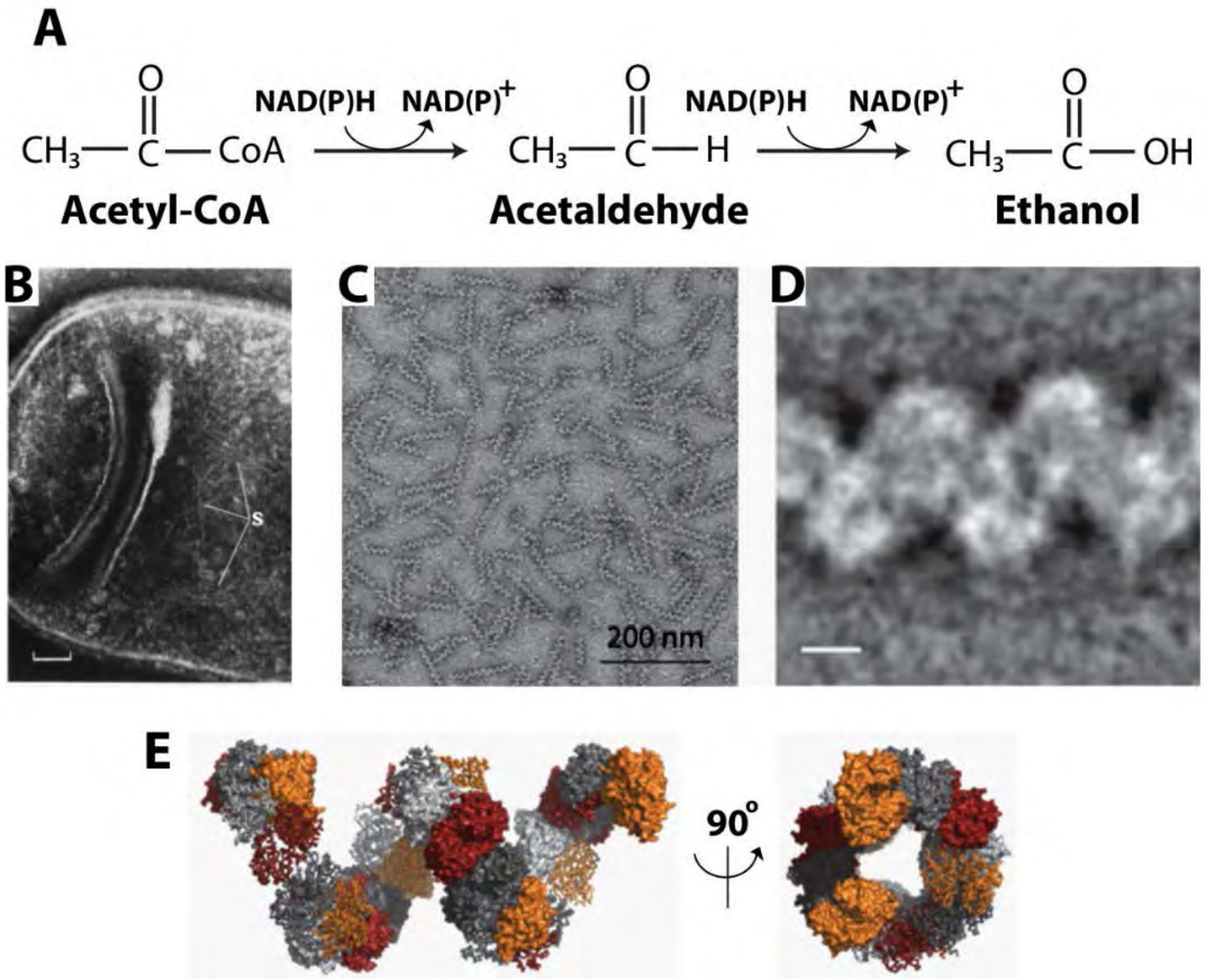

**Figure 10.** Reactions catalyzed and structures formed by AdhE. **A.** An example reaction carried out by AdhE (EC:1.2.1.10). This enzyme can follow the same pathway with longer aliphatic chains including proprionyl-CoA, valeryl-CoA, butyryl-CoA, and hexanoyl-CoA. **B.** Formation of helical structures in cells viewed using negative stain EM. Portion of a partially lysed cell found in the cytoplasmic membrane fraction obtained from *L. casei Var. casei* 1001. Fine spirals (S) are seen in a cluster within the degraded cytoplasm. Scale bar is 100 nm. (adapted with permission from Kawata, *et al.* (1975)[157]). **C.** Negative stain EM micrographs of immunopurified *Streptococcus pneumoniae* **"**spiralosomes", helical structures formed by AdhE. (adapted with permission from Laurenceau *et al.* (2015)[158]). **D.** A representative class average of the *S. pneumoniae* spiralosome. Scale bar is 5 nm (adapted with permission from Laurenceau *et al.* (2015)[158]). **E.** Two views of a spiralosome model built with the x-ray crystal structure of *G. thermoglucosidasius* spirosome[159]. The color of the individual protomers alternate between grayscale and domain-coded color representation, and successive AdhE dimers alternate between ribbon and surface representation (adapted with permission from Laurenceau *et al.* (2015)[158]).

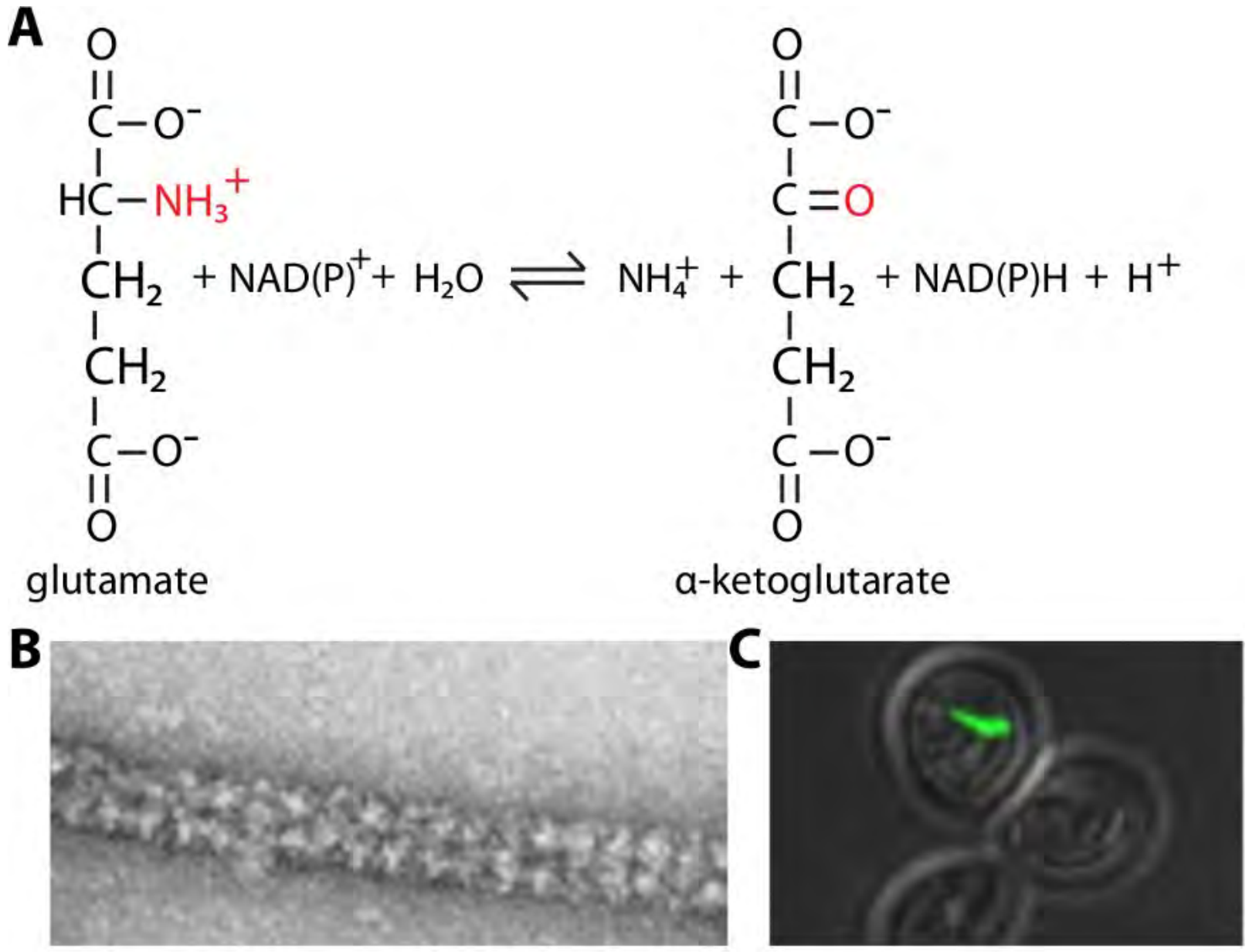

**Figure 11**. **Reactions catalyzed and structures formed by glutaraldehyde dehydrogenase (GDH). A.** Reaction catalyzed by glutamate dehydrogenase (EC 1.4.1.2). **B.** Helical tubes of bovine GDH in vitro by negative stain EM (adapted with permission from Josephs and Borisy (1972)[14]). **C.** GFP labeled GDH form a rod-like structure or "cytoophidia" in cells (yeast) under nutrient starvation conditions (adapted with permission from Shen *et al.* (2016)[26]).

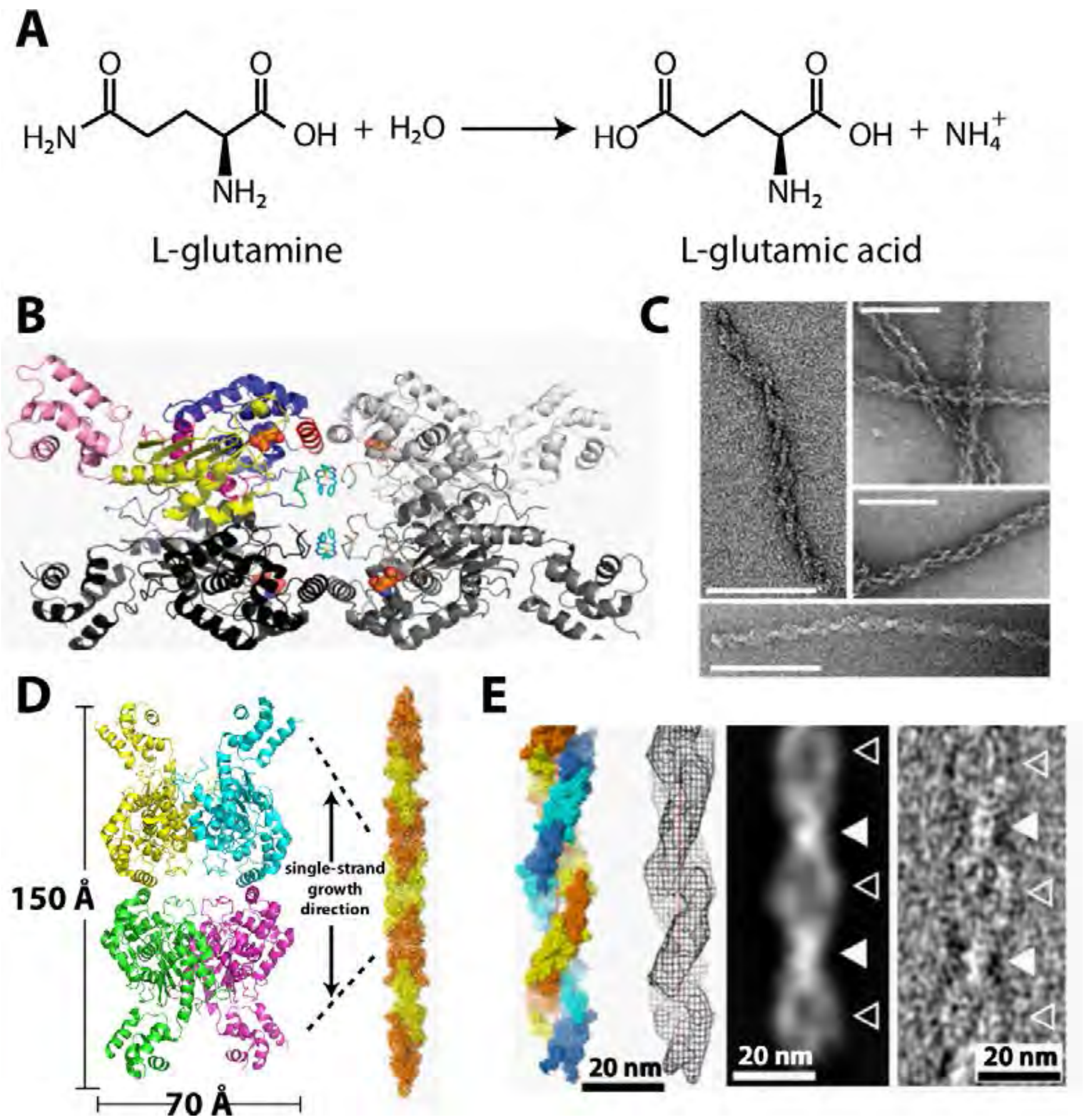

**Figure 12**. **Reaction catalyzed and structures formed from glutaminase. A**. Reaction catalyzed by glutaminase (EC 3.5.1.2). **B.** Structure of full length human glutaminase bound to the inhibitory molecule BPTES (adapted with permission from DeLaBarre *et al.* (2011)[185]). **C.** Negative Stain TEM of crosslinked glutaminase filaments (scale bars represents 100 nm)**(**adapted with permission from Ferreira *et al.* (2013)[195]). **D.** A model of one of the two strands of the double helical glutaminase filaments **(**adapted with permission from Ferreira *et al.* (2013)[195]). **E.** A model of the double helical filament of glutaminase **(**adapted with permission from Ferreira *et al.* (2013)[195]).

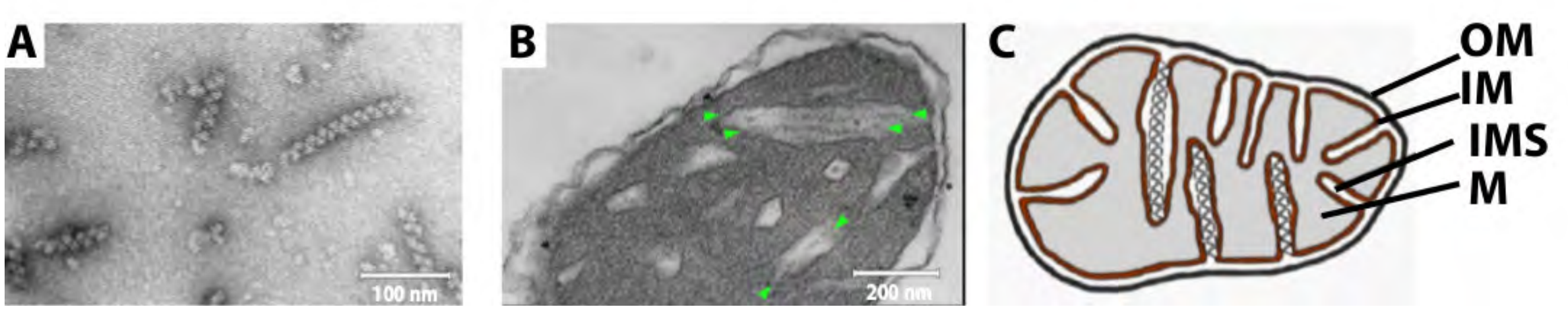

**Figure 13. A.** Filaments of purified LACTB viewed by negative stain EM (adapted with permission from Plianskyte *et al.* (2009)[199]). **B.** Thin section of a rat liver mitochondrion with filaments in the cristal part of the intermembrane space. (adapted with permission from Plianskyte *et al.* (2009)[199]). **C.** Schematic of location of LACTB filaments in the intra-cristal part of the intermembrane space of mitochondria. IM, inner mitochondrial membrane, OM, outer mitochondrial membrane, IMS, intermembrane space, M, matrix. (adapted with permission from Plianskyte *et al.* (2009)[199]).

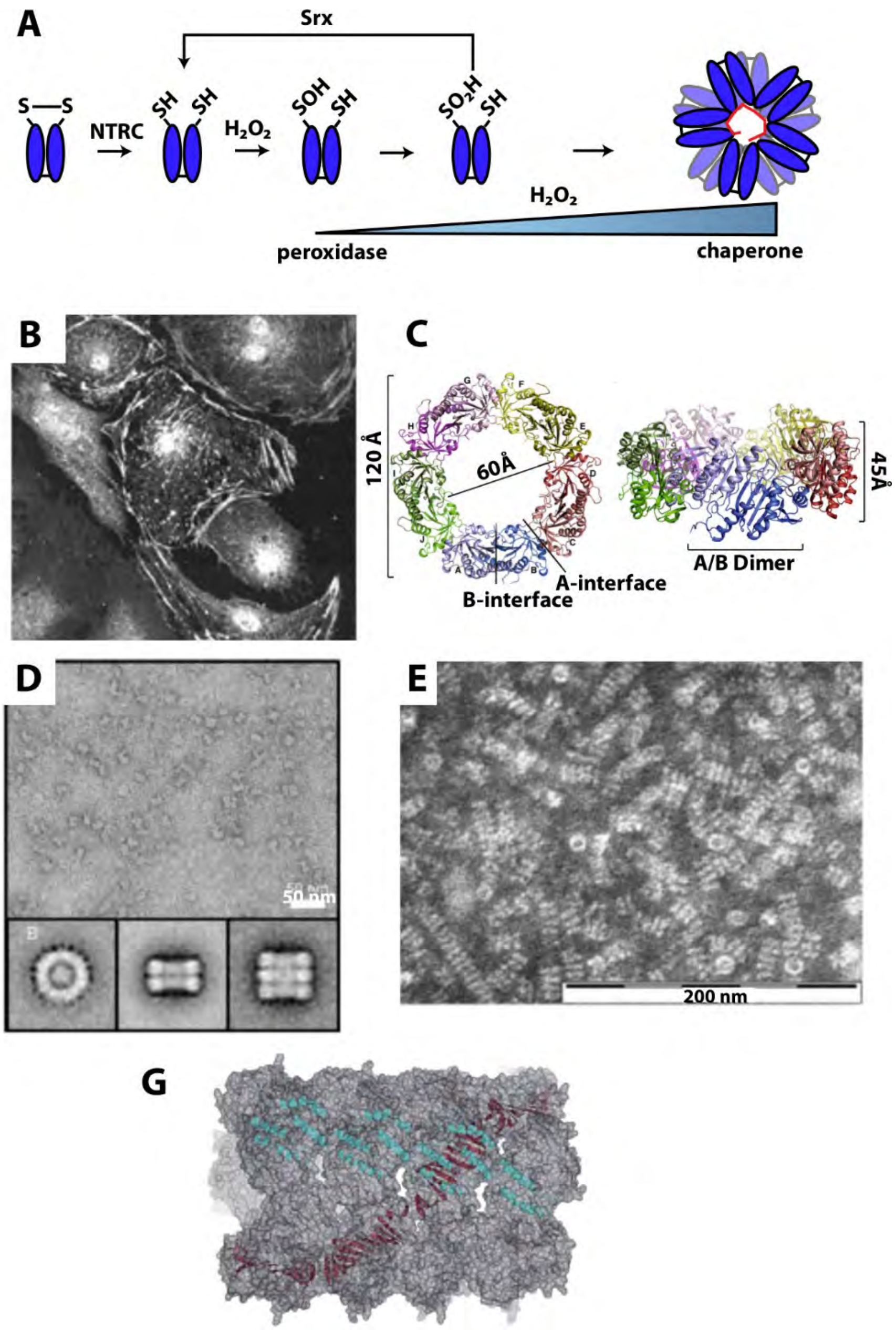

**Figure 14.** Reactions catalyzed and structures formed from 2-Cys peroxiredoxin (2CP). **A.** Under oxidant conditions, the sulfenic acid intermediate of the peroxidatic cysteine residue may be further oxidized to sulfinic acid. The reduction of the enzyme, which is most efficiently performed by NADPH-dependent thioredoxin reductase C (NTRC), is a pre-requisite for sulfenic acid formation and, thus, for overoxidation. Srx is able to catalyze the reversion of the overoxidized to the reduced form of the enzyme. Therefore, the redox status of chloroplast 2-Cys Prxs is highly dependent of NTRC and Srx. The quaternary structure of 2-Cys Prxs determines the activity of these enzymes. In the reduced form the enzyme is a dimer and shows peroxidase activity; overoxidation as occurs in oxidative stress favors the formation of the decameric form, which lacks peroxidase activity, can stack into filaments, and shows chaperone activity. (adapted with permission from Puerto-Galán *et al.* (2013)[201]). **B.** Synchronized C19 mouse epithelial cells treated with glucose oxidase then stained for 2CPrx. (adapted with permission from Phalen *et al.* (2006)[324]). **C.** Toroidal decamer of *Schistosoma mansoni* 2CPrx. (adapted with permission from Saccoccia *et al.* (2012)[207]). **D.** Negative stain EM micrograph of bovine CPrx showing stacked rings. (adapted with permission from Gourlay *et al.* (2003)[325]). **E.** Negative stain TEM of recombinant C48S mutant of CPrx from *Schistosoma mansoni* that results in constitutive chaperone activity (and constitutive filament formation)(adapted with permission from Angelucci *et al.* (2013)[202]). **F.** Negative stain EM micrograph of bovine C47S CPrx

showing filaments (adapted with permission from Gourlay *et al.* (2003)[325]). **G.** Theoretical model of filamentous 2CPrx from from stacked decameric rings. Cyan and red show alpha and beta structures, respectively, from subunits of each ring and their relative orientations (adapted with permission from Saccoccia *et al.* (2012)[207].

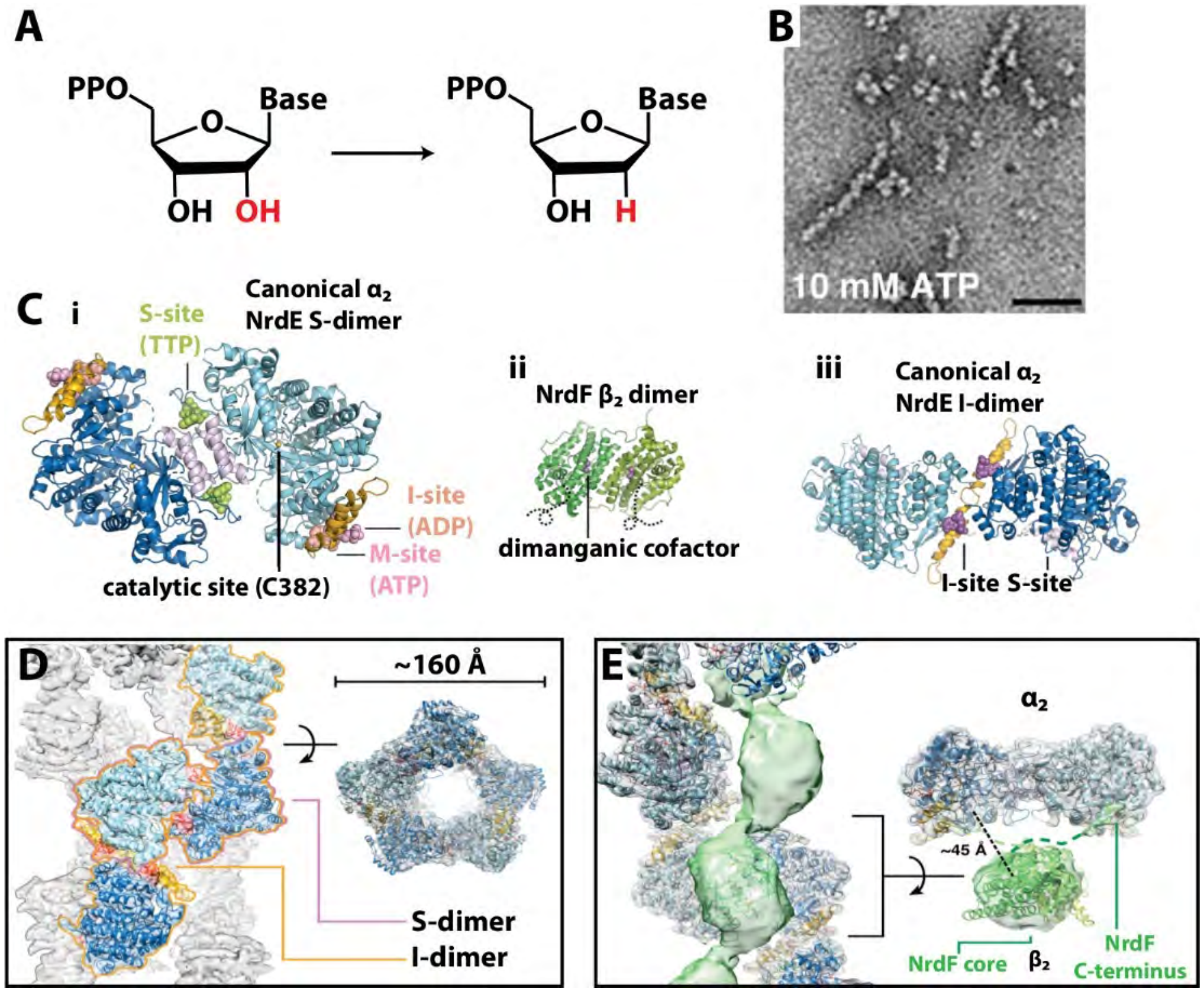

**Figure 15. A.** Reaction catalyzed by ribonucleotide reductase (EC 1.17.4.1). **B.** Negative stain TEM showing filaments of human RNR. Scale bar represents 50 nm (adapted with permission from Ando *et al.* (2016)[209]). **C.** X-ray crystallographic structures of *B. subtilis* RNR dimers. **i.** S-type dimer of NrdE (α subunit of RNR). **ii.** I-type dimer of NrdE (α subunit of RNR). **iii.** Dimer of NrdF (β subunit r RNR)(adapted with permission from Thomas *et al.* (2018)[210]). **D.** Cryo-EM structure of *B. subtilis* RNR filament of NrdE bound to dATP (adapted with permission from Thomas *et al.* (2018)[210]). **E.** Cryo-EM structure of *B. subtilis* RNR filament of NrdE and NrdF with dATP (adapted with permission from Thomas *et al.* (2018)[210]).

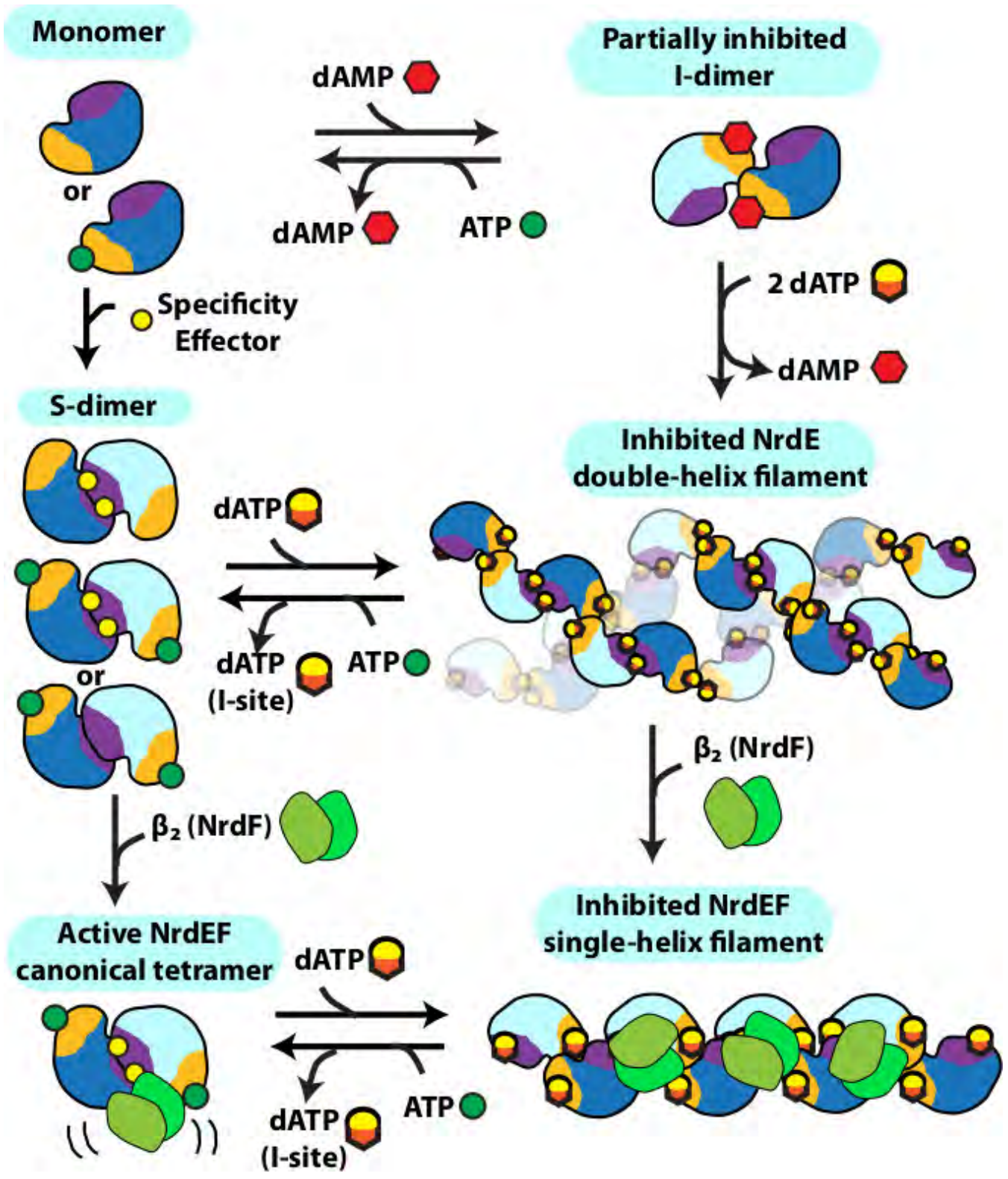

**Figure 16. Model for allosteric regulation of *B. subtilis* RNR**. Without nucleotides, NrdE is a monomer, but dAMP and dATP can bind the I-site and induce a partially inhibited I-dimer. Addition of specificity effectors (dATP, TTP, or dGTP) instead induces the monomer to form an S-dimer. When specificity effectors (including dATP) bind to the I-dimer, they induce formation of an inhibited double-helical NrdE filament composed of alternating S- and I-dimer interfaces. NrdF competes for the NrdE double-helical interface, and thus NrdF binding leads to the dissociation of the NrdE double-helix into an individual helical structure. Binding of NrdF to the NrdE filament leads to an inter-subunit gap that is too large to allow for radical transfer. Both the NrdE and NrdEF filaments are reversible by addition of ATP, which can displace dATP from the I-site and induce dissociation of the I-dimer interface. Finally, addition of NrdF to the S-dimer leads to formation of an active but asymmetric $\alpha 2\beta 2$ tetramer in which a hinge motion between the two subunits plays an important role in activity (adapted with permission from Thomas *et al.* (2018)[210]).

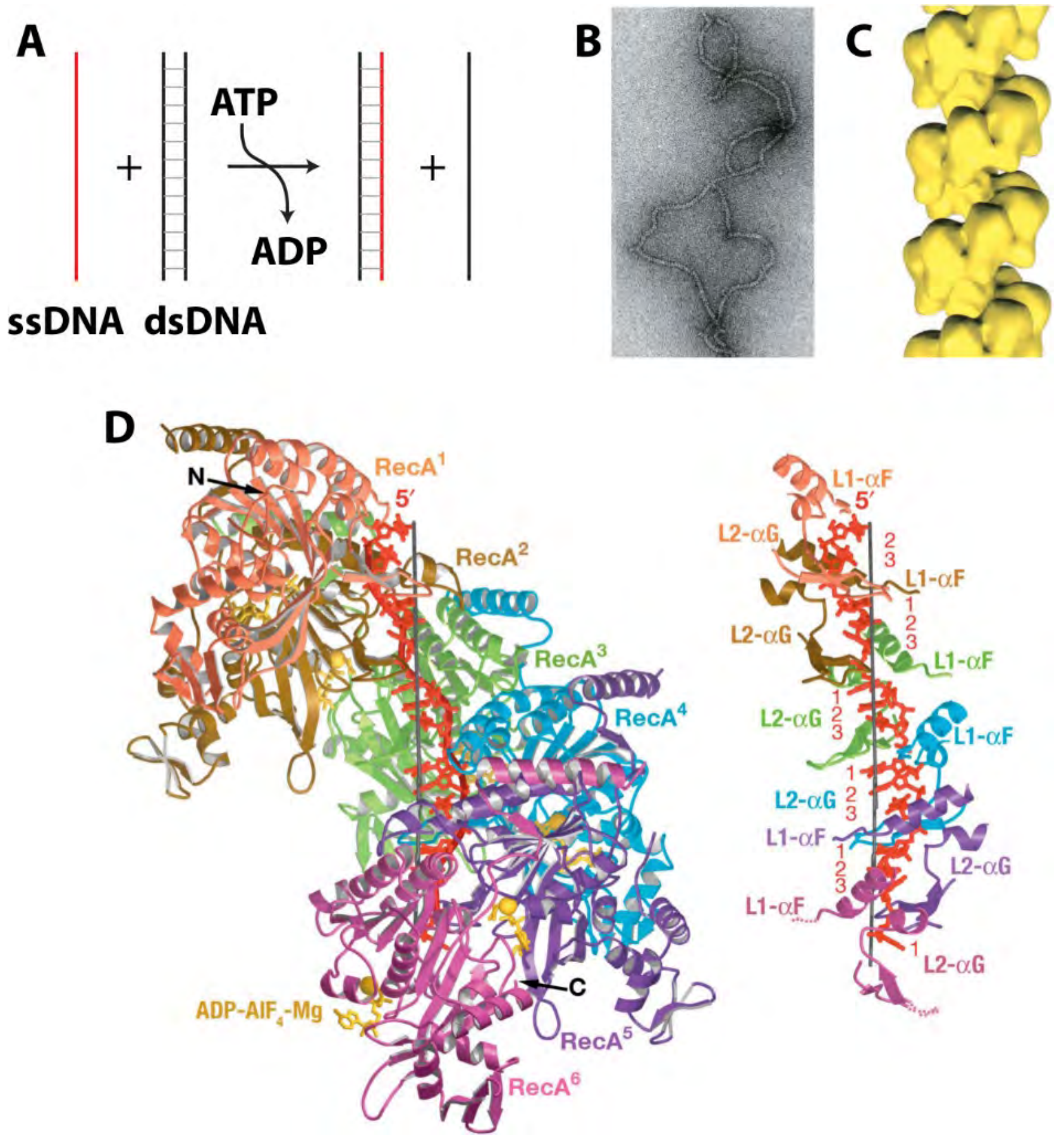

**Figure 17. Reaction catalyzed and structures formed by RecA/Rad51. A.** Reaction of DNA strand exchange catalyzed by RecA/Rad51. **B.** Negative stain electron micrograph of human Rad51 with single stranded DNA, ADP and $AlF_4^-$ (adapted with permission from Yu *et al.* (2001)[326]). **C.** Helical reconstruction of filament shown in B (adapted with permission from Yu *et al.* (2001)[326]). **D.** X-ray crystal structure of *E. coli* RecA bound to single stranded DNA, ADP, and $AlF_4^-$ (adapted with permission from Chen *et al.* (2008)[214]) **.**

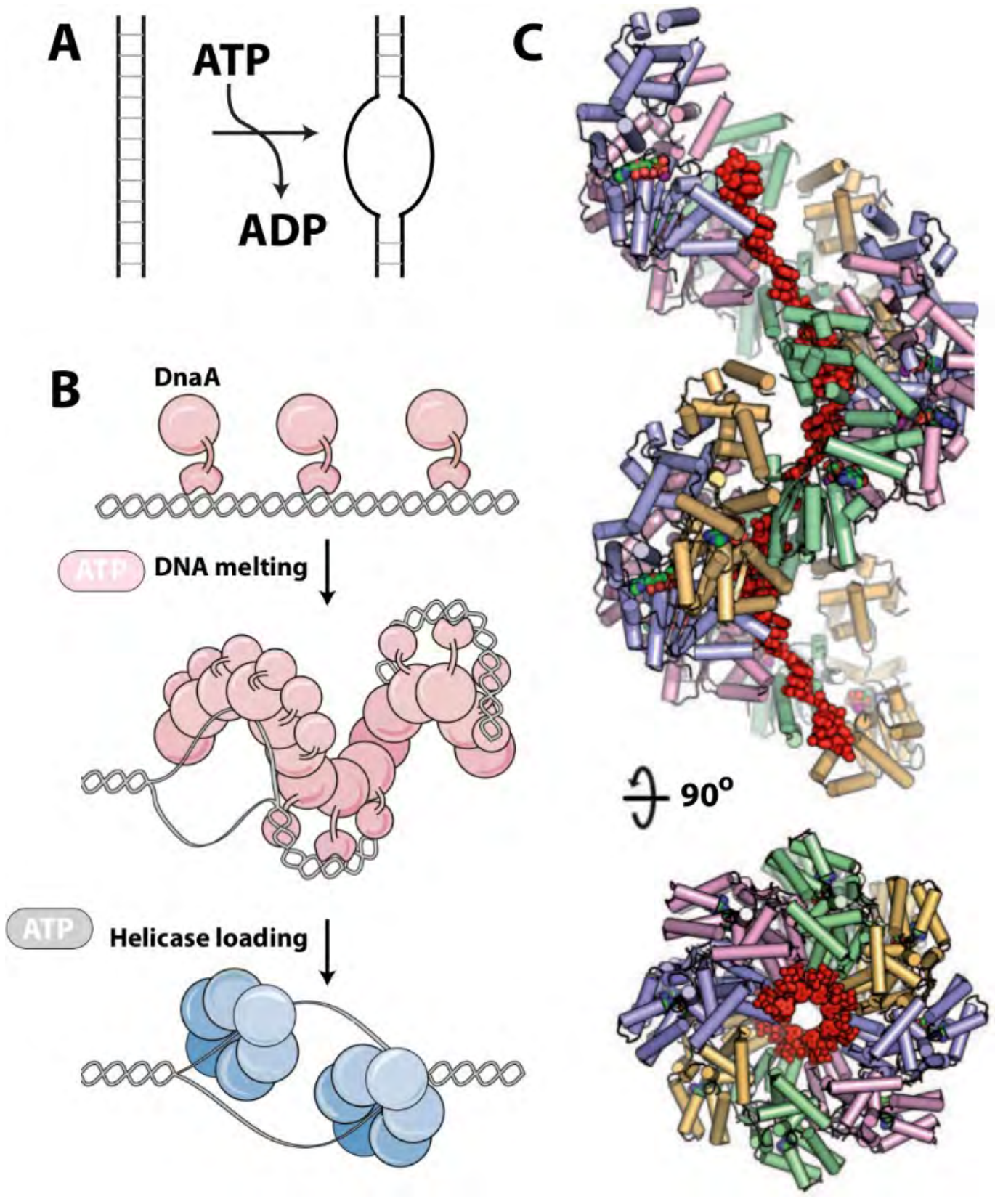

**Fig. 18. Reaction catalyzed and structures formed by *E. coli* DnaA. A.** ATP dependent DNA duplex opening reaction catalyzed by DnaA. **B.** Filament formation by DnaA on the E. coli origin of replication results in strand separation facilitating replicative helicase loading (adapted with permission from Bleichert *et al* (2017)[327]). **C.** X-ray crystal structure of *Aquifex aeolicus* DnaA bound to single stranded DNA (adapted with permission from Duderstadt *et al* (2011)[218]).

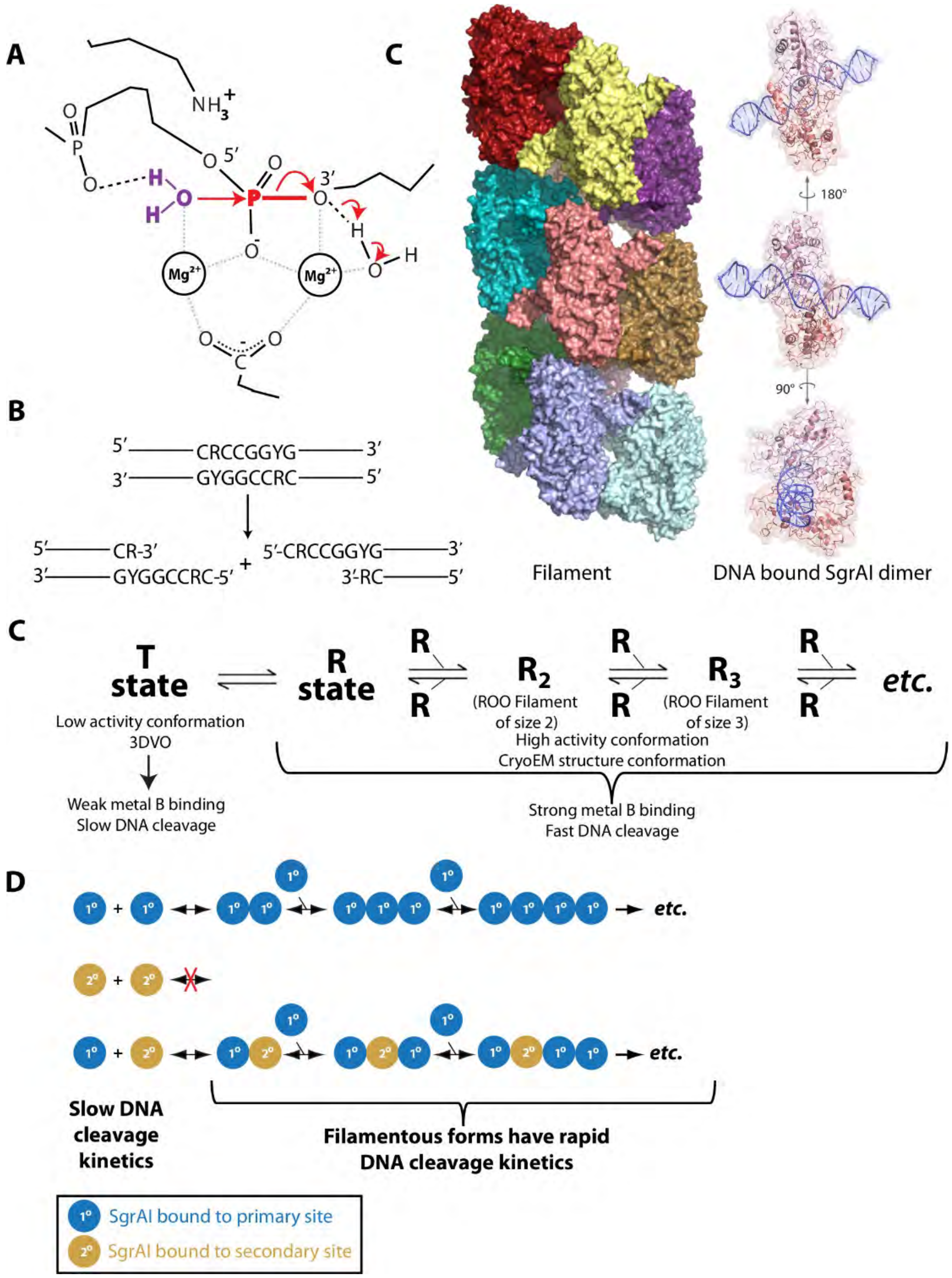

**Figure 19. Reactions catalyzed and structures formed by SgrAI. A-B.** Cleavage (hydrolysis) of double stranded DNA catalyzed by SgrAI, a type II restriction endonuclease (EC:3.1.21.4). **C.** (Left) Filament formed by SgrAI bound to primary site DNA, also known as a run-on oligomer. (Right) Three different views of a single SgrAI/DNA complex, the basic building block of the run-on oligomer filament. **C.** Working model of activation via filament formation by SgrAI. **D.** Working model for expansion of DNA sequence specificity of SgrAI in recruiting SgrAI bound to secondary site DNA sequences (gold balls) by filaments formed by SgrAI bound to primary sites (blue balls).

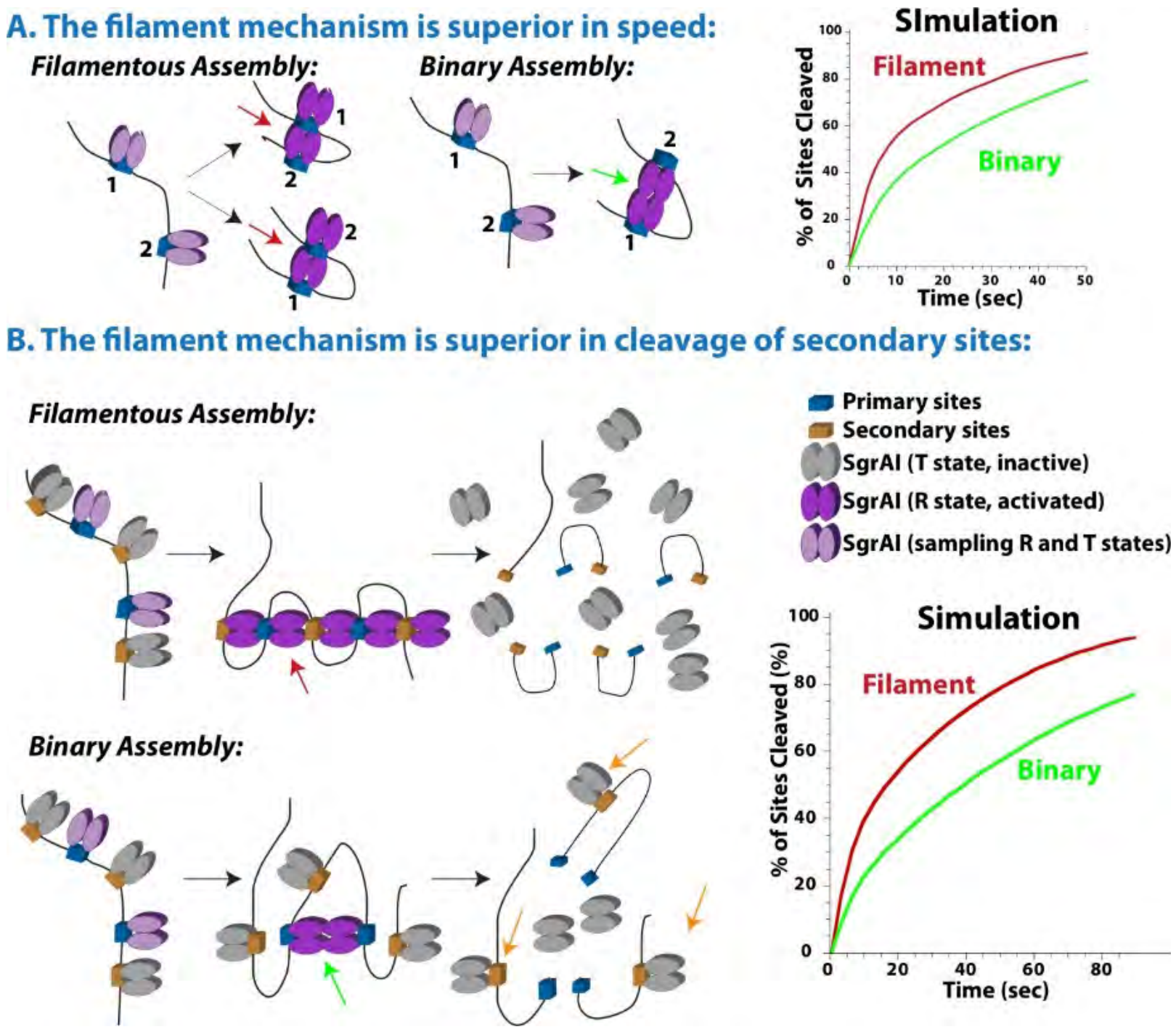

**Figure 20. The filament mechanism of SgrAI is superior to a non-filament, binary mechanism (forming a complex of 2 enzymes) in both speed and cleavage of secondary sites. A.** The greater speed derives from multiple ways enzymes can associate in the filament (left, red arrows) compared to a finite, discrete, binary non-filament model (middle, green arrow). Simulations show the greater enzymatic activity of the filament mechanism (red) over the non-filament or binary model (green)(right). **B.** The filament mechanism is superior in cleavage of secondary sites as well, since these are out-competed compared to primary sites in the binary model (lower, green and gold arrows), but do not compete in the filament model as more enzymes can always add to either end (top, red arrow).

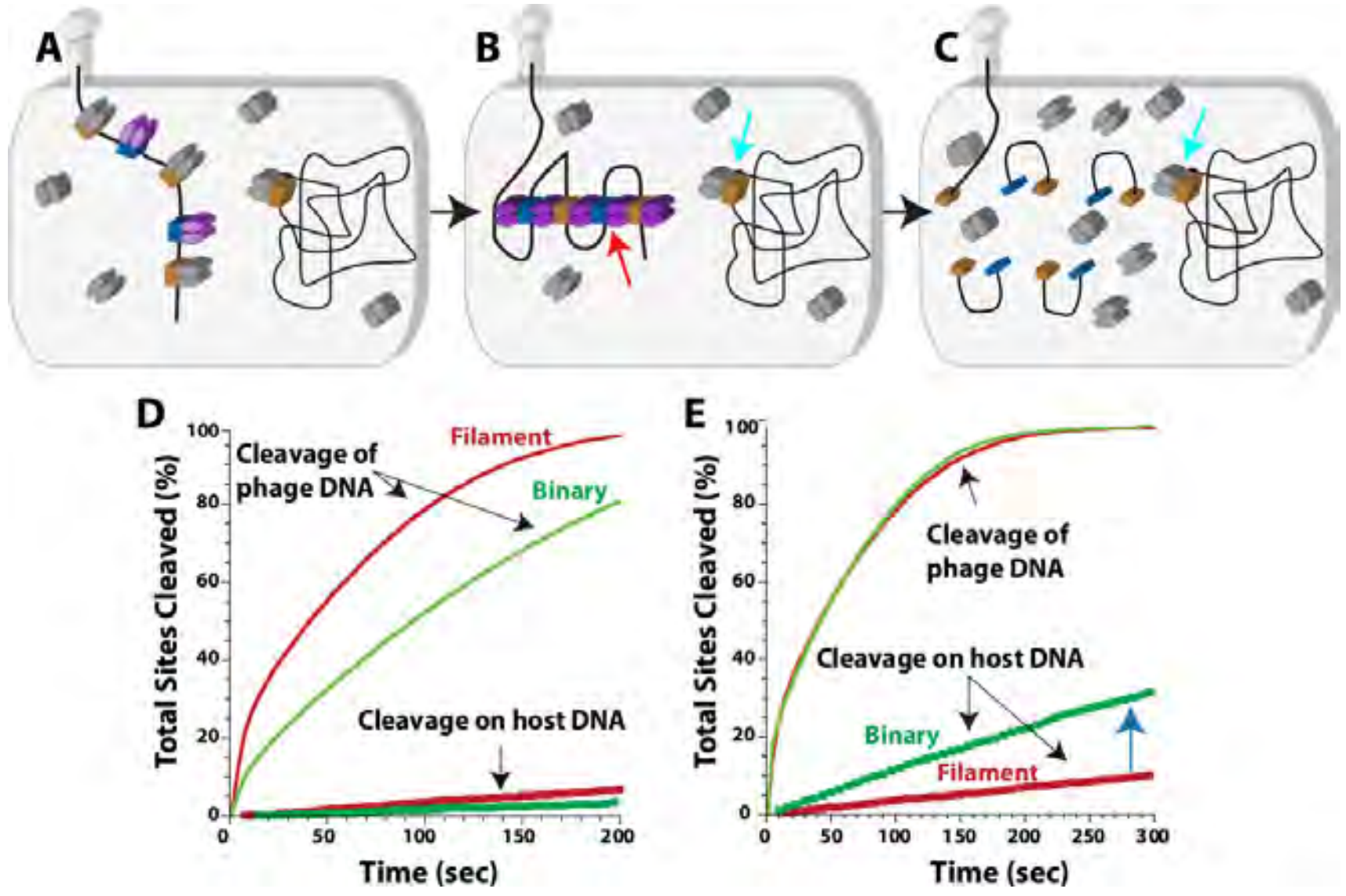

**Figure 21. The filament mechanism is superior in sequestration of secondary site cleavage activity**. **A.** SgrAI is inactive (grey) and binds to secondary sites (gold boxes) in the host DNA (right) and both primary (blue boxes) and secondary sites in the invading DNA. While bound to primary sites, SgrAI equilibrates between active and inactive states (partial purple color). **B.** SgrAI in the active conformation (purple) associates into a filament (red arrow), drawing in SgrAI bound to secondary sites, inducing their activation (purple color), and resulting in DNA cleavage. **C.** Both primary and secondary sites are cleaved on the invading DNA, but secondary sites on the host DNA are uncleaved (cyan arrow). **D.** Simulation using kinetic model derived from global data fitting showing the superior DNA cleavage kinetics of the filament mechanism (red solid line) over the non-filamentous binary model (green solid line). Cleavage of the host DNA is minimal in both models (thicker dotted lines). **E.** Simulation with increased association rate constant in the binary model only such that its DNA cleavage rate matches that of the filament mechanism on the invading DNA, however, damaging cleavage of the host DNA is now much larger in the binary model, indicating loss of sequestration (blue arrow).

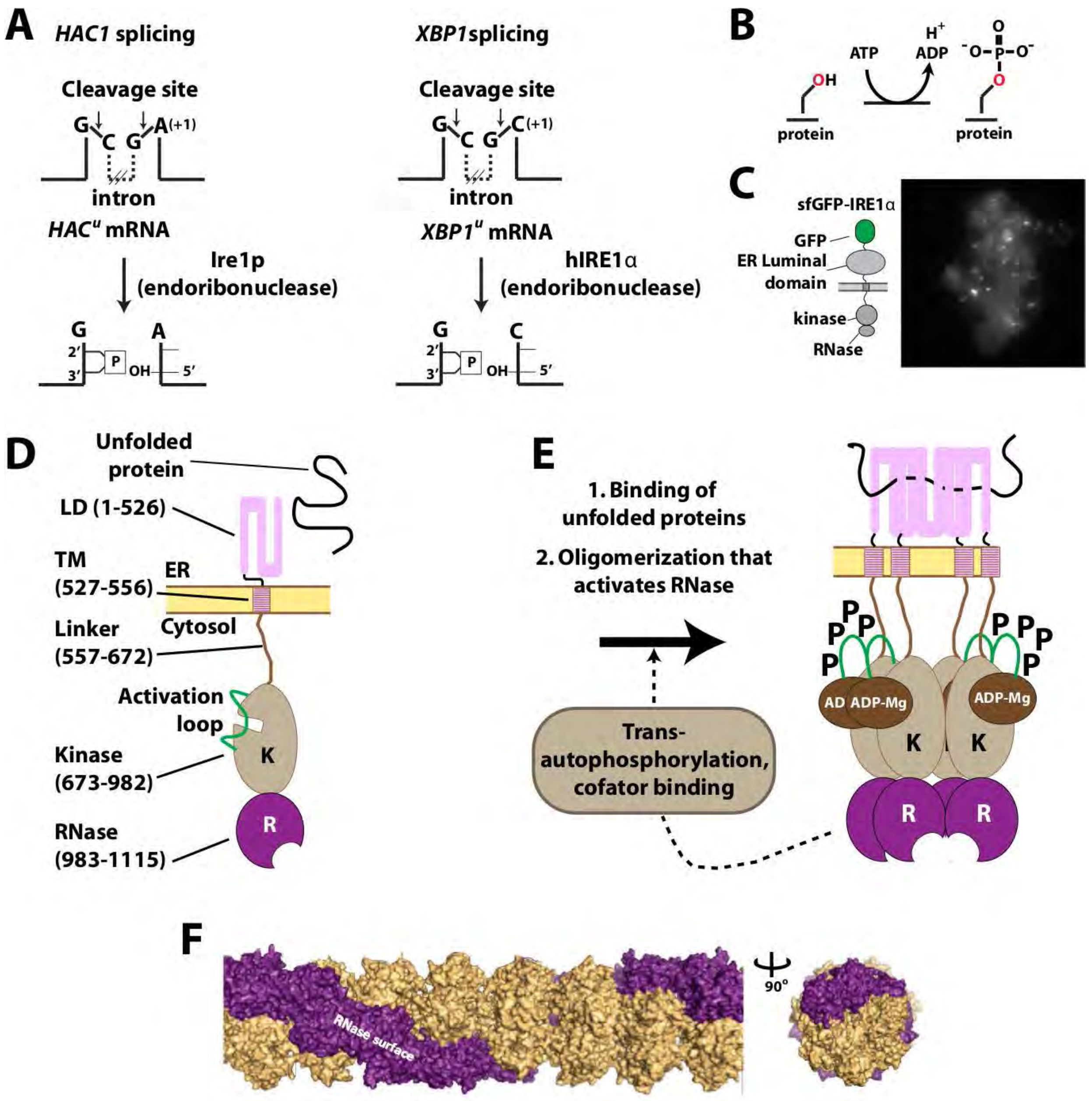

**Figure 22. Reactions catalyzed and structures formed by Ire1 (or hIRE1) A.** Splicing/RNase activity of Ire1/IRE1α (adapted with permission from Poothong *et al.* (2018)[328]). **B.** ATP dependent phosphorylation of protein side chains via kinase activity (IRE1 autophosphorylates by kinasing adjacent IRE1 chains when oligomerized). **C.** Human IRE1α forms foci in cells upon addition of the ER stress agent DTT (adapted with permission from Ghosh *et al* (2014)[329]). **D.** Schematic of IRE1. The kinase domain of IRE1 is colored light brown; the RNase domain is colored purple. TM, transmembrane domain (adapted with permission from Korennykh *et al.* (2009)[18]). **E.** The Splicing/RNase activity of IRE1 is activated upon oligomerization, induced via unfolded protein binding in the ER lumen (adapted with permission from Korennykh *et al.* (2009)[18]). **F.** Two orthogonal views of the filament formed by IRE1, determined via x-ray crystallography (adapted with permission from Korennykh *et al.* (2009)[18]).

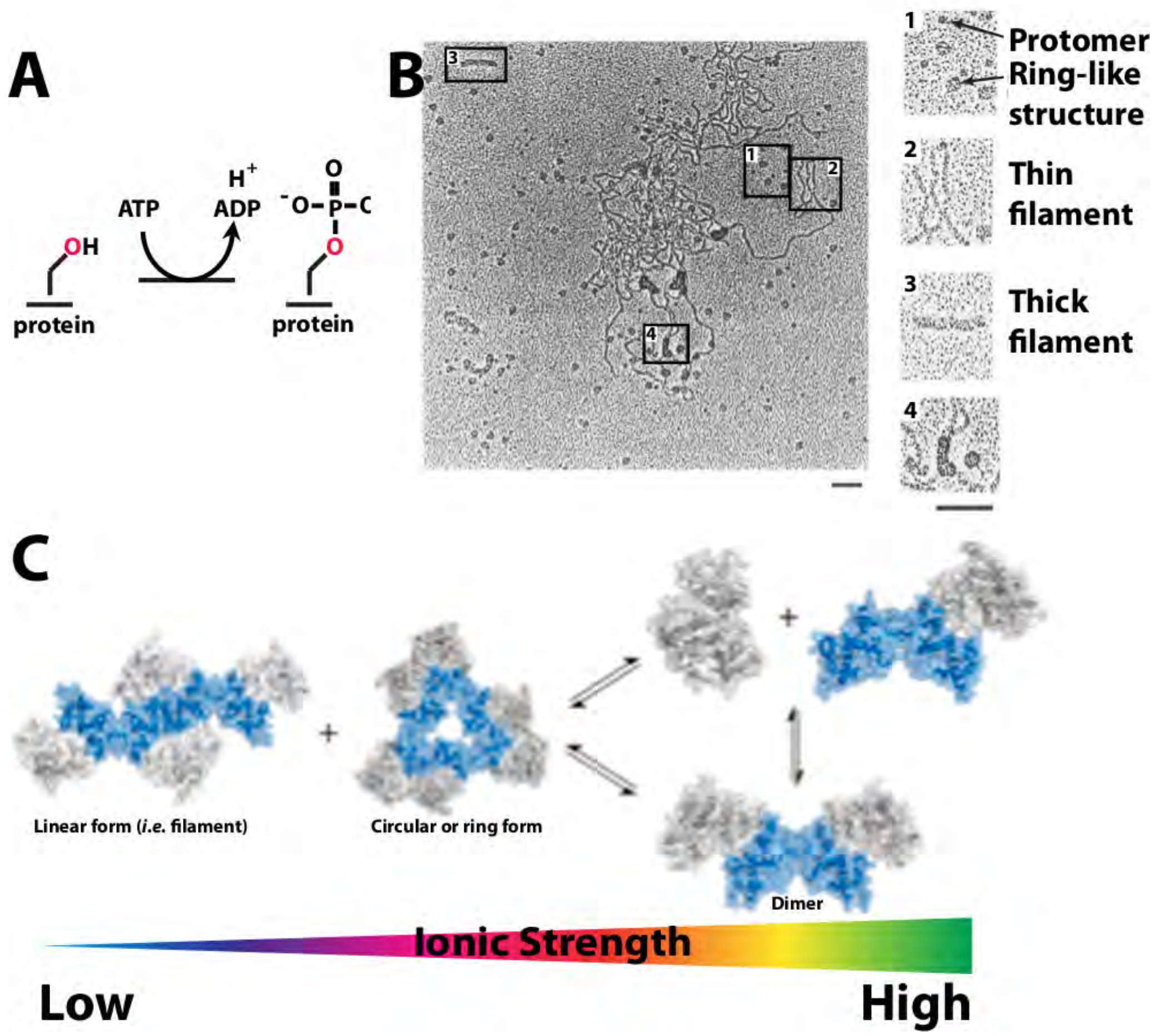

**Figure 23.** Reaction catalyzed and structures formed by casein kinase 2 (CK2). **A.** Protein kinase activity of CK2. **B.** Negative stain EM of CK2 showing different oligomeric and filamentous forms (adapted with permission from Valero *et al.* (1995)[266]). Scale bar corresponds to 100 nm. **C.** Schematic of interconversion of oligomeric forms of CK2 and their dependence on ionic strength (adapted with permission from Seetoh *et al.* (2016)[269]).

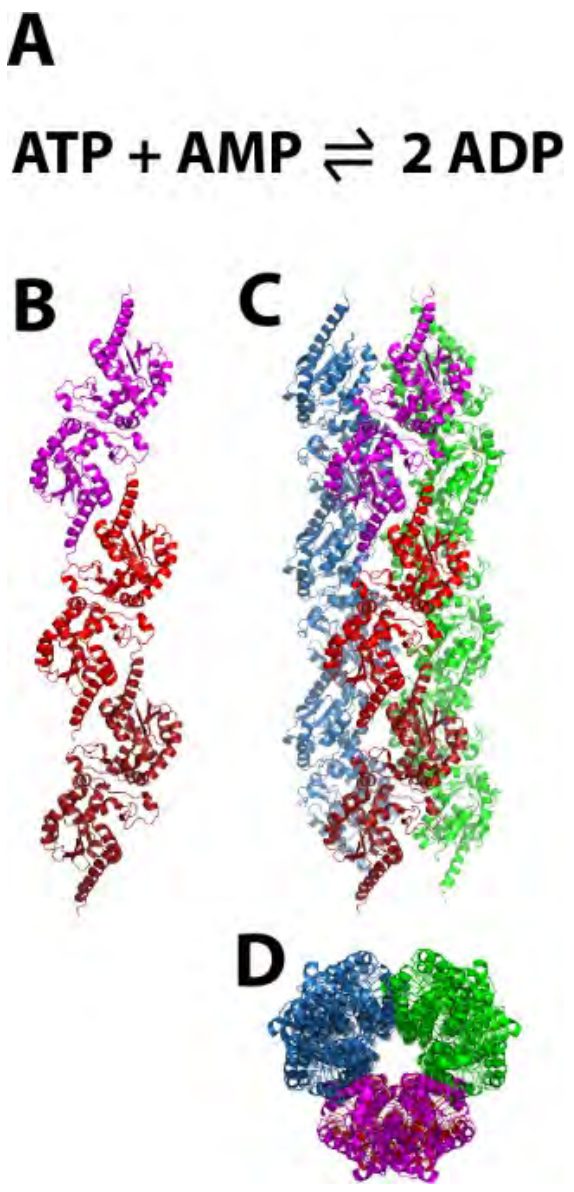

**Figure 24.** Reaction catalyzed by and **A.** Reaction catalyzed by maize adenylate kinase (EC 2.7.4.3). **B-D.** Crystallographic packing in crystals of AK from maize (PDB code 1ZAK) resembling a single infinite filament, three protomers shown and colored distinctly (**B**), three filaments assembled around a crystallographic 3-fold (**C**) and viewed from the top (**D**).

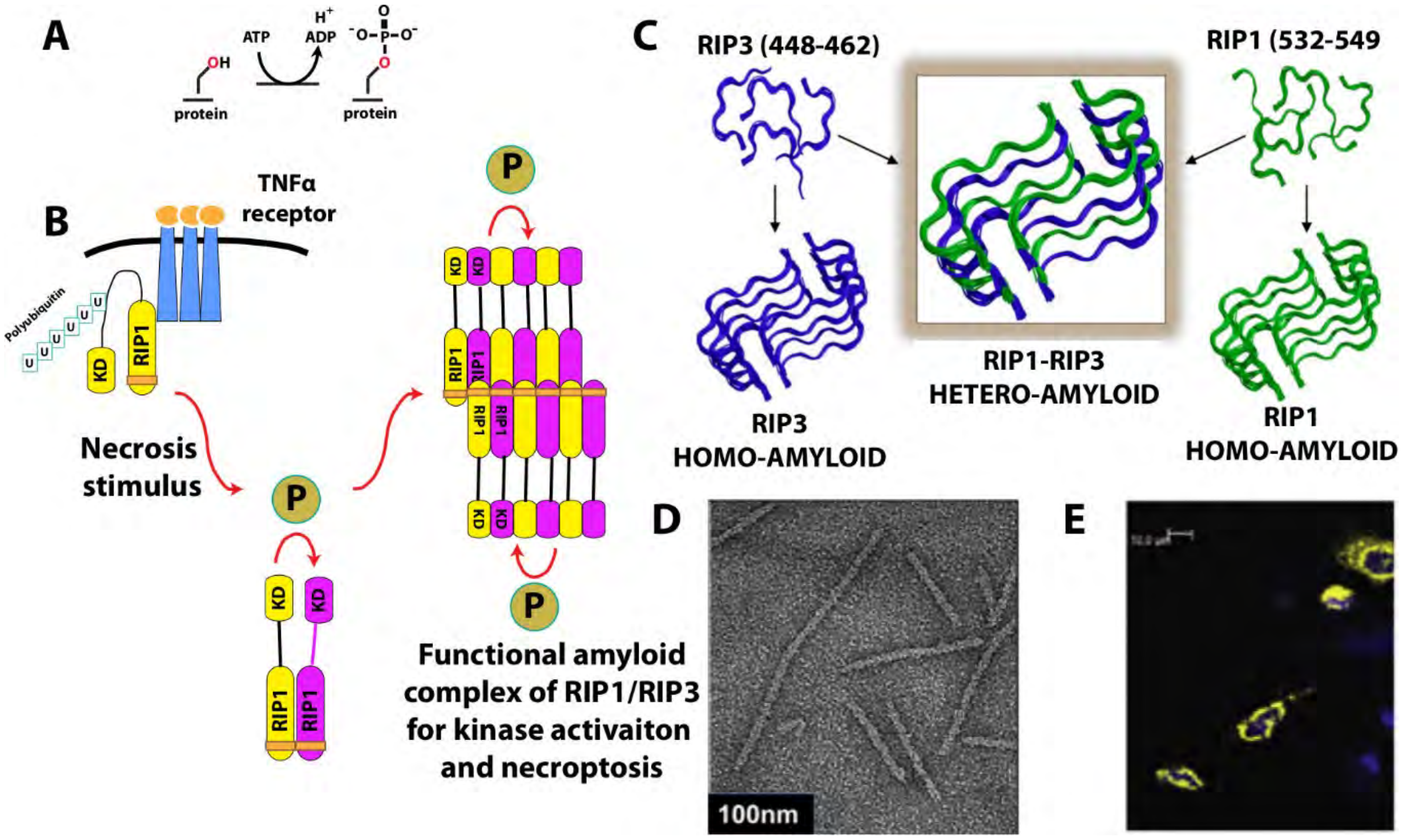

**Figure 25. Reactions catalyzed by and structures formed by RIP1/RIP3 kinases. A.** Protein phosphorylation reaction catalyzed by the kinase domain of RIP1 and RIP3. **B.** Formation of amyloid via the RHIM domain (brown) activates its kinase activity (adapted with permission from Li, *et al.* (2012)). **C.** Proposed structure of the RHIM or amyloid forming domains of RIP1 and RIP3 kinases (adapted with permission from Mompean, *et al.* (2018)[282]). **D.** Negative stain EM of filaments formed by RIP1 and the RHIM sequence of RIP3 (adapted with permission from Li, *et al.* (2012)). **E.** Foci formed in HeLa cells of YFP labeled RIP3 (adapted with permission from Li, *et al.* (2012)).

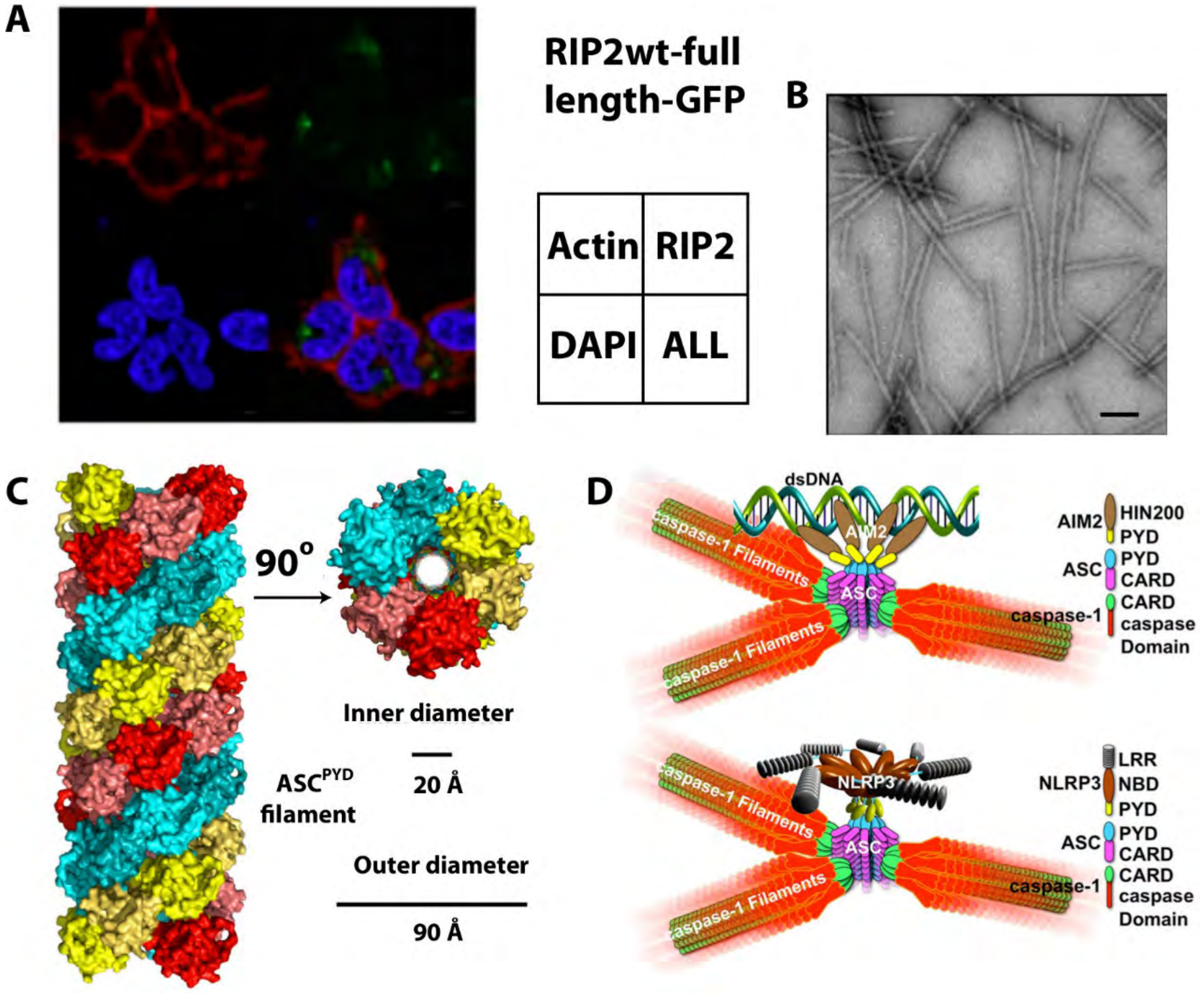

**Figure 26. A.** RIP2 forms foci in cells (adapted with permission from Gong *et al.* (2018)[289]). **B.** Negative stain EM image of RIP2-CARD filament (adapted with permission from Gong *et al.* (2018)[289]). **C.** Filament of the ASC PYD with C3 symmetry. The filament can be thought of as composed of 3 single strands in a right-handed helix (adapted with permission from Lu *et al.* (2014)[293]). **D.** Models of inflammasome signaling involving protein filamentation (adapted with permission from Lu *et al.* (2014)[293]).

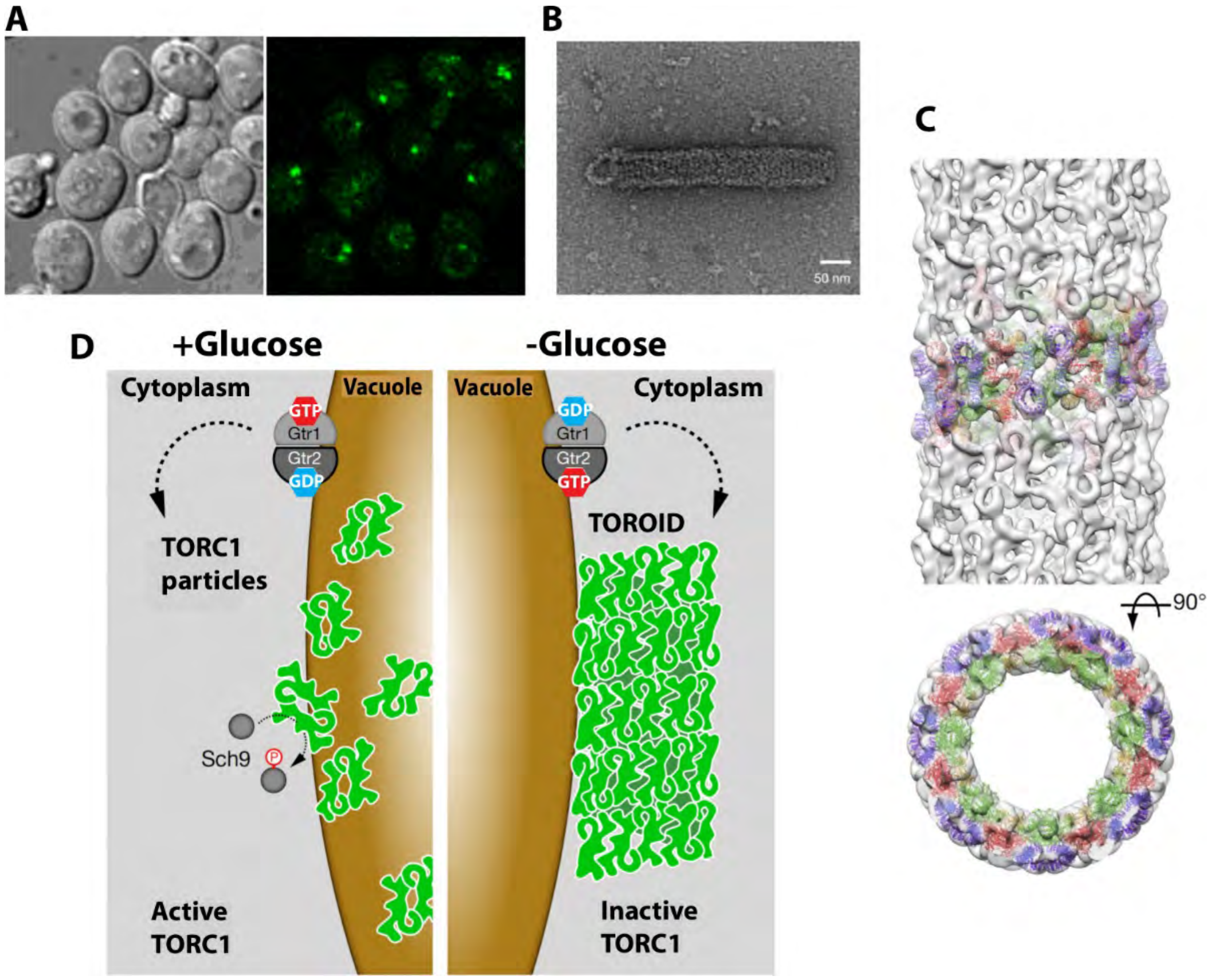

**Figure 27. A.** Foci formed by GFP labeled TORC1 (GFP-Kog1) under glucose starvation. Left, DIC image, right, confocal image (adapted with permission from Prouteau *et al.* (2017)[296]**). B.** Negative stain EM micrograph of TORC1 filament (adapted with permission from Prouteau *et al.* (2017)[296]). **C.** Single particle reconstruction at 27 Å resolution of TORC1 filament with constituent coordinate files fit into density (adapted with permission from Prouteau *et al.* (2017)[296]). **D.** Model of the biological role of TORC1 filaments (adapted with permission from Prouteau *et al.* (2017)[296]).

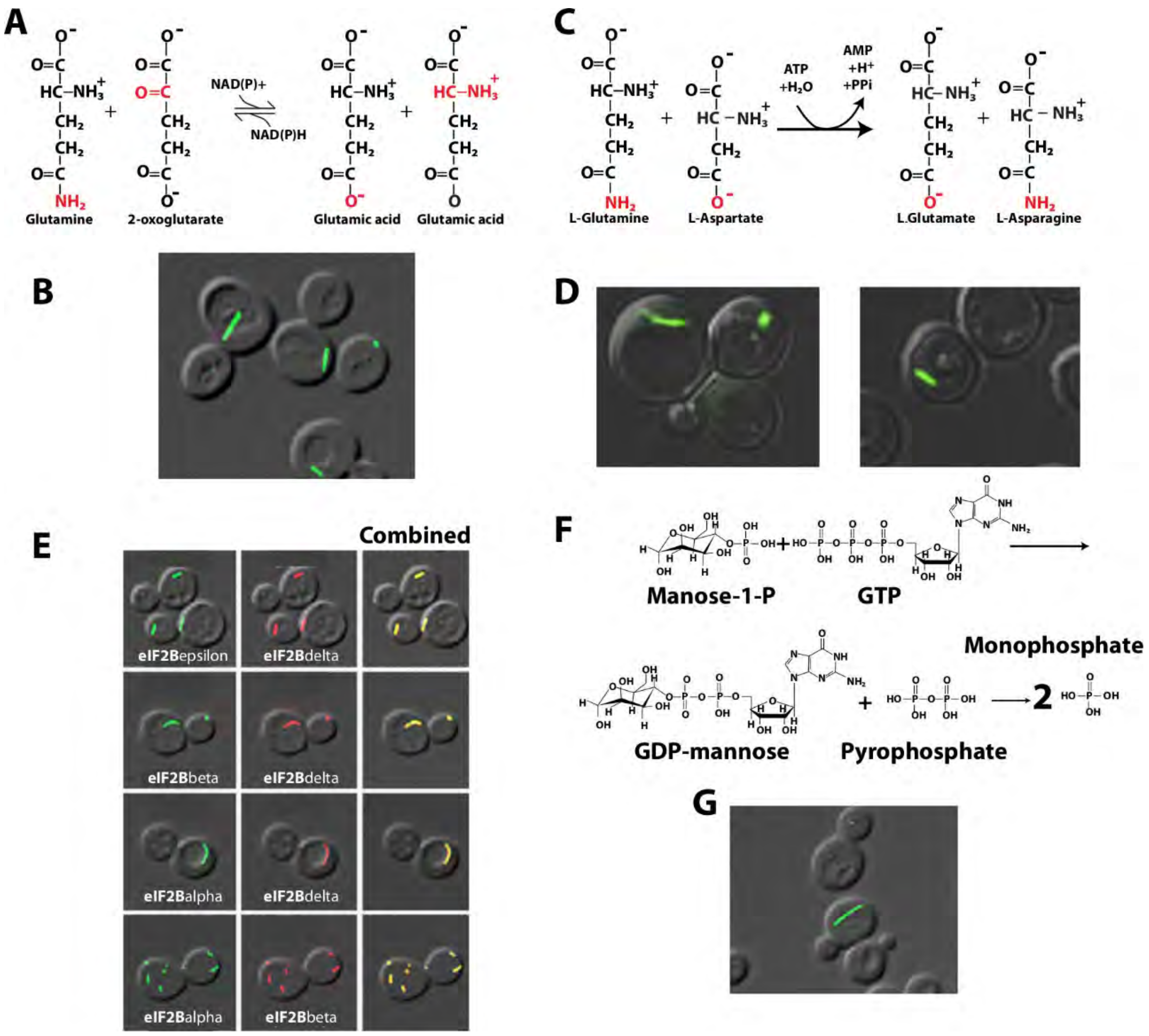

**Figure 28**. **A**. Reaction catalyzed by glutamate synthase (Glt1 in yeast)(EC 1.4.1.13). **B.** Cytoophidia formed by GFP labeled Glt1p in *S. cerevisiae* (adapted with permission from Noree *et al.* (2010)[24]). **C.** Reaction catalyzed by asparagine synthetase (EC 6.3.5.4). **D.** Cytoophidia formed by GFP labeled asparagine synthetase (adapted with permission from Shen *et al.* (2016)[26]). **E.** Colocalization of various eIF2 and eIF2B proteins in cytoophidia, and lack of colocalization with other, unrelated proteins (adapted with permission from Noree *et al.* (2010)[24]). **F.** Reaction catalyzed by GDP-mannose pyrophosphorylase (EC 2.7.7.13).

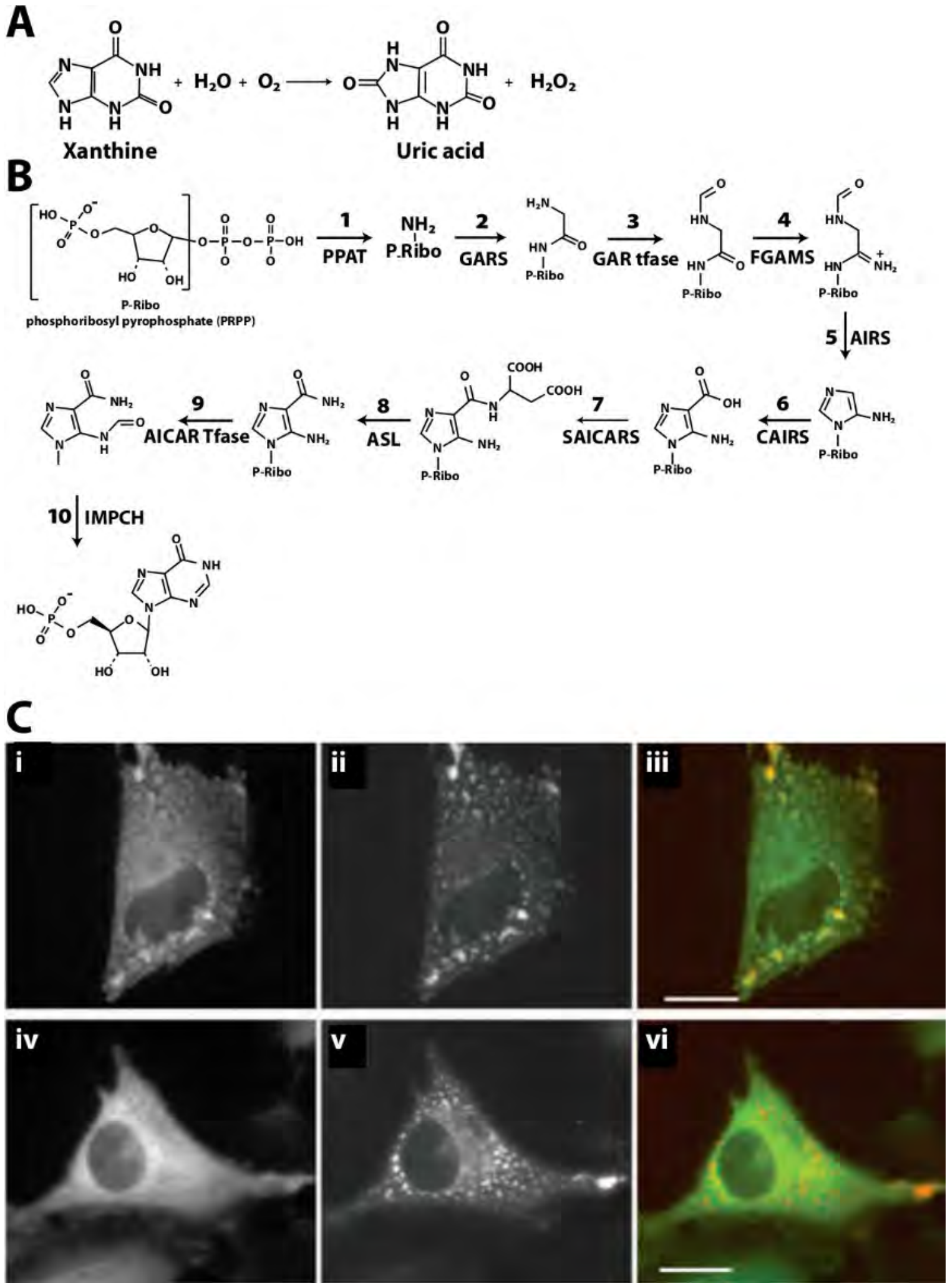

**Figure 29. A.** Reaction catalyzed by xanthene oxidase (EC 1.1.3.22). **B**. Reactions catalyzed by enzymes of the purinosome (adapted with permission from An *et al.* (2008)[111]). **C.** Foci formed by members of the purinosome in purine depleted media. (i and iii) hTrifGART-GFP [(i) and green in (iii)] coclusters with hFGAMS-OFP [(ii) and red in (vi)] in HeLa cells under purine depleted conditions. (iv to vi) The control hC1THF-GFP [(iv) and green in (vi)], which does not participate directly in *de novo* purine biosynthesis, does not form clusters or colocalize with hFGAMS OFP [(v) and red in (vi)] in HeLa cells grown in purine-depleted media. Scale bar, 10 mm (adapted with permission from An *et al.* (2008)[111]).

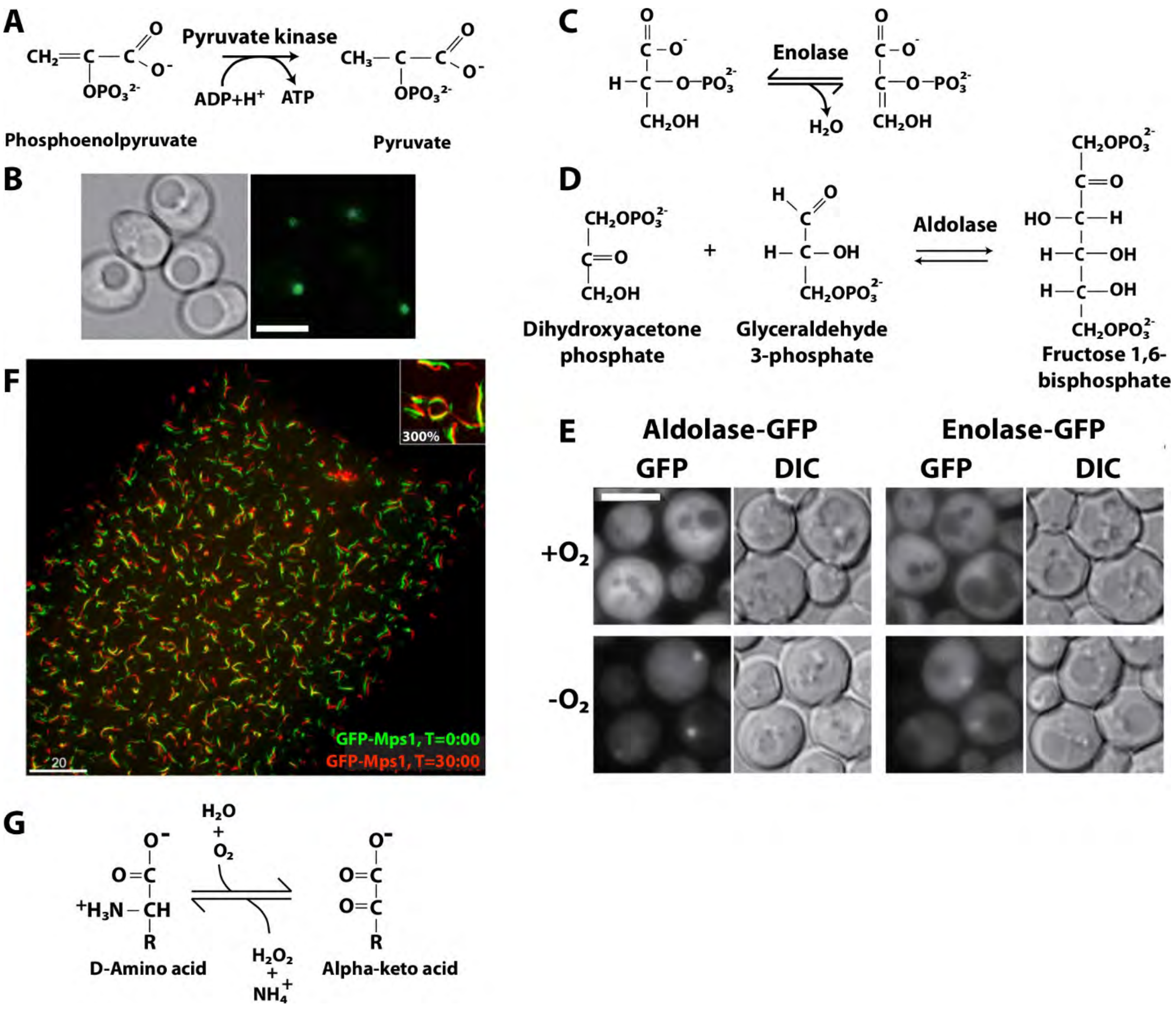

**Figure 30. A.** Reaction catalyzed by pyruvate kinase (EC 2.7.1.40). **B.** Large assembly formed by GFP labeled Cdc19 in yeast cells under hypoxic conditions (adapted with permission from Jin *et al.* (2017)[33]). **C.** Reaction catalyzed by enolase (EC 4.2.1.11). **D.** Reaction catalyzed by fructose bisphosphate aldolase (EC 4.1.2.13). **E.** G bodies (assemblies) of enolase and fructose bisphosphate aldolase in yeast under hypoxic conditions (adapted with permission from Jin *et al.* (2017)[33]). **F.** Assemblies of Mps1 kinase in *Drosophila* a single oocyte. Time=0 (green) and Time=30 minutes (red) show assemblies with exposure to $CO_2$ (time=0, green), followed by exposure to ambient air whereupon the assemblies dissociate, then reintroduction of $CO_2$ (time=30, red). The assemblies reappear in roughly the same locations and with nearly the same geometry. Foci also appear at the meiotic spindle (brighter spots near upper right)( adapted with permission from Gilliland *et al.* (2009)[305]). **G.** Reaction catalyzed by D-amino acid oxidase (EC 1.4.3.3).

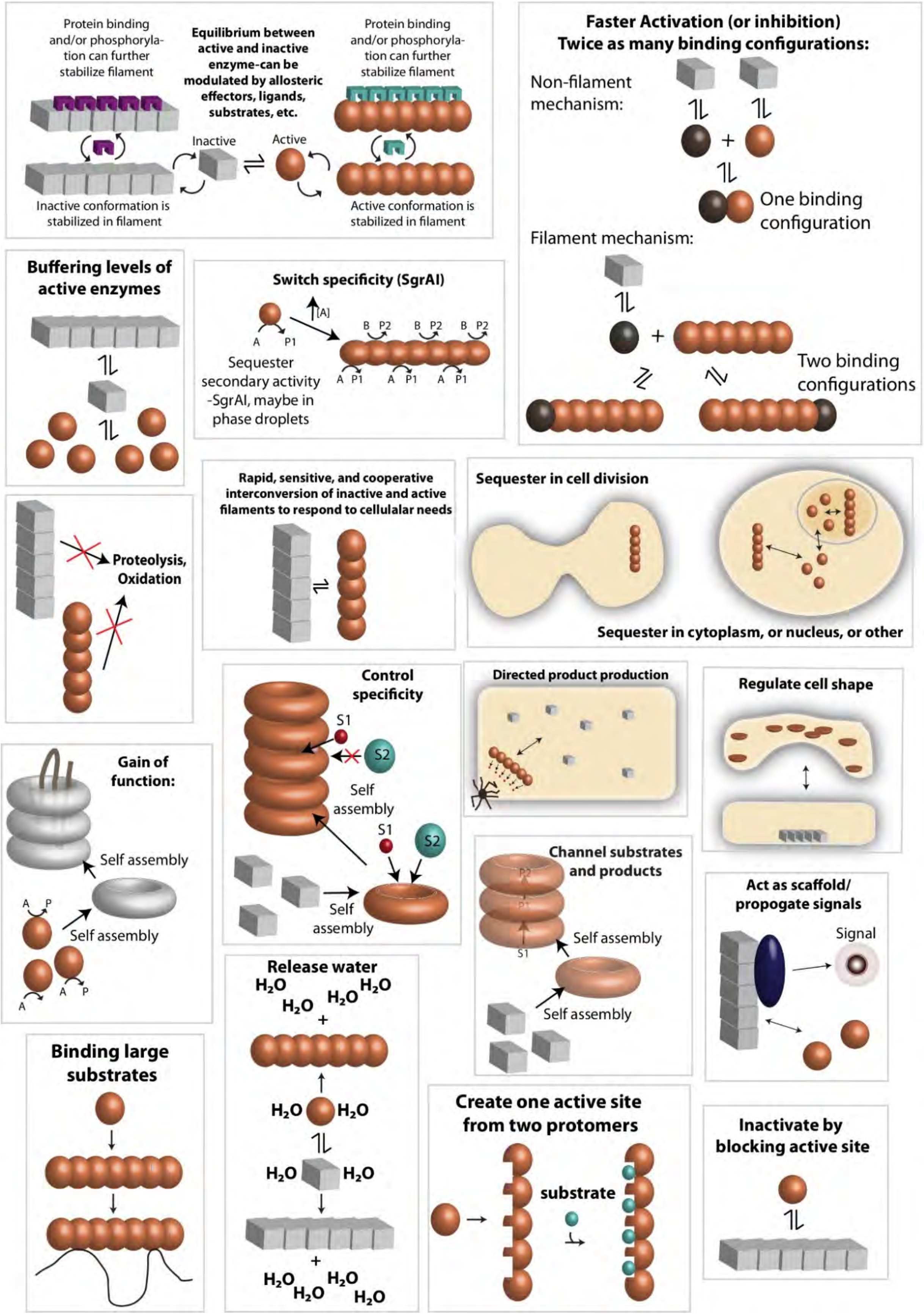

**Figure 31. Biological roles of filament forming enzymes.** Grey squares, inactive enzyme; copper spheres, active enzyme.

## Table 1. Summary of Activities and Filament Structures for Selected Enzymes

| Enzyme | Type of Structure | Regulation of Enzyme Activity by Assembly | Conditions Promoting Assembly | References |
|---|---|---|---|---|
| **Metabolic Enzymes** | | | | |
| Acetyl-CoA carboxylase 1 and 2 (ACC1 and ACC2) (yeast, chicken, mice, human) Pathway: fatty acid biosynthesis | Filaments 1-3 determined from single particle reconstructions using cryo-EM, 4-6 Å resolution. Filament 1: left-handed helix, 3 enzyme dimers per turn (120° between dimers), 154 Å between adjacent dimers[71]. Filament 2: left-handed helical backbone similar to Filament 1 with 3 dimers per turn but thinner, BC domain juts out[71]. Filament 3: Two left-handed filaments with 3 dimers per turn (120° between dimers), and 190 Å rise[71]. ACC2 also forms filament with citrate (negative stain TEM)[20, 70]. | Filament 1: Citrate but no ACC phosphorylation, active enzyme[71]. Filament 2: Citrate, no ACC phosphorylation, palmitoyl-CoA, inactive enzyme[71]. Filament 3: BRCT domains of BRCA1 bound to phosphorylated ACC1, inactive enzyme[71]. Filament of ACC1 with MIG12: 60x activation[20]. | Filament 1: 10 mM citrate and dephosphorylated ACC1[71]. Filament 2: 10-fold molar excess of palmitoyl-CoA to Filament 1[71]. Filament 3: addition of BRCT domains of BRCA1 to phosphorylated ACC1 (they bind 1:1 BRCT domain:ACC1 dimer)[71]. >5 mM Citrate induces filament formation by both ACC1 and ACC2[1, 20, 70]. MIG12 promotes filamentation of ACC1 and ACC2[20]. SPOT14 and MIG12 inhibit filament formation by ACC1 and ACC2[70]. Forms foci when GFP labeled in yeast cells under nutrient starvation conditions[24, 26]. | Kleinschmidt *et al.* (1969)[1] Meredith & Lane (1978)[2] Beaty *et al.* (1983)[3] Brownsey *et al.* (2006)[66] Kim *et al.* (2010)[20] Noree *et al.* (2010)[24] Park, *et al.* (2013)[70] Shen *et al.* (2016)[26] Jin *et al.* (2017)[33] Hunkeler *et al.* (2018)[71] |
| ATP dependent 6-phosphofructokinase (yeast genes Pfk1, Pfk2) (mammals, *C. elegans*) Pathway: glycolysis | Single particle reconstruction from negative stain TEM, 25Å resolution. Filament of stacked tetramers which can assemble either of two ways resulting in either straight or kinked filament. Right handed helix with stacked PFKL tetramers related by a rotation of 221° and translation of 83 Å between subunits[76]. Average size is 6 tetramers. Kink leads to junction of ~130°. | Effect on enzyme activity not yet known. | Filaments form in response to substrate, 6-phosphofuctose (*in vitro*) and citrate (in cells)[76]. Forms foci when GFP labeled in yeast cells under nutrient starvation or hypoxic conditions[24, 26, 33]. Formed foci at nerve cell synapses where energy demands are likely high[75]. Found in G-bodies[33]. | Kemp (1971)[4] Telford *et al.* (1975)[330] Trujillo & Deal (1977)[5] Reinhart & Lardy (1980)[6] Foe & Trujillo (1980)[77] Schwock *et al.* (2004)[331] Ranjit *et al.* (2014)[332] Jang *et al.* (2016)[75] Shen *et al.* (2016)[26] Webb *et al.* (2017)[76] Jin *et al.* (2017)[33] |
| Glucokinase (yeast gene Glk1) (fungi) Pathway: glycolysis | Filament of S. cerevisiae Glk1 was resolved to 3.8 Å by cryo-EM and single particle reconstruction[90] and forms a right-handed, anti-parallel double helix. | Enzyme is inhibited in the filament[90]. | Bundles visible in cells form of Glk1-GFP fusions in response to glucose[90]. Filaments form in response to substrates (ATP and glucose, mannose, or glucosamine) or its products (ADP and sugar-6-phosphate)[90]. Modest filament formation in the presence of inhibitors N-acetylglucosamine-6-phosphate and N-acetylglucosamine[90]. Filaments do not form in vitro in the presence of fructose or galactose[90]. | Stoddard *et al.* (2019)[90] |
| CTP Synthase (CTPS) (bacterial, yeast genes Ura7 and Ura8, *Drosophila,* human) Pathway: pyrimidine biosynthesis (CTP) | Filaments from bacteria and from human are composed of stacked tetrameric CTPS, but they assemble differently. The bacterial CTPS filament was resolved to 4.6 Å by cryo-EM and single particle reconstruction and shows interdigitating tetramers with a rotation of ~49° between and ~3.7 per turn[98]. The human CTPS filament was resolved to 6.1 Å by cryo-EM and single particle | Bacterial CTPS is inhibited when in filament[32, 96]. Human CTPS is activated when in filament[98]. | In bacteria, product (CTP) drives filamentation[19]. In human, filaments are favored in presence of substrates UTP and ATP, not products CTP and ADP[98]. Forms foci when GFP labeled in yeast cells under nutrient starvation conditions[24, 26]. Forms cytoophidia in *Drosophila*[27]. | Werner *et al.* (2009)[23] Noree *et al.* (2010)[24] Liu *et al.* (2010)[27] Ingerson-Mahar *et al.* (2010)[19] Chen *et al.* (2011)[41] Carcamo *et al.* (2011)[40] Barry *et al.* (2014)[96] Aughey *et al.* (2014)[97] Noree *et al.* (2014)[32] Shen *et al.* (2016)[26] |
| | reconstruction and shows stacked tetramers with less interdigitation than the bacterial form: ~30° per subunit and ~3 per turn[98]. | | | Lynch *et al.* (2017)[98] |
| Inosine monophosphate dehydrogenase (IMPDH) (yeast, *Drosophila*, human)<br>Pathway: nucleotide synthesis (GTP) | Filaments were characterized by single particle or helical reconstructions and negative stain TEM (20 Å resolution), and cryo-EM of Y12A IMPDH with GTP (8.7 Å resolution), and were found to be composed of stacked IMPDH octamers with a slight rotation between them (due to high symmetry, this could be considered either right or left-handed)[128].<br>Octamers resemble balls with holes in side, which can be open (assumed to be the active state) or collapsed (assumed to be the inactive)[128].<br>Filaments of open filaments have a rotation of 30° between adjacent octamers, with each octamer spanning a distance of 111 Å.<br>Filaments of collapsed octamers have a rotation of about 35.5° between adjacent octamers, and one octamer spans a distance of 94 Å[128].<br>Both open and collapsed octamers appear to be able to coexist in the same filament[128].<br>Also forms large structures known as “cytoophidia” in cells which may be bundles of filaments[40, 122]. | Though it’s not known, the open form is considered the active form and the collapsed the inactive form[128].<br>IMPDH was shown to be stimulated 2.7-fold in filaments[127] | ATP induces both types of filament[128].<br>GTP, a downstream product (or guanonsine) induces the collapsed type of filament[128].<br>IMP, a substrate, and $NAD^+$, the cofactor consumed in the reaction, induce the open form[128].<br>Mycophenolic acid (an inhibitor) induces filaments in cells, GTP disperses filaments in cells, suggesting an inactive state[333-334].<br>MgATP stimulates filamentation *in vitro* and stimulates enzyme 2.7-fold[127]. | Carcamo *et al.* (2011)[40]<br>Liu *et al.* (2016)[31]<br>Anthony *et al.* (2017)[128]<br>Keppekke *et al.* (2018)[124] |
| Glutamine (Q) synthetase (yeast gene Gln1) | Forms pentameric rings that stack into a decamer (two rings with 30° twist between them) which further stack into filaments with ~10° twist between them[101].<br>Forms foci in cells as a result of nutrient starvation and heat stress[24].<br>Correlative light microscopy and negative stain TEM shows these to be composed of aligned filaments ~1 micron long[101]. | Glutamine synthetase is inactive in the filament[101]. | Filaments form at low pH[101].<br>Zinc induces decameric rings into filaments[7].<br>Filamentation in cells is reversible in the absence and presence of glucose[25]. | Miller *et al.* (1974)[7]<br>Frey *et al.* (1975)[8]<br>Narayanaswamy *et al.* (2009)[25]<br>Noree *et al.* (2010)[24]<br>O’Connell *et al.* (2014)[130]<br>Petrovska *et al.* (2014)[101]<br>Shen *et al.* (2016)[26] |
| β-glucosidase (plants including oat)<br>Pathway: host defense | Negative stain TEM shows trimeric rings that dimerize into hexamers and larger oligomers[140-141].<br>Single particle reconstructions suggest a difference in the twist between neighboring trimers forming a “tunnel” housing the active sites. In the multimers, the twist between trimers is 38°. In the hexamers this is 60°. The smaller twist seems to result in smaller side fenestrations presumably changing access to the active site[140-141].<br>Forms fibrils in cells termed stromacenters[9]. | Two-fold lower $K_M$ and 1.6 lower $k_{cat}$ in the filament[141].<br>This effect may be related to steric hindrance of substrates and products entering and leaving the enzyme through openings in the filament vs. the hexamer (tighter binding but slower turnover)[141]. | Activated upon fungal infection. Disassembled by $CaCl_2$[141]. | Gunning *et al.* (1965)[9]<br>Nisius & Ruppel (1987)[142]<br>Nisius *et al.* (1988)[335]<br>Kim *et al* (2000)[140]<br>Kim *et al.* (2005)[141] |
| $CO_2$ reductase (bacteria, *Acetobacterium woodii*) Pathway: carbon fixation and energy conservation | Filaments are 10-15 nm, with an apparent helical structure[148] | Filamentous form is 2-fold more active[148]. | Filaments are induced by 5-20 mM $MgSO_4$, 20 mM $MnCl_2$ or $CaCl_2$[148]. | Schuchmann *et al.* (2016)[148] |
| Nitrilase (bacteria, fungi, plants) Pathway: detoxification | Left-handed helical filament with 4-5 enzyme dimers per turn[315]. | The filament is more active[155].<br>The helical twist of filaments of nitrilases from different organisms appears to correlate with substrate specificity, hypothesized to be due to substrate binding cleft size[315]. | 10 mM benzonitrile, but not aliphatic nitriles, induced filamentation and increased activity of the enzyme[155].<br>10% saturated ammonium sulfate and 50% (v/v) glycerol, increased temperatures, or enzyme concentrations induces the activated filament[155].<br>Wounding of plants (cutting or with herbicides) induces foci formation in plant cells[21]. | Harper *et al.* (1977)[10-11]<br>Cutler *et al.* (2005)[21]<br>Thuku *et al.* (2007)[152]<br>Thuku *et al.* (2009)[156]<br>Chan *et al.* (2011)[323]<br>Woodward *et al.* (2018)[315] |
| CoA-Dependent acetaldehyde and alcohol dehydrogenase (AdhE) (bacteria, photosynthesizing unicellular organisms) Pathway: fermentation | Negative stain TEM showing "spiralosomes" of 20-60 subunits in rods of 45-120 nm in length with left-handed geometry[22, 158].<br>Model building with 7 copies per turn with right-handed helical geometry[159]. | The helical filaments extend and become slimmer upon $Fe^{2+}$ and $NAD^+$ addition. Authors speculate that this is the active form, because it is more open, allowing substrates and products access to the enzymes, whereas the wider, shorter, more compact forms may be the inactive form of the enzyme[22]. | Filaments found in cells, however ligands ($Fe^{3+}$, $NAD^+$) can induce structural changes in helical filaments[22]. | Kessler *et al.* (1991)[336]<br>Kessler *et al.* (1992)[22]<br>Extance *et al.* (2013)[159]<br>Laurenceau *et al.* (2015)[158] |
| Glutamate (E) dehydrogenase (yeast gene Gdh2, mammals) Pathway: nitrogen and glutamate metabolism, energy homeostasis, amino acid synthesis, and in plants tolerance to herbicide, water deficit, pathogen infections | Helical tubes *in vitro* visualized by negative stain TEM (9-9.3 subunits per turn)[14]. | Though the effect on enzyme activity of filamentation is not known, conditions activating the enzyme also induce filamentation, and conditions deactivating the enzyme dissociate the filament, suggesting that filamentation increases enzyme activity[180-181, 183]. | Stationary phase (*i.e.* nutrient starvation) induces cytoophidia in yeast[26].<br>GTP and NADH induce depolymerization (and are inhibitors) *in vitro*[13, 15].<br>ADP favors polymerization (and is an activator) *in vitro*[13, 15]. | Olson & Anfinsen (1952)[178]<br>Eisenbergy & Reisler (1971)[12]<br>Huang *et al.* (1972)[13]<br>Josephs & Borisy (1972)[14]<br>Zeiri & Reisler (1978)[15]<br>Shen *et al.* (2016)[26] |
| Glutaminase (mammals) Pathway: nitrogen metabolism | Negative stain TEM (35 Å resolution) with single molecule reconstruction and model fitting showing right-handed double helix with 53±2 nm rise per turn, strand inclination of 25°, and an average width for a single strand of 6.6±0.7 nm[195]. | Filaments have increased catalytic activity[195]. | Formed *in vitro* with purified protein, phosphate and phosphate-borate added to protein in Tris-HCl induces polymers[16]. | Olsen *et al.* (1970)[16]<br>Ferreira *et al* (2013)[195]<br>Petrovska *et al.* (2014)[101] |
| β-lactamase-like protein (LACTB) (mammals) Pathway: unknown | Appears as a braided filament in negative stain TEM[199]. | Enzyme activity has not been investigated, and therefore it is not known how filamentation affects enzyme activity. | Filaments are found in the intra-cristal part of the intermembrane space of mitochondria, spanning the cristae. | Plianskyte *et al.* (2009)[199] |
| 2-Cys peroxiredoxins (archea, bacteria, eukaryotes) Pathway: $H_2O_2$ metabolisom and detoxification, protein folding stress | Filament has hollow core that binds unfolded proteins. Formed from stacks of decameric rings viewed by negative stain TEM[325]. | Filaments are inactive in peroxidase activity, but active in chaperone activity (binding unfolded proteins, sequestering them to prevent aggregation). | Redox stress, high hydrogen peroxide levels, heat shock, low pH, and site-specific phosphorylation induce filamentation. | Wood *et al.* (2002)[205]<br>Gourlay, *et al.* (2003)[202, 325]<br>Phalen *et al.* (2006)[324]<br>Saccoccia *et al.* (2012)[207]<br>Angelucci *et al.* (2013)[202]<br>Puerto-Galán *et al.* (2013)[201]<br>Noichri *et al* (2015)[200]<br>Teixeira *et al.* (2019)[206] |
| Ribonucleotide reductase (archea, bacteria, eukaryotes; yeast genes Rnr2 and Rnr4) Pathway: nucleotide metabolism | cryo-EM, 2 types of filaments, 4.7-4.8 Å resolution, one a double helix (right-handed), the other having a filament of NrdE entwinned with a filament of NrdF (also right-handed)[210]. | Enzyme is inhibited in both types of filaments[210]. | High dATP (100 μM) induces filaments[210].<br>Filaments differ with respect to the presence or absence of NrdF protein[210].<br>Punctate foci seen in yeast with nutrient starvation[25]. | Noree *et al.* (2010)[25]<br>Ando *et al.* (2016)[209]<br>Thomas *et al.* (2018)[210]<br>Thomas et al. (2019)[211] |
| **Nucleic Acid Enzymes** | | | | |
| SgrAI (bacterial: *Streptomyces griseus*) Pathway: host defense | Helical reconstruction and cryo-EM (3.5 Å resolution) shows left handed helix with ~4 DNA bound enzyme dimers per turn (~90°)[39]. | The enzyme is 200-1000x activated in the filament[17, 38]. | Binding to double-stranded DNA containing the primary site sequence CRCCGGYG, and at least 7 base pairs on either side, induce filamentation.<br>Double-stranded DNA containing secondary site. sequences CCCCGGYG or XRCCGGYG will be bound by the enzyme, and will not induce filamentation but will join filaments formed by enzyme bound to primary site DNA. | Park *et al.* (2010)[17]<br>Shah *et al.* (2015)[38]<br>Lyumkis *et al.* (2013)[37]<br>Park *et al.* (2018)[51-52]<br>Barahona *et al.* (2019)[337]<br>Polley *et al.* (2019)[39] |
| IRE1 (eukaryotes) Pathway: unfolded protein response, response to stress | X-ray crystal structure showed right-handed helical filament with 7 dimeric forms of the IRE1 chain per turn (~50°)[18].<br>Also forms foci in live cells[18, 234-236]. | Filamentation activates splicing of TF mRNA 100,000-fold and possibly degradation of other mRNAs[18].<br>Upon dimerization via binding unfolded proteins, the kinase domains trans-autophosphorylate lead to phosphorylated sites that likely stabilize the filamentous form[18].<br>Binding to ADP also favors filamentation[18]. | Binding to unfolded proteins induces kinase activity that induces oligomerization, which then binds and splices particular mRNAs[18, 230-234].<br>In the filamentous form, the enzyme may cleave many mRNAs nonspecifically (*i.e.* RIDD activity), while in the dimeric form it cleaves only the particular mRNAs for transcription factors that induce the unfolded protein response pathway. | Korennykh *et al.* (2009)[18]<br>Li *et al.* (2010)[236]<br>Ghosh *et al.* (2014)[329] |
| RecA (*E. coli*) Pathway: Homologous recombination | Right-handed helical filaments on DNA have been characterized by electron microscopy and x-ray crystallography (2.8-4.3 Å resolution) with ~60° per turn, 94 Å pitch, ~6 RecA proteins per turn[214]. This filament has polarity (the two ends are not equivalent). | Binding to ssDNA and ATP stimulates its filamentation and binding to dsDNA, which is stretched, causing duplex melting, allowing invasion by the single stranded DNA, good base pairing results in ATP hydrolysis and dissociation from the DNA | Binding to ssDNA and ATP stimulates its filamentation which can then bind to dsDNA and perform its activity. | Chen *et al.* (2008)[214] |
| DnaA (*E. coli*) Pathway: DNA replication initiation | Right handed helical filaments on DNA have been characterized by x-ray crystallography with 45° per turn, 178 Å pitch, 8 copies per turn[218, 338]. This filament has polarity (the two ends are not equivalent). | Binding to ATP and specific sites in the origin of replication induces filamentation, which stretches and/or supercoils the DNA inducing strand melting (that allows the replicative helices to be loaded) | Binding to ATP and specific sites stimulates filamentation which gives it its activity | Erzberger *et al.* (2006)[338]<br>Duderstadt *et al.* (2011)[218] |
| **Kinases and Innate Immunity Signaling Filaments** | | | | |
| Casein kinase CK2 (human) Pathway: cell cycle control, cellular differentiation, and proliferation, circadian rhythm, apoptosis, and gene expression | Protomer is heterotetramer ($\alpha_2\beta_2$). Oligomers are deduced from native MS, IM-MS, and H/D exchange[269]. X-ray crystal structure shows linear form[272]. | Filament is likely the inactive form of the enzyme[266]. | High ionic strength (*e.g.* >0.5 M NaCl) disrupts filaments[266].<br>Ring form is favored in saturating concentration of substrates[266]. | Valero, *et al.* (1995)[266]<br>Seetoh *et al.* (2016)[269] |
| Adenylate kinase (bacteria, plants) Pathway: nucleotide balance | X-ray crystal structure shows linear filament with no helicity form from stacked hexamers[274]. | Forms linear polymer in crystal packing.<br>Not clear this occurs *in vivo*.<br>Speculated to inhibit the activity based on steric/conformation arguments. | Thought to occur at night, when enzyme activity is not needed, and to release water, though not understood how this is regulated[274]. | Wild *et al.* (1997)[274] |
| RIP1/RIP3 kinase (mammals)<br>Pathway: host defense, innate immunity | Negative stain TEM. RIP homotypic interaction motif (RHIM) is a 5-6 residue segment that forms cross-beta amyloid structure upon self-association or association of RIP1 with RIP3[281-282]. | Kinase activity activated by assembly[276-278]. | Signaling induces association.<br>The purpose of amyloid/fibril formation may be for feed-forward gain of function in which kinase activation and RIP1/RIP3 necrosome formation are mutually reinforcing[281].<br>Upon amyloid formation, transphosphorylation occurs which activates RIP1/RIP3[281]. | Cho *et al.* (2009)[276]<br>Li *et al.* (2012)[281]<br>Mompean *et al.* (2016)[282] |
| Death domain containing (vertebrates)<br>Pathway: host defense, cell death, innate immunity | Structures solved via helical reconstruction cryo-EM, single particle reconstruction cryo-EM and x-ray crystallography. Filaments assemble via death domain (CARD, PYD, DED). Some right-handed, others left-handed. Can appear as a triple helix. | Filaments are the activated form of these signaling proteins. | Proteins in this category have CARD or Death Domains that form filaments. Part of innate immunity in humans. Form in response to signal, often a ligand that induces dimerization of a death-domain containing protein, leading to seeding filament that is formed by another death-domain containing protein, which amplifies the signal and creates a scaffold for activation of downstream proteins, such as caspase-1 or kinases, leading to outcome such as increased transcription of inflammatory genes or induction of apoptosis or necrosis. | Lin *et al.* (2010)[290]<br>Wu *et al.* (2014)[288]<br>Lu *et al.* (2014)[293]<br>Diebolder *et al.* (2015)[291]<br>Lu *et al.* (2016)[292]<br>Gong *et al.* (2018)[289] |
| TORC1 (yeast)<br>Pathway: nutrient/energy/redox sensor and controls protein synthesis | Negative stain EM single particle reconstruction (27 Å resolution). Hollow tube or cylinder called TOROID, right handed with ~9 copies per turn[296]. | Kinase activity is inactive in TOROID filament[296]. | Glucose starvation[296].<br>Dependent on Rag GTPases for assembly[296]. | Prouteau *et al.* (2017)[296] |

**Table 2. Enzymes that form assemblies (in cells or *in vitro*), but structure and/or regulatory properties not yet determined.**

| Enzyme/organism | Type of Assembly Observed or Image | Conditions triggering assembly in cells | Is the assembly reversible? | References |
|---|---|---|---|---|
| **Amino acid metabolism, protein synthesis, protein folding, protein degradation** | | | | |
| Glutamate (E) synthase<br>(yeast gene Glt1)<br>Pathway: ammonia assimilation cycle/glutamate biosynthetic process | Cytoophidia, rods, and punctate foci | Similar in stationary phase, diauxic phase, and log phase[24], more in stationary phase[26] | Shown to be motile, undergo sub-diffusion[26] | Noree *et al.* (2010)[24]<br>Shen *et al.*, (2016)[26] |
| Asparagine synthetase<br>(yeast gene Asn1, Asn2)<br>Pathway: amino acid metabolism, nitrogen metabolism | Cytoophidia and rods | Increasing on going from log, to diauxic and stationary phases, glucose deprivation | Yes[25] | Narayanaswamy *et al.* (2009)[25]<br>Shen *et al.* (2016)[26]<br>Zhang *et al.* (2018)[297] |
| Alanyl-tRNA synthetase<br>(yeast gene Ala1)<br>Pathway: protein synthesis, ala-tRNA aminoacylation | Punctate foci | Stationary phase | Yes[25] | Narayanaswamy *et al.* (2009)[25] |
| Leucyl tRNA synthetase<br>(yeast gene Cdc60)<br>Pathway: protein synthesis, Leu-tRNA aminoacylation | Punctate foci | Stationary phase | Yes[25] | Narayanaswamy *et al.* (2009)[25] |
| Cystathionine β-synthase<br>(yeast gene Cys4)<br>Pathway: amino acid metabolism, cysteine biosynthesis | Punctate foci | Stationary phase | Yes[25] | Narayanaswamy *et al.* (2009)[25] |
| eIF2B, eIF2<br>(yeast genes Gcd1 (eIF2b-γ, Gcd2 (eIF2B-δ), Gcd7 (eIF2B-β), Gcd6 (eIF2B-ε), and Sui2 (eIF2-α), Gcn3 (eIF2B-α)) | Cytoophidia, rods, and punctate foci | Log and stationary phases, cycloheximide, acidification | Yes[25] | Noree *et al.* (2010)[32]<br>Shen *et al.* (2016)[26]<br>Nuske *et al.* (2018)[300] |
| Peptidyl-prolyl cis-trans isomerase (cyclophilin)<br>(yeast gene Cpr6)<br>Pathway: Protein folding | Punctate foci | Stationary phase | Yes[25] | Narayanaswamy *et al.* (2009)[25] |
| Glutamine tRNA synthetase<br>(yeast gene Gln4)<br>Pathway: protein synthesis, aminoacylating appropriate tRNA with glutamine | Punctate foci | Stationary phase | Yes[25] | Narayanaswamy *et al.* (2009)[25] |
| Histidine tRNA synthetase<br>(yeast gene Hts1)<br>Pathway: protein synthesis, aminoacylating appropriate tRNA with histidine | Punctate foci | Stationary phase | Yes[25] | Narayanaswamy *et al.* (2009)[25] |
| Isoleucine tRNA synthetase<br>(yeast gene Ils1)<br>Pathway: protein synthesis, aminoacylating appropriate tRNA with isoleucine | Punctate foci | Stationary phase | Yes[25] | Narayanaswamy *et al.* (2009)[25] |
| Ribosome-associated molecular chaperones, ATPases<br>(yeast genes Ssb1, Ssb2)<br>Pathway: protein biosynthesis | Punctate foci | Stationary phase | Yes[25] | Narayanaswamy *et al.* (2009)[25] |
| Threonyl tRNA synthetase<br>(yeast gene Ths1)<br>Pathway: protein synthesis, aminoacylating appropriate tRNA with threonine | Punctate foci | Stationary phase | Yes[25] | Narayanaswamy *et al.* (2009)[25] |
| Valyl-tRNA synthetase<br>(yeast gene Vas1)<br>Pathway: protein synthesis, aminoacylating appropriate tRNA with threonine | Punctate foci | Stationary phase | Yes[25] | Narayanaswamy *et al.* (2009)[25] |
| Threonine aldolase | Punctate foci | Stationary phase | Not known | Noree *et al.* (2010)[24] |
| (yeast gene Gly1)<br>Pathway: glycine biosynthesis | | | | |
| Multifunctional enzyme containing phosphoribosyl-ATP pyrophosphatase; phosphoribosyl-AMP cyclohydrolase, and histidinol dehydrogenase activities<br>(yeast gene His4)<br>Pathway: histidine biosynthesis | Punctate foci | Stationary phase | Not known | Noree *et al.* (2010)[24] |
| S-adenosylmethionine synthetase<br>(yeast genes Sam1 and Sam2)<br>Pathway: methionine metabolism, S-adenosylmethionine biosynthesis | Punctate foci | Stationary phase | Not known | Noree *et al.* (2010)[24] |
| Disaggregase (ATPase)<br>(yeast gene Hsp104)<br>Pathway: protein folding | Punctate foci | Stationary phase | Not known | Noree *et al.* (2010)[24] |
| ATPase involved in protein folding and NLS-directed nuclear transport<br>(yeast gene Ssa1)<br>Pathway: protein folding | Punctate foci | Stationary phase | Not known | Noree *et al.* (2010)[24] |
| HSP70 family ATP-binding protein<br>(yeast gene Ssa2)<br>Pathway: protein folding | Punctate foci | Stationary phase | Not known | Noree *et al.* (2010)[24] |
| Homoserine kinase<br>(yeast gene Thr1)<br>Pathway: threonine biosynthesis | Punctate foci | Stationary phase | Not known | Noree *et al.* (2010)[24]<br>Thayer *et al.* (2014) |
| Non-ATPase regulatory subunit of the 26S proteasome<br>(yeast gene Rpn9)<br>Pathway: protein degradation | Punctate foci | Stationary phase | Not known | Noree *et al.* (2010)[24], |
| **Cell wall biosynthesis** | | | | |
| GDP-mannose pyrophosphorylase<br>(yeast gene Psa1)<br>Pathway: cell wall biosynthesis, synthesizes GDP-mannose from GTP and mannose-1-phosphate in cell wall biosynthesis; required for normal cell wall structure | Cytoophidia, rods and foci | Stationary phase, sodium azide, staurosporine | Not known | Noree *et al.* (2010)[24]<br>Shen *et al.*, (2016)[26] |
| UDP-N-acetylmuramate-alanine ligase<br>(the bacterium *Caulobacter crescentus*)<br>Pathway: cell wall biosynthesis | Cytoophidia, rods | Normal cell growth | Not known | Werner *et al.* (2009)[23] |
| 5-phospho-ribosyl-1(alpha)-pyrophosphate synthetase<br>(yeast gene Prs4)<br>Pathway: synthesizes PRPP; which is required for nucleotide, histidine, and tryptophan biosynthesis | Punctate foci | Stationary phase | Not known | Noree *et al.* (2010)[24] |
| **Purine metabolism** | | | | |
| Xanthene oxidase<br>(mammals)<br>Pathway: purine catabolism, nitrogen metabolism, important role in the catabolism of purines | *In vitro* forms homotetrameric rings which can stack into fibers to form crystalline cores | Formed *in vitro* | Not known | Angermuller *et al.* (1987)[301] |
| **Purinosome** | | | | |
| Phosphoribosyl-<br>pyrophosphate amidotransferase<br>(PRPPAT)<br>(yeast gene Ade4, mammals)<br>Pathway: Purine biosynthesis | Colocalizes with other enzymes of the purinosome in punctate foci | Stationary phase[25], purine starvation[111], female *Drosophila* germ line[29] | Yes[25, 111] | An *et al.* (2008)[111]<br>Narayanaswamy *et al.* (2009)[25]<br>Deng *et al.* (2012)[112] |
| Trifunctional GART enzyme<br>(yeast gene Ade5,7, *Drosophila* gene Ade3, mammals)<br>Pathway: Purine nucleotide biosynthesis | Colocalizes with other enzymes of the purinosome in punctate foci,<br>in *Drosophila* forms cytoophidia | Stationary phase[25],<br>purine starvation[111],<br>female *Drosophila* germ line[29] | Yes[25, 111] | An *et al.* (2008)[111]<br>Narayanaswamy *et al.* (2009)[25]<br>Deng *et al.* (2012)[112]<br>Lowe *et al.* (2014)[29] |
| FGAMS/PFAS enzyme, phosphoribosylformylglycinamidine synthase<br>(mammals)<br>Pathway: purine biosynthesis, reaction step 4 of purinosome | Colocalizes with other enzymes of the purinosome in punctate foci | Purine starvation[111] | Yes[111] | An *et al.* (2008)[111]<br>Deng *et al.* (2012)[112] |
| Bifunctional PAICS enzyme<br>(yeast gene Ade2, *Drosophila* gene Ade5, mammals)<br>Pathway: Purine nucleotide biosynthesis | Colocalizes with other enzymes of the purinosome in punctate foci,<br>in *Drosophila* forms cytoophidia | Stationary phase[25],<br>purine starvation[111],<br>female *Drosophila* germ line[29] | Yes[25, 111] | An *et al.* (2008)[111]<br>Narayanaswamy *et al.* (2009)[25]<br>Deng *et al.* (2012)[112]<br>Lowe *et al.* (2014)[29] |
| ADSL enzyme<br>(mammals)<br>Pathway: Purine biosynthesis | Colocalizes with other enzymes of the purinosome in punctate foci | Purine starvation[111] | Yes[111] | An *et al.* (2008)[111]<br>Deng *et al.* (2012)[112] |
| Bifunctional ATIC enzyme<br>(yeast gene Ade17, mammals)<br>Pathway: Purine biosynthesis, | Colocalizes with other enzymes of the purinosome in punctate foci | Stationary phase[25],<br>purine starvation[111] | Yes[25, 111] | An *et al.* (2008)[111]<br>Narayanaswamy *et al.* (2009)[25]<br>Deng *et al.* (2012)[112] |
| **Pyrimidine biosynthesis** | | | | |
| Trifunctional CAD enzyme<br>(carbamoyl-phosphate synthetase 2, aspartate transcarbamylase, and dihydroorotase)<br>(yeast gene Ura2)<br>Pathway: Pyrimidine synthesis | Punctate foci | Stationary phase | Yes[25] | Narayanaswamy *et al.* (2009)[25] |
| Adenylosuccinate synthetase<br>(yeast gene Ade12)<br>Pathway: Pyrimidine biosynthesis | Punctate foci | Stationary phase | Yes[25] | Narayanaswamy *et al.* (2009)[25] |
| **Carbohydrate metabolism** | | | | |
| Glycogen debranching enzyme<br>(yeast gene Gdb1)<br>Pathway: carbohydrate metabolism, contains glucanotranferase and alpha-1,6-amyloglucosidase activities; required for glycogen degradation | Cytoophidia | Stationary phase | Not known | Noree *et al.* (2010)[24]<br>Shen *et al.* (2016)[26] |
| Pyruvate kinase<br>(yeast gene Cdc19, mammals)<br>Pathway: carbohydrate metabolism | Punctate foci | Stationary phase[25, 302],<br>glucose starvation[302],<br>heat shock[302] | Yes[25] | Narayanaswamy *et al.* (2009)[25]<br>Saad *et al.* (2017)[302]<br>Jin *et al.* (2017)[33] |
| Enolase I<br>(yeast gene Eno1)<br>Pathway: carbohydrate metabolism | G bodies | Hypoxia | Not known | Jin *et al.* (2017)[33] |
| Enolase II<br>(yeast-gene Eno2)<br>Pathway: carbohydrate metabolism | G bodies | Hypoxia | Yes[303] | Miura *et al.* (2013)[303]<br>Jin *et al.* (2017)[33] |
| Fructose bisphosphate aldolase<br>(yeast gene Fba1)<br>Pathway: carbohydrate metabolism | G bodies | Hypoxia | Not known | Jin *et al.* (2017)[33] |
| Glyceraldehyde-3-phosphate dehydrogenase (GAPDH)<br>(yeast gene Tdh3)<br>Pathway: carbohydrate metabolism | G bodies | Hypoxia | Not known | Jin *et al.* (2017)[33] |
| Phosphoglucose isomerase<br>(yeast gene Pgi1) | G bodies | Hypoxia | Not known | Jin *et al.* (2017)[33] |
| Pathway: carbohydrate metabolism | | | | |
| Tetrameric phosphoglycerate mutase<br>(yeast gene Gpm1)<br>Pathway: carbohydrate metabolism | G bodies | Hypoxia | Not known | Jin *et al.* (2017)[33] |
| Trehalose-6-P synthase/phosphatase complex (synthesis subunit)<br>(yeast gene Tps1)<br>Pathway: carbohydrate metabolism, response to desication | G bodies | Hypoxia | Not known | Jin *et al.* (2017)[33] |
| Trehalose-6-P synthase/phosphatase complex<br>(yeast gene Tps2)<br>Pathway: response to stress, synthesis of the storage carbohydrate trehalose | Punctate foci | Stationary phase | Yes[25] | Narayanaswamy *et al.* (2009)[25] |
| Alcohol dehydrogenase<br>(yeast gene Adh2)<br>Pathway: carbohydrate metabolism | Punctate foci | Stationary phase | Yes[25] | Narayanaswamy *et al.* (2009)[25] |
| UDP-Glucose-pyrophosphorylase<br>(yeast gene Ugp1)<br>Pathway: carbohydrate metabolism | Punctate foci | Stationary phase | Yes[25] | Narayanaswamy *et al.* (2009)[25] |
| Glycogen synthase<br>(yeast gene Gsy1)<br>Pathway: glycogen biosynthesis | Punctate foci | Stationary phase | Not known | Noree *et al.* (2010)[24] |
| Glycogen phosphorylase<br>(yeast gene Gph1)<br>Pathway: glycogen catabolism | Punctate foci | Stationary phase | Not known | Noree *et al.* (2010)[24] |
| **Cell cycle** | | | | |
| Mps1/Polo kinases<br>(*Drosophila*)<br>Pathway: cell cycle regulation | Punctate foci | Hypoxia | Yes[305] | Gilliland *et al.* (2009)[305]<br>Pandey *et al.* (2007)[306] |
| **Fatty acid and sterol metabolism** | | | | |
| Fatty acid synthase complex<br>(yeast genes Fas1 and Fas2)<br>Pathway: fatty acids and sterol metabolism | Punctate foci | Glucose starvation | Yes[28] | Suresh *et al.* (2016)[28]<br>Jin *et al.* (2017)[33] |
| Sterol 3-β-glucosyltransferase<br>(yeast gene Atg26/Ugt51)<br>Pathway: synthesis of sterol glucoside membrane lipids | Punctate foci | Stationary phase | Yes[25] | Narayanaswamy *et al.* (2009)[25] |
| **Other** | | | | |
| D-Amino acid oxidase<br>(mammal)<br>Pathway: detoxification (mammals), energy production (microorganisms) | Light scattering indicated polymerization | Purified protein *in vitro* | Yes[308] | Antonini *et al.* (1966)[308] |
| Protein kinase of the PAK/Ste20 family<br>(yeast gene Kic1)<br>Pathway: cell wall organization | Punctate foci | Stationary phase | Yes[25] | Narayanaswamy *et al.* (2009)[25] |
| Gamma-aminobutyrate (GABA) transaminase<br>(yeast gene Uga1)<br>Pathway: Nitrogen utilization | Punctate foci | Stationary phase | Yes[25] | Narayanaswamy *et al.* (2009)[25] |
| Serine hydroxymethyltransferase<br>(yeast gene Shm2)<br>Pathway: one-carbon metabolism | Punctate foci | Stationary phase | Yes[25] | Narayanaswamy *et al.* (2009)[25] |
| Thioredoxin peroxidase<br>(yeast gene Tsa1) | Cytoophidia | Stationary phase | Not known | Shen *et al.* (2016)[26] |
| Pathway: cell redox homeostasis | | | | |
| Glutamine amidotransferase (GATase II)<br>(yeast gene Dug2)<br>Pathway: glutathione catabolic process | Punctate foci | Stationary phase | Not known | Noree *et al.* (2010)[24] |
| Aminolevulinate dehydratase<br>(yeast gene Hem2)<br>Pathway: heme biosynthesis | Punctate foci | Stationary phase | Not known | Noree *et al.* (2010)[24] |
| Protein of unknown function with glutathione-S-transferase domain<br>(*Drosophila* gene Fax)<br>Pathway: unknown | Punctate foci | Female *Drosophila* germ line | Not known | Lowe *et al.* (2014)[29] |

1. Kleinschmidt, A. K.; Moss, J.; Lane, D. M., Acetyl coenzyme A carboxylase: filamentous nature of the animal enzymes. *Science* **1969,** *166* (3910), 1276-8.
2. Meredith, M. J.; Lane, M. D., Acetyl-CoA carboxylase. Evidence for polymeric filament to protomer transition in the intact avian liver cell. *J Biol Chem* **1978,** *253* (10), 3381-3.
3. Beaty, N. B.; Lane, M. D., Kinetics of activation of acetyl-CoA carboxylase by citrate. Relationship to the rate of polymerization of the enzyme. *J Biol Chem* **1983,** *258* (21), 13043-50.
4. Kemp, R. G., Rabbit liver phosphofructokinase. Comparison of some properties with those of muscle phosphofructokinase. *J Biol Chem* **1971,** *246* (1), 245-52.
5. Trujillo, J. L.; Deal, W. C., Jr., Pig liver phosphofructokinase: asymmetry properties, proof of rapid association--dissociation equilibria, and effect of temperature and protein concentration on the equilibria. *Biochemistry* **1977,** *16* (14), 3098-104.
6. Reinhart, G. D.; Lardy, H. A., Rat liver phosphofructokinase: kinetic and physiological ramifications of the aggregation behavior. *Biochemistry* **1980,** *19* (7), 1491-5.
7. Miller, R. E.; Shelton, E.; Stadtman, E. R., Zinc-induced paracrystalline aggregation of glutamine synthetase. *Archives of biochemistry and biophysics* **1974,** *163* (1), 155-71.
8. Frey, T. G.; Eisenberg, D.; Eiserling, F. A., Glutamine synthetase forms three- and seven-stranded helical cables. *Proc Natl Acad Sci U S A* **1975,** *72* (9), 3402-6.
9. Gunning, B. E., The Fine Structure of Chloroplast Stroma Following Aldehyde Osmium-Tetroxide Fixation. *The Journal of cell biology* **1965,** *24*, 79-93.
10. Harper, D. B., Microbial metabolism of aromatic nitriles. Enzymology of C-N cleavage by Nocardia sp. (Rhodochrous group) N.C.I.B. 11216. *The Biochemical journal* **1977,** *165* (2), 309-19.
11. Harper, D. B., Fungal degradation of aromatic nitriles. Enzymology of C-N cleavage by Fusarium solani. *The Biochemical journal* **1977,** *167* (3), 685-92.
12. Eisenberg, H.; Reisler, E., Angular dependence of scattered light, rotary frictional coefficients, and distribution of sizes of associated oligomers in solutions of bovine liver glutamate dehydrogenase. *Biopolymers* **1971,** *10* (12), 2363-76.
13. Huang, C. Y.; Frieden, C., The mechanism of ligand-induced structural changes in glutamate dehydrogenase. Studies of the rate of depolymerization and isomerization effected by coenzymes and guanine nucleotides. *J Biol Chem* **1972,** *247* (11), 3638-46.
14. Josephs, R.; Borisy, G., Self-assembly of glutamic dehydrogenase into ordered superstructures: multichain tubes formed by association of single molecules. *J Mol Biol* **1972,** *65* (1), 127-55.
15. Zeiri, L.; Reisler, E., Uncoupling of the catalytic activity and the polymerization of beef liver glutamate dehydrogenase. *J Mol Biol* **1978,** *124* (1), 291-5.
16. Olsen, B. R.; Svenneby, G.; Kvamme, E.; Tveit, B.; Eskeland, T., Formation and ultrastructure of enzymically active polymers of pig renal glutaminase. *J Mol Biol* **1970,** *52* (2), 239-45.
17. Park, C. K.; Stiteler, A. P.; Shah, S.; Ghare, M. I.; Bitinaite, J.; Horton, N. C., Activation of DNA cleavage by oligomerization of DNA-bound SgrAI. *Biochemistry* **2010,** *49* (41), 8818-30.
18. Korennykh, A. V.; Egea, P. F.; Korostelev, A. A.; Finer-Moore, J.; Zhang, C.; Shokat, K. M.; Stroud, R. M.; Walter, P., The unfolded protein response signals through high-order assembly of Ire1. *Nature* **2009,** *457* (7230), 687-93.
19. Ingerson-Mahar, M.; Briegel, A.; Werner, J. N.; Jensen, G. J.; Gitai, Z., The metabolic enzyme CTP synthase forms cytoskeletal filaments. *Nat Cell Biol* **2010,** *12* (8), 739-46.
20. Kim, C. W.; Moon, Y. A.; Park, S. W.; Cheng, D.; Kwon, H. J.; Horton, J. D., Induced polymerization of mammalian acetyl-CoA carboxylase by MIG12 provides a tertiary level of regulation of fatty acid synthesis. *Proc Natl Acad Sci U S A* **2010,** *107* (21), 9626-31.
21. Cutler, S. R.; Somerville, C. R., Imaging plant cell death: GFP-Nit1 aggregation marks an early step of wound and herbicide induced cell death. *BMC Plant Biol* **2005,** *5*, 4.
22. Kessler, D.; Herth, W.; Knappe, J., Ultrastructure and pyruvate formate-lyase radical quenching property of the multienzymic AdhE protein of Escherichia coli. *J Biol Chem* **1992,** *267* (25), 18073-9.
23. Werner, J. N.; Chen, E. Y.; Guberman, J. M.; Zippilli, A. R.; Irgon, J. J.; Gitai, Z., Quantitative genome-scale analysis of protein localization in an asymmetric bacterium. *Proc Natl Acad Sci U S A* **2009,** *106* (19), 7858-63.

24. Noree, C.; Sato, B. K.; Broyer, R. M.; Wilhelm, J. E., Identification of novel filament-forming proteins in Saccharomyces cerevisiae and Drosophila melanogaster. *The Journal of cell biology* **2010,** *190* (4), 541-51.
25. Narayanaswamy, R.; Levy, M.; Tsechansky, M.; Stovall, G. M.; O'Connell, J. D.; Mirrielees, J.; Ellington, A. D.; Marcotte, E. M., Widespread reorganization of metabolic enzymes into reversible assemblies upon nutrient starvation. *Proc Natl Acad Sci U S A* **2009,** *106* (25), 10147-52.
26. Shen, Q.-J.; Kassim, H.; Huang, Y.; Li, H.; Zhang, J.; Li, G.; Wang, P.-Y.; Ye, F.; Liu, J.-L., Filamentation of Metabolic Enzymes in *Saccharomyces cerevisiae*. *Journal of Genetics and Genomics* **2016,** *43*, 393-404.
27. Liu, J. L., Intracellular compartmentation of CTP synthase in Drosophila. *J Genet Genomics* **2010,** *37* (5), 281-96.
28. Suresh, H. G.; da Silveira Dos Santos, A. X.; Kukulski, W.; Tyedmers, J.; Riezman, H.; Bukau, B.; Mogk, A., Prolonged starvation drives reversible sequestration of lipid biosynthetic enzymes and organelle reorganization in Saccharomyces cerevisiae. *Molecular biology of the cell* **2015,** *26* (9), 1601-15.
29. Lowe, N.; Rees, J. S.; Roote, J.; Ryder, E.; Armean, I. M.; Johnson, G.; Drummond, E.; Spriggs, H.; Drummond, J.; Magbanua, J. P.; Naylor, H.; Sanson, B.; Bastock, R.; Huelsmann, S.; Trovisco, V.; Landgraf, M.; Knowles-Barley, S.; Armstrong, J. D.; White-Cooper, H.; Hansen, C.; Phillips, R. G.; Consortium, U. K. D. P. T. S.; Lilley, K. S.; Russell, S.; St Johnston, D., Analysis of the expression patterns, subcellular localisations and interaction partners of Drosophila proteins using a pigP protein trap library. *Development* **2014,** *141* (20), 3994-4005.
30. Ibstedt, S.; Sideri, T. C.; Grant, C. M.; Tamas, M. J., Global analysis of protein aggregation in yeast during physiological conditions and arsenite stress. *Biol Open* **2014,** *3* (10), 913-23.
31. Liu, J. L., The Cytoophidium and Its Kind: Filamentation and Compartmentation of Metabolic Enzymes. *Annu Rev Cell Dev Biol* **2016,** *32*, 349-372.
32. Noree, C.; Monfort, E.; Shiau, A. K.; Wilhelm, J. E., Common regulatory control of CTP synthase enzyme activity and filament formation. *Molecular biology of the cell* **2014,** *25* (15), 2282-90.
33. Jin, M.; Fuller, G. G.; Han, T.; Yao, Y.; Alessi, A. F.; Freeberg, M. A.; Roach, N. P.; Moresco, J. J.; Karnovsky, A.; Baba, M.; Yates, J. R., 3rd; Gitler, A. D.; Inoki, K.; Klionsky, D. J.; Kim, J. K., Glycolytic Enzymes Coalesce in G Bodies under Hypoxic Stress. *Cell Rep* **2017,** *20* (4), 895-908.
34. Woodward, J. D.; Weber, B. W.; Scheffer, M. P.; Benedik, M. J.; Hoenger, A.; Sewell, B. T., Helical structure of unidirectionally shadowed metal replicas of cyanide hydratase from Gloeocercospora sorghi. *J Struct Biol* **2008,** *161* (2), 111-9.
35. Bitinaite, J.; Schildkraut, I., Self-generated DNA termini relax the specificity of SgrAI restriction endonuclease. *Proc Natl Acad Sci U S A* **2002,** *99* (3), 1164-9.
36. Ma, X.; Shah, S.; Zhou, M.; Park, C. K.; Wysocki, V. H.; Horton, N. C., Structural Analysis of Activated SgrAI-DNA Oligomers Using Ion Mobility Mass Spectrometry. *Biochemistry* **2013,** *52* (25), 4373-81.
37. Lyumkis, D.; Talley, H.; Stewart, A.; Shah, S.; Park, C. K.; Tama, F.; Potter, C. S.; Carragher, B.; Horton, N. C., Allosteric regulation of DNA cleavage and sequence-specificity through run-on oligomerization. *Structure* **2013,** *21* (10), 1848-58.
38. Shah, S.; Sanchez, J.; Stewart, A.; Piperakis, M. M.; Cosstick, R.; Nichols, C.; Park, C. K.; Ma, X.; Wysocki, V.; Bitinaite, J.; Horton, N. C., Probing the Run-On Oligomer of Activated SgrAI Bound to DNA. *PLoS One* **2015,** *10* (4), e0124783.
39. Polley, S.; Lyumkis, D.; Horton, N. C., Mechanism of Activation by Filament Formation of a Sequence Specific Endonuclease. *Submitted for publication, Biorxiv doi: https://doi.org/10.1101/585943* **2019**.
40. Carcamo, W. C.; Satoh, M.; Kasahara, H.; Terada, N.; Hamazaki, T.; Chan, J. Y.; Yao, B.; Tamayo, S.; Covini, G.; von Muhlen, C. A.; Chan, E. K., Induction of cytoplasmic rods and rings structures by inhibition of the CTP and GTP synthetic pathway in mammalian cells. *PLoS One* **2011,** *6* (12), e29690.
41. Chen, K.; Zhang, J.; Tastan, O. Y.; Deussen, Z. A.; Siswick, M. Y.; Liu, J. L., Glutamine analogs promote cytoophidium assembly in human and Drosophila cells. *J Genet Genomics* **2011,** *38* (9), 391-402.
42. O'Connell, J. D.; Zhao, A.; Ellington, A. D.; Marcotte, E. M., Dynamic reorganization of metabolic enzymes into intracellular bodies. *Annu Rev Cell Dev Biol* **2012,** *28*, 89-111.

43. Barry, R. M.; Gitai, Z., Self-assembling enzymes and the origins of the cytoskeleton. *Current opinion in microbiology* **2011,** *14* (6), 704-11.
44. Cabeen, M. T.; Jacobs-Wagner, C., A metabolic assembly line in bacteria. *Nat Cell Biol* **2010,** *12* (8), 731-3.
45. Prouteau, M.; Loewith, R., Regulation of Cellular Metabolism through Phase Separation of Enzymes. *Biomolecules* **2018,** *8* (4).
46. Carcamo, W. C.; Calise, S. J.; von Muhlen, C. A.; Satoh, M.; Chan, E. K., Molecular cell biology and immunobiology of mammalian rod/ring structures. *Int Rev Cell Mol Biol* **2014,** *308,* 35-74.
47. Garcia-Seisdedos, H.; Empereur-Mot, C.; Elad, N.; Levy, E. D., Proteins evolve on the edge of supramolecular self-assembly. *Nature* **2017,** *548* (7666), 244-247.
48. Selwood, T.; Jaffe, E. K., Dynamic dissociating homo-oligomers and the control of protein function. *Archives of biochemistry and biophysics* **2012,** *519* (2), 131-43.
49. Griffin, M. D. W.; Gerrard, J. A., The Relationship Between Oligomeric State and Protein Function. In *Protein Dimerization and Oligomerization in Biology*, Matthews, J. M., Ed. Landes Bioscience and Springer Science+Business Media: 2012; pp 74-90.
50. Barahona, C. J.; Basantes, L. E.; Tompkins, K. J.; Heitman, D. M.; Chukwu, B. I.; Sanchez, J.; Sanchez, J. L.; Ghadirian, N.; Park, C. K.; Horton, N. C., The Need for Speed: Run-On Oligomer Filament Formation Provides Maximum Speed with Maximum Sequestration of Activity. *J Virol* **2019,** *93* (5).
51. Park, C. K.; Sanchez, J. L.; Barahona, C.; Basantes, L. E.; Sanchez, J.; Hernandez, C.; Horton, N. C., The run-on oligomer filament enzyme mechanism of SgrAI: Part 1. Assembly kinetics of the run-on oligomer filament. *J Biol Chem* **2018,** *293* (38), 14585-14598.
52. Park, C. K.; Sanchez, J. L.; Barahona, C.; Basantes, L. E.; Sanchez, J.; Hernandez, C.; Horton, N. C., The run-on oligomer filament enzyme mechanism of SgrAI: Part 2. Kinetic modeling of the full DNA cleavage pathway. *J Biol Chem* **2018,** *293* (38), 14599-14615.
53. Tong, L., Structure and function of biotin-dependent carboxylases. *Cell Mol Life Sci* **2013,** *70* (5), 863-91.
54. Harwood, H. J., Jr., Acetyl-CoA carboxylase inhibition for the treatment of metabolic syndrome. *Curr Opin Investig Drugs* **2004,** *5* (3), 283-9.
55. Stiede, K.; Miao, W.; Blanchette, H. S.; Beysen, C.; Harriman, G.; Harwood, H. J., Jr.; Kelley, H.; Kapeller, R.; Schmalbach, T.; Westlin, W. F., Acetyl-coenzyme A carboxylase inhibition reduces de novo lipogenesis in overweight male subjects: A randomized, double-blind, crossover study. *Hepatology* **2017,** *66* (2), 324-334.
56. Swinnen, J. V.; Brusselmans, K.; Verhoeven, G., Increased lipogenesis in cancer cells: new players, novel targets. *Curr Opin Clin Nutr Metab Care* **2006,** *9* (4), 358-65.
57. Guri, Y.; Colombi, M.; Dazert, E.; Hindupur, S. K.; Roszik, J.; Moes, S.; Jenoe, P.; Heim, M. H.; Riezman, I.; Riezman, H.; Hall, M. N., mTORC2 Promotes Tumorigenesis via Lipid Synthesis. *Cancer Cell* **2017,** *32* (6), 807-823 e12.
58. Svensson, R. U.; Parker, S. J.; Eichner, L. J.; Kolar, M. J.; Wallace, M.; Brun, S. N.; Lombardo, P. S.; Van Nostrand, J. L.; Hutchins, A.; Vera, L.; Gerken, L.; Greenwood, J.; Bhat, S.; Harriman, G.; Westlin, W. F.; Harwood, H. J., Jr.; Saghatelian, A.; Kapeller, R.; Metallo, C. M.; Shaw, R. J., Inhibition of acetyl-CoA carboxylase suppresses fatty acid synthesis and tumor growth of non-small-cell lung cancer in preclinical models. *Nat Med* **2016,** *22* (10), 1108-1119.
59. Abu-Elheiga, L.; Jayakumar, A.; Baldini, A.; Chirala, S. S.; Wakil, S. J., Human acetyl-CoA carboxylase: characterization, molecular cloning, and evidence for two isoforms. *Proc Natl Acad Sci U S A* **1995,** *92* (9), 4011-5.
60. Saggerson, D., Malonyl-CoA, a key signaling molecule in mammalian cells. *Annu Rev Nutr* **2008,** *28*, 253-72.
61. Wakil, S. J.; Abu-Elheiga, L. A., Fatty acid metabolism: target for metabolic syndrome. *J Lipid Res* **2009,** *50 Suppl*, S138-43.
62. Abu-Elheiga, L.; Almarza-Ortega, D. B.; Baldini, A.; Wakil, S. J., Human acetyl-CoA carboxylase 2. Molecular cloning, characterization, chromosomal mapping, and evidence for two isoforms. *J Biol Chem* **1997,** *272* (16), 10669-77.

63. Abu-Elheiga, L.; Matzuk, M. M.; Abo-Hashema, K. A.; Wakil, S. J., Continuous fatty acid oxidation and reduced fat storage in mice lacking acetyl-CoA carboxylase 2. *Science* **2001,** *291* (5513), 2613-6.
64. Abu-Elheiga, L.; Oh, W.; Kordari, P.; Wakil, S. J., Acetyl-CoA carboxylase 2 mutant mice are protected against obesity and diabetes induced by high-fat/high-carbohydrate diets. *Proc Natl Acad Sci U S A* **2003,** *100* (18), 10207-12.
65. Ha, J.; Daniel, S.; Broyles, S. S.; Kim, K. H., Critical phosphorylation sites for acetyl-CoA carboxylase activity. *J Biol Chem* **1994,** *269* (35), 22162-8.
66. Brownsey, R. W.; Boone, A. N.; Elliott, J. E.; Kulpa, J. E.; Lee, W. M., Regulation of acetyl-CoA carboxylase. *Biochem Soc Trans* **2006,** *34* (Pt 2), 223-7.
67. Vagelos, P. R.; Alberts, A. W.; Martin, D. B., Activation of acetyl-CoA carboxylase and associated alteration of sedimentation characteristics of the enzyme. *Biochem Biophys Res Commun* **1962,** *8*, 4-8.
68. Moss, J.; Lane, M. D., Acetyl coenzyme A carboxylase. IV. Biotinyl prosthetic group-independent malonyl coenzyme A decarboxylation and carbosyl transfer: generalization to other biotin enzymes. *J Biol Chem* **1972,** *247* (16), 4952-9.
69. Ashcraft, B. A.; Fillers, W. S.; Augustine, S. L.; Clarke, S. D., Polymer-protomer transition of acetyl-CoA carboxylase occurs in vivo and varies with nutritional conditions. *J Biol Chem* **1980,** *255* (21), 10033-5.
70. Park, S.; Hwang, I. W.; Makishima, Y.; Perales-Clemente, E.; Kato, T.; Niederlander, N. J.; Park, E. Y.; Terzic, A., Spot14/Mig12 heterocomplex sequesters polymerization and restrains catalytic function of human acetyl-CoA carboxylase 2. *J Mol Recognit* **2013,** *26* (12), 679-88.
71. Hunkeler, M.; Hagmann, A.; Stuttfeld, E.; Chami, M.; Guri, Y.; Stahlberg, H.; Maier, T., Structural basis for regulation of human acetyl-CoA carboxylase. *Nature* **2018,** *558* (7710), 470-474.
72. Ray, H.; Suau, F.; Vincent, A.; Dalla Venezia, N., Cell cycle regulation of the BRCA1/acetyl-CoA-carboxylase complex. *Biochem Biophys Res Commun* **2009,** *378* (3), 615-9.
73. Magnard, C.; Bachelier, R.; Vincent, A.; Jaquinod, M.; Kieffer, S.; Lenoir, G. M.; Venezia, N. D., BRCA1 interacts with acetyl-CoA carboxylase through its tandem of BRCT domains. *Oncogene* **2002,** *21* (44), 6729-39.
74. Shen, Y.; Tong, L., Structural evidence for direct interactions between the BRCT domains of human BRCA1 and a phospho-peptide from human ACC1. *Biochemistry* **2008,** *47* (21), 5767-73.
75. Jang, S.; Nelson, J. C.; Bend, E. G.; Rodriguez-Laureano, L.; Tueros, F. G.; Cartagenova, L.; Underwood, K.; Jorgensen, E. M.; Colon-Ramos, D. A., Glycolytic Enzymes Localize to Synapses under Energy Stress to Support Synaptic Function. *Neuron* **2016,** *90* (2), 278-91.
76. Webb, B. A.; Dosey, A. M.; Wittmann, T.; Kollman, J. M.; Barber, D. L., The glycolytic enzyme phosphofructokinase-1 assembles into filaments. *The Journal of cell biology* **2017,** *216* (8), 2305-2313.
77. Foe, L. G.; Trujillo, J. L., Quaternary structure of pig liver phosphofructokinase. *J Biol Chem* **1980,** *255* (21), 10537-41.
78. Webb, B. A.; Forouhar, F.; Szu, F. E.; Seetharaman, J.; Tong, L.; Barber, D. L., Structures of human phosphofructokinase-1 and atomic basis of cancer-associated mutations. *Nature* **2015,** *523* (7558), 111-4.
79. Moreno-Sanchez, R.; Marin-Hernandez, A.; Gallardo-Perez, J. C.; Quezada, H.; Encalada, R.; Rodriguez-Enriquez, S.; Saavedra, E., Phosphofructokinase type 1 kinetics, isoform expression, and gene polymorphisms in cancer cells. *J Cell Biochem* **2012,** *113* (5), 1692-703.
80. Yi, W.; Clark, P. M.; Mason, D. E.; Keenan, M. C.; Hill, C.; Goddard, W. A., 3rd; Peters, E. C.; Driggers, E. M.; Hsieh-Wilson, L. C., Phosphofructokinase 1 glycosylation regulates cell growth and metabolism. *Science* **2012,** *337* (6097), 975-80.
81. Hanahan, D.; Weinberg, R. A., Hallmarks of cancer: the next generation. *Cell* **2011,** *144* (5), 646-74.
82. Ristow, M.; Carlqvist, H.; Hebinck, J.; Vorgerd, M.; Krone, W.; Pfeiffer, A.; Muller-Wieland, D.; Ostenson, C. G., Deficiency of phosphofructo-1-kinase/muscle subtype in humans is associated with impairment of insulin secretory oscillations. *Diabetes* **1999,** *48* (8), 1557-61.
83. Ristow, M.; Vorgerd, M.; Mohlig, M.; Schatz, H.; Pfeiffer, A., Insulin resistance and impaired insulin secretion due to phosphofructo-1-kinase-deficiency in humans. *J Mol Med (Berl)* **1999,** *77* (1), 96-103.

84. Yang, Z.; Matteson, E. L.; Goronzy, J. J.; Weyand, C. M., T-cell metabolism in autoimmune disease. *Arthritis Res Ther* **2015,** *17*, 29.
85. Bruser, A.; Kirchberger, J.; Kloos, M.; Strater, N.; Schoneberg, T., Functional linkage of adenine nucleotide binding sites in mammalian muscle 6-phosphofructokinase. *J Biol Chem* **2012,** *287* (21), 17546-53.
86. Kloos, M.; Bruser, A.; Kirchberger, J.; Schoneberg, T.; Strater, N., Crystallization and preliminary crystallographic analysis of human muscle phosphofructokinase, the main regulator of glycolysis. *Acta Crystallogr F Struct Biol Commun* **2014,** *70* (Pt 5), 578-82.
87. Kloos, M.; Bruser, A.; Kirchberger, J.; Schoneberg, T.; Strater, N., Crystal structure of human platelet phosphofructokinase-1 locked in an activated conformation. *The Biochemical journal* **2015,** *469* (3), 421-32.
88. Schoneberg, T.; Kloos, M.; Bruser, A.; Kirchberger, J.; Strater, N., Structure and allosteric regulation of eukaryotic 6-phosphofructokinases. *Biological chemistry* **2013,** *394* (8), 977-93.
89. Maitra, P. K., A glucokinase from Saccharomyces cerevisiae. *J Biol Chem* **1970,** *245* (9), 2423-31.
90. Stoddard, P. R.; Lynch, E. M.; Farrell, D. P.; Justman, Q. A.; Dosey, A. M.; DiMaio, F.; Williams, T. A.; Kollman, J. M.; Murray, A. W.; Garner, E. C., Independent evolution of polymerization in the Actin ATPase clan regulates hexokinase activity. *bioRxiv* **2019**.
91. Martin, E.; Palmic, N.; Sanquer, S.; Lenoir, C.; Hauck, F.; Mongellaz, C.; Fabrega, S.; Nitschke, P.; Esposti, M. D.; Schwartzentruber, J.; Taylor, N.; Majewski, J.; Jabado, N.; Wynn, R. F.; Picard, C.; Fischer, A.; Arkwright, P. D.; Latour, S., CTP synthase 1 deficiency in humans reveals its central role in lymphocyte proliferation. *Nature* **2014,** *510* (7504), 288-92.
92. Hofer, A.; Steverding, D.; Chabes, A.; Brun, R.; Thelander, L., Trypanosoma brucei CTP synthetase: a target for the treatment of African sleeping sickness. *Proc Natl Acad Sci U S A* **2001,** *98* (11), 6412-6.
93. Hindenburg, A. A.; Taub, R. N.; Grant, S.; Chang, G.; Baker, M. A., Effects of pyrimidine antagonists on sialic acid regeneration in HL-60 cells. *Cancer Res* **1985,** *45* (7), 3048-52.
94. Kang, G. J.; Cooney, D. A.; Moyer, J. D.; Kelley, J. A.; Kim, H. Y.; Marquez, V. E.; Johns, D. G., Cyclopentenylcytosine triphosphate. Formation and inhibition of CTP synthetase. *J Biol Chem* **1989,** *264* (2), 713-8.
95. Politi, P. M.; Xie, F.; Dahut, W.; Ford, H., Jr.; Kelley, J. A.; Bastian, A.; Setser, A.; Allegra, C. J.; Chen, A. P.; Hamilton, J. M.; et al., Phase I clinical trial of continuous infusion cyclopentenyl cytosine. *Cancer Chemother Pharmacol* **1995,** *36* (6), 513-23.
96. Barry, R. M.; Bitbol, A. F.; Lorestani, A.; Charles, E. J.; Habrian, C. H.; Hansen, J. M.; Li, H. J.; Baldwin, E. P.; Wingreen, N. S.; Kollman, J. M.; Gitai, Z., Large-scale filament formation inhibits the activity of CTP synthetase. *Elife* **2014,** *3*, e03638.
97. Aughey, G. N.; Grice, S. J.; Shen, Q. J.; Xu, Y.; Chang, C. C.; Azzam, G.; Wang, P. Y.; Freeman-Mills, L.; Pai, L. M.; Sung, L. Y.; Yan, J.; Liu, J. L., Nucleotide synthesis is regulated by cytoophidium formation during neurodevelopment and adaptive metabolism. *Biol Open* **2014,** *3* (11), 1045-56.
98. Lynch, E. M.; Hicks, D. R.; Shepherd, M.; Endrizzi, J. A.; Maker, A.; Hansen, J. M.; Barry, R. M.; Gitai, Z.; Baldwin, E. P.; Kollman, J. M., Human CTP synthase filament structure reveals the active enzyme conformation. *Nat Struct Mol Biol* **2017,** *24* (6), 507-514.
99. Calise, S. J.; Carcamo, W. C.; Krueger, C.; Yin, J. D.; Purich, D. L.; Chan, E. K., Glutamine deprivation initiates reversible assembly of mammalian rods and rings. *Cell Mol Life Sci* **2014,** *71* (15), 2963-73.
100. Strochlic, T. I.; Stavrides, K. P.; Thomas, S. V.; Nicolas, E.; O'Reilly, A. M.; Peterson, J. R., Ack kinase regulates CTP synthase filaments during Drosophila oogenesis. *EMBO reports* **2014,** *15* (11), 1184-91.
101. Petrovska, I.; Nuske, E.; Munder, M. C.; Kulasegaran, G.; Malinovska, L.; Kroschwald, S.; Richter, D.; Fahmy, K.; Gibson, K.; Verbavatz, J. M.; Alberti, S., Filament formation by metabolic enzymes is a specific adaptation to an advanced state of cellular starvation. *Elife* **2014,** *3*, e02409.
102. Chang, Y. F.; Martin, S. S.; Baldwin, E. P.; Carman, G. M., Phosphorylation of human CTP synthetase 1 by protein kinase C: identification of Ser(462) and Thr(455) as major sites of phosphorylation. *J Biol Chem* **2007,** *282* (24), 17613-22.

103. Choi, M. G.; Park, T. S.; Carman, G. M., Phosphorylation of Saccharomyces cerevisiae CTP synthetase at Ser424 by protein kinases A and C regulates phosphatidylcholine synthesis by the CDP-choline pathway. *J Biol Chem* **2003,** *278* (26), 23610-6.
104. Han, G. S.; Sreenivas, A.; Choi, M. G.; Chang, Y. F.; Martin, S. S.; Baldwin, E. P.; Carman, G. M., Expression of Human CTP synthetase in Saccharomyces cerevisiae reveals phosphorylation by protein kinase A. *J Biol Chem* **2005,** *280* (46), 38328-36.
105. Endrizzi, J. A.; Kim, H.; Anderson, P. M.; Baldwin, E. P., Mechanisms of product feedback regulation and drug resistance in cytidine triphosphate synthetases from the structure of a CTP-inhibited complex. *Biochemistry* **2005,** *44* (41), 13491-9.
106. Goto, M.; Omi, R.; Nakagawa, N.; Miyahara, I.; Hirotsu, K., Crystal structures of CTP synthetase reveal ATP, UTP, and glutamine binding sites. *Structure* **2004,** *12* (8), 1413-23.
107. Kursula, P.; Flodin, S.; Ehn, M.; Hammarstrom, M.; Schuler, H.; Nordlund, P.; Stenmark, P., Structure of the synthetase domain of human CTP synthetase, a target for anticancer therapy. *Acta Crystallogr Sect F Struct Biol Cryst Commun* **2006,** *62* (Pt 7), 613-7.
108. Lauritsen, I.; Willemoes, M.; Jensen, K. F.; Johansson, E.; Harris, P., Structure of the dimeric form of CTP synthase from Sulfolobus solfataricus. *Acta Crystallogr Sect F Struct Biol Cryst Commun* **2011,** *67* (Pt 2), 201-8.
109. Trudel, M.; Van Genechten, T.; Meuth, M., Biochemical characterization of the hamster thy mutator gene and its revertants. *J Biol Chem* **1984,** *259* (4), 2355-9.
110. Rudolph, F. B., The biochemistry and physiology of nucleotides. *J Nutr* **1994,** *124* (1 Suppl), 124S-127S.
111. An, S.; Kumar, R.; Sheets, E. D.; Benkovic, S. J., Reversible compartmentalization of de novo purine biosynthetic complexes in living cells. *Science* **2008,** *320* (5872), 103-6.
112. Deng, Y.; Gam, J.; French, J. B.; Zhao, H.; An, S.; Benkovic, S. J., Mapping protein-protein proximity in the purinosome. *J Biol Chem* **2012,** *287* (43), 36201-7.
113. Nair, V.; Ma, X.; Shu, Q.; Zhang, F.; Uchil, V.; Cherukupalli, G. R., IMPDH as a biological probe for RNA antiviral drug discovery: synthesis, enzymology, molecular docking, and antiviral activity of new ribonucleosides with surrogate bases. *Nucleosides Nucleotides Nucleic Acids* **2007,** *26* (6-7), 651-4.
114. Nair, V.; Shu, Q., Inosine monophosphate dehydrogenase as a probe in antiviral drug discovery. *Antivir Chem Chemother* **2007,** *18* (5), 245-58.
115. Chen, L.; Pankiewicz, K. W., Recent development of IMP dehydrogenase inhibitors for the treatment of cancer. *Curr Opin Drug Discov Devel* **2007,** *10* (4), 403-12.
116. Ratcliffe, A. J., Inosine 5'-monophosphate dehydrogenase inhibitors for the treatment of autoimmune diseases. *Curr Opin Drug Discov Devel* **2006,** *9* (5), 595-605.
117. Hedstrom, L., IMP dehydrogenase: structure, mechanism, and inhibition. *Chem Rev* **2009,** *109* (7), 2903-28.
118. Carr, S. F.; Papp, E.; Wu, J. C.; Natsumeda, Y., Characterization of human type I and type II IMP dehydrogenases. *J Biol Chem* **1993,** *268* (36), 27286-90.
119. Gunter, J. H.; Thomas, E. C.; Lengefeld, N.; Kruger, S. J.; Worton, L.; Gardiner, E. M.; Jones, A.; Barnett, N. L.; Whitehead, J. P., Characterisation of inosine monophosphate dehydrogenase expression during retinal development: differences between variants and isoforms. *Int J Biochem Cell Biol* **2008,** *40* (9), 1716-28.
120. Thomas, E. C.; Gunter, J. H.; Webster, J. A.; Schieber, N. L.; Oorschot, V.; Parton, R. G.; Whitehead, J. P., Different characteristics and nucleotide binding properties of inosine monophosphate dehydrogenase (IMPDH) isoforms. *PLoS One* **2012,** *7* (12), e51096.
121. Kozhevnikova, E. N.; van der Knaap, J. A.; Pindyurin, A. V.; Ozgur, Z.; van Ijcken, W. F.; Moshkin, Y. M.; Verrijzer, C. P., Metabolic enzyme IMPDH is also a transcription factor regulated by cellular state. *Molecular cell* **2012,** *47* (1), 133-9.
122. Ji, Y.; Gu, J.; Makhov, A. M.; Griffith, J. D.; Mitchell, B. S., Regulation of the interaction of inosine monophosphate dehydrogenase with mycophenolic Acid by GTP. *J Biol Chem* **2006,** *281* (1), 206-12.
123. Chang, C. C.; Keppeke, G. D.; Sung, L. Y.; Liu, J. L., Interfilament interaction between IMPDH and CTPS cytoophidia. *The FEBS journal* **2018,** *285* (20), 3753-3768.

124. Keppeke, G. D.; Chang, C. C.; Peng, M.; Chen, L. Y.; Lin, W. C.; Pai, L. M.; Andrade, L. E. C.; Sung, L. Y.; Liu, J. L., IMP/GTP balance modulates cytoophidium assembly and IMPDH activity. *Cell Div* **2018,** *13*, 5.
125. Keppeke, G. D.; Calise, S. J.; Chan, E. K.; Andrade, L. E., Assembly of IMPDH2-based, CTPS-based, and mixed rod/ring structures is dependent on cell type and conditions of induction. *J Genet Genomics* **2015,** *42* (6), 287-99.
126. Chang, C. C.; Lin, W. C.; Pai, L. M.; Lee, H. S.; Wu, S. C.; Ding, S. T.; Liu, J. L.; Sung, L. Y., Cytoophidium assembly reflects upregulation of IMPDH activity. *J Cell Sci* **2015,** *128* (19), 3550-5.
127. Labesse, G.; Alexandre, T.; Vaupre, L.; Salard-Arnaud, I.; Him, J. L.; Raynal, B.; Bron, P.; Munier-Lehmann, H., MgATP regulates allostery and fiber formation in IMPDHs. *Structure* **2013,** *21* (6), 975-85.
128. Anthony, S. A.; Burrell, A. L.; Johnson, M. C.; Duong-Ly, K. C.; Kuo, Y. M.; Simonet, J. C.; Michener, P.; Andrews, A.; Kollman, J. M.; Peterson, J. R., Reconstituted IMPDH polymers accommodate both catalytically active and inactive conformations. *Molecular biology of the cell* **2017**.
129. Eisenberg, D.; Gill, H. S.; Pfluegl, G. M.; Rotstein, S. H., Structure-function relationships of glutamine synthetases. *Biochim Biophys Acta* **2000,** *1477* (1-2), 122-45.
130. O'Connell, J. D.; Tsechansky, M.; Royall, A.; Boutz, D. R.; Ellington, A. D.; Marcotte, E. M., A proteomic survey of widespread protein aggregation in yeast. *Mol Biosyst* **2014,** *10* (4), 851-61.
131. He, Y. X.; Gui, L.; Liu, Y. Z.; Du, Y.; Zhou, Y.; Li, P.; Zhou, C. Z., Crystal structure of Saccharomyces cerevisiae glutamine synthetase Gln1 suggests a nanotube-like supramolecular assembly. *Proteins* **2009,** *76* (1), 249-54.
132. Orij, R.; Postmus, J.; Ter Beek, A.; Brul, S.; Smits, G. J., In vivo measurement of cytosolic and mitochondrial pH using a pH-sensitive GFP derivative in Saccharomyces cerevisiae reveals a relation between intracellular pH and growth. *Microbiology* **2009,** *155* (Pt 1), 268-78.
133. Dechant, R.; Binda, M.; Lee, S. S.; Pelet, S.; Winderickx, J.; Peter, M., Cytosolic pH is a second messenger for glucose and regulates the PKA pathway through V-ATPase. *EMBO J* **2010,** *29* (15), 2515-26.
134. Orij, R.; Brul, S.; Smits, G. J., Intracellular pH is a tightly controlled signal in yeast. *Biochim Biophys Acta* **2011,** *1810* (10), 933-44.
135. Bernstein, B. W.; Chen, H.; Boyle, J. A.; Bamburg, J. R., Formation of actin-ADF/cofilin rods transiently retards decline of mitochondrial potential and ATP in stressed neurons. *Am J Physiol Cell Physiol* **2006,** *291* (5), C828-39.
136. Bernstein, B. W.; Bamburg, J. R., ADF/cofilin: a functional node in cell biology. *Trends Cell Biol* **2010,** *20* (4), 187-95.
137. Peters, L. Z.; Hazan, R.; Breker, M.; Schuldiner, M.; Ben-Aroya, S., Formation and dissociation of proteasome storage granules are regulated by cytosolic pH. *The Journal of cell biology* **2013,** *201* (5), 663-71.
138. Isom, D. G.; Sridharan, V.; Baker, R.; Clement, S. T.; Smalley, D. M.; Dohlman, H. G., Protons as second messenger regulators of G protein signaling. *Molecular cell* **2013,** *51* (4), 531-8.
139. Orij, R.; Urbanus, M. L.; Vizeacoumar, F. J.; Giaever, G.; Boone, C.; Nislow, C.; Brul, S.; Smits, G. J., Genome-wide analysis of intracellular pH reveals quantitative control of cell division rate by pH(c) in Saccharomyces cerevisiae. *Genome Biol* **2012,** *13* (9), R80.
140. Kim, Y. W.; Kang, K. S.; Kim, S. Y.; Kim, I. S., Formation of fibrillar multimers of oat beta-glucosidase isoenzymes is mediated by the As-Glu1 monomer. *J Mol Biol* **2000,** *303* (5), 831-42.
141. Kim, S. Y.; Kim, Y. W.; Hegerl, R.; Cyrklaff, M.; Kim, I. S., Novel type of enzyme multimerization enhances substrate affinity of oat beta-glucosidase. *J Struct Biol* **2005,** *150* (1), 1-10.
142. Nisius, A.; Ruppel, H. G., Immunocytological and chemical studies on the stromacentre-forming protein from Avena plastids. *Planta* **1987,** *171* (4), 443-52.
143. Maia, L. B.; Moura, J. J.; Moura, I., Molybdenum and tungsten-dependent formate dehydrogenases. *J Biol Inorg Chem* **2015,** *20* (2), 287-309.
144. Sawers, G., The hydrogenases and formate dehydrogenases of Escherichia coli. *Antonie Van Leeuwenhoek* **1994,** *66* (1-3), 57-88.
145. Schuchmann, K.; Muller, V., Direct and reversible hydrogenation of CO2 to formate by a bacterial carbon dioxide reductase. *Science* **2013,** *342* (6164), 1382-5.

146. Poehlein, A.; Schmidt, S.; Kaster, A. K.; Goenrich, M.; Vollmers, J.; Thurmer, A.; Bertsch, J.; Schuchmann, K.; Voigt, B.; Hecker, M.; Daniel, R.; Thauer, R. K.; Gottschalk, G.; Muller, V., An ancient pathway combining carbon dioxide fixation with the generation and utilization of a sodium ion gradient for ATP synthesis. *PLoS One* **2012,** *7* (3), e33439.
147. Schuchmann, K.; Muller, V., Autotrophy at the thermodynamic limit of life: a model for energy conservation in acetogenic bacteria. *Nat Rev Microbiol* **2014,** *12* (12), 809-21.
148. Schuchmann, K.; Vonck, J.; Muller, V., A bacterial hydrogen-dependent CO2 reductase forms filamentous structures. *The FEBS journal* **2016,** *283* (7), 1311-22.
149. Fernandes, B. C. M.; Mateo, C.; Kiziak, C.; Chmura, A.; Wacker, J.; van Rantwijk, F.; Stolz, A.; Sheldon, R. A., Nitrile hydratase activity of a recombinant nitrilase. *Adv. Synth. Catal.* **2006,** *348*, 2597-2603.
150. Pace, H. C.; Brenner, C., The nitrilase superfamily: classification, structure and function. *Genome Biol* **2001,** *2* (1), REVIEWS0001.
151. Sewell, B. T.; Berman, M. N.; Meyers, P. R.; Jandhyala, D.; Benedik, M. J., The cyanide degrading nitrilase from Pseudomonas stutzeri AK61 is a two-fold symmetric, 14-subunit spiral. *Structure* **2003,** *11* (11), 1413-22.
152. Thuku, R. N.; Weber, B. W.; Varsani, A.; Sewell, B. T., Post-translational cleavage of recombinantly expressed nitrilase from Rhodococcus rhodochrous J1 yields a stable, active helical form. *The FEBS journal* **2007,** *274* (8), 2099-108.
153. Dent, K. C.; Weber, B. W.; Benedik, M. J.; Sewell, B. T., The cyanide hydratase from Neurospora crassa forms a helix which has a dimeric repeat. *Appl Microbiol Biotechnol* **2009,** *82* (2), 271-8.
154. Vejvoda, V.; Kaplan, O.; Bezouska, K.; Pompach, P.; Sulc, M.; Cantarella, M.; Benada, O.; Uhnakov, B.; Rinagelov, A.; Lutz-Wahl, S.; Fischer, L.; Kren, V.; Martınkova, L., Purification and characterization of a nitrilase from Fusarium solani O1. *Journal of Molecular Catalysis B: Enzymatic* **2008,** *50*, 99-106.
155. Nagasawa, T.; Wieser, M.; Nakamura, T.; Iwahara, H.; Yoshida, T.; Gekko, K., Nitrilase of Rhodococcus rhodochrous J1. Conversion into the active form by subunit association. *European journal of biochemistry / FEBS* **2000,** *267* (1), 138-44.
156. Thuku, R. N.; Brady, D.; Benedik, M. J.; Sewell, B. T., Microbial nitrilases: versatile, spiral forming, industrial enzymes. *J Appl Microbiol* **2009,** *106* (3), 703-27.
157. Kawata, T.; Masuda, K.; Yoshino, K., Presence of fine spirals (spirosomes) in Lactobacillus fermenti and Lactobacillus casei. *Jpn J Microbiol* **1975,** *19* (3), 225-7.
158. Laurenceau, R.; Krasteva, P. V.; Diallo, A.; Ouarti, S.; Duchateau, M.; Malosse, C.; Chamot-Rooke, J.; Fronzes, R., Conserved Streptococcus pneumoniae spirosomes suggest a single type of transformation pilus in competence. *PLoS Pathog* **2015,** *11* (4), e1004835.
159. Extance, J.; Crennell, S. J.; Eley, K.; Cripps, R.; Hough, D. W.; Danson, M. J., Structure of a bifunctional alcohol dehydrogenase involved in bioethanol generation in Geobacillus thermoglucosidasius. *Acta Crystallogr D Biol Crystallogr* **2013,** *69* (Pt 10), 2104-15.
160. Hudson, R. C.; Daniel, R. M., L-glutamate dehydrogenases: distribution, properties and mechanism. *Comp Biochem Physiol B* **1993,** *106* (4), 767-92.
161. Lightfoot, D. A.; Baron, A. J.; Wootton, J. C., Expression of the Escherichia coli glutamate dehydrogenase gene in the cyanobacterium Synechococcus PCC6301 causes ammonium tolerance. *Plant Mol Biol* **1988,** *11* (3), 335-44.
162. Mungur, R.; Glass, A. D.; Goodenow, D. B.; Lightfoot, D. A., Metabolite fingerprinting in transgenic Nicotiana tabacum altered by the Escherichia coli glutamate dehydrogenase gene. *J Biomed Biotechnol* **2005,** *2005* (2), 198-214.
163. Grabowska, A.; Nowicki, M.; Kwinta, J., Glutamate dehydrogenase of the germinating triticale seeds: gene expression, activity distribution and kinetic characteristics. *Acta Physiol. Plant* **2011,** *33* (5), 1981-1990.
164. Lightfoot, D. A.; Bernhardt, K.; Mungur, R.; Nolte, S.; Ameziane, R.; Colter, A.; Jones, K.; Iqbal, M. J.; Varsa, E.; Young, B., Improved drought tolerance of transgenic Zea mays plants that express the glutamate dehydrogenase gene (gdhA) of E. coli. *Euphytica* **2007,** *156*, 103-116.

165. Frieden, C., Glutamate Dehydrogenase. Vi. Survey of Purine Nucleotide and Other Effects on the Enzyme from Various Sources. *J Biol Chem* **1965,** *240*, 2028-35.
166. Frieden, C., Glutamic dehydrogenase. I. The effect of coenzyme on the sedimentation velocity and kinetic behavior. *J Biol Chem* **1959,** *234* (4), 809-14.
167. Tomkins, G. M.; Yielding, K. L.; Curran, J. F., The influence of diethylstilbestrol and adenosine diphosphate on pyridine nucleotide coenzyme binding by glutamic dehydrogenase. *J Biol Chem* **1962,** *237*, 1704-8.
168. Bailey, J.; Bell, E. T.; Bell, J. E., Regulation of bovine glutamate dehydrogenase. The effects of pH and ADP. *J Biol Chem* **1982,** *257* (10), 5579-83.
169. Sener, A.; Malaisse, W. J., L-leucine and a nonmetabolized analogue activate pancreatic islet glutamate dehydrogenase. *Nature* **1980,** *288* (5787), 187-9.
170. Yielding, K. L.; Tomkins, G. M., An effect of L-leucine and other essential amino acids on the structure and activity of glutamic dehydrogenase. *Proc Natl Acad Sci U S A* **1961,** *47*, 983-9.
171. Dieter, H.; Koberstein, R.; Sund, H., Studies of glutamate dehydrogenase. The interaction of ADP, GTP, and NADPH in complexes with glutamate dehydrogenase. *European journal of biochemistry / FEBS* **1981,** *115* (1), 217-26.
172. Stanley, C. A.; Lieu, Y. K.; Hsu, B. Y.; Burlina, A. B.; Greenberg, C. R.; Hopwood, N. J.; Perlman, K.; Rich, B. H.; Zammarchi, E.; Poncz, M., Hyperinsulinism and hyperammonemia in infants with regulatory mutations of the glutamate dehydrogenase gene. *N Engl J Med* **1998,** *338* (19), 1352-7.
173. Stanley, C. A.; Fang, J.; Kutyna, K.; Hsu, B. Y.; Ming, J. E.; Glaser, B.; Poncz, M., Molecular basis and characterization of the hyperinsulinism/hyperammonemia syndrome: predominance of mutations in exons 11 and 12 of the glutamate dehydrogenase gene. HI/HA Contributing Investigators. *Diabetes* **2000,** *49* (4), 667-73.
174. Li, M.; Li, C.; Allen, A.; Stanley, C. A.; Smith, T. J., The structure and allosteric regulation of glutamate dehydrogenase. *Neurochem Int* **2011,** *59* (4), 445-55.
175. Li, M.; Li, C.; Allen, A.; Stanley, C. A.; Smith, T. J., The structure and allosteric regulation of mammalian glutamate dehydrogenase. *Archives of biochemistry and biophysics* **2012,** *519* (2), 69-80.
176. Smith, H. Q.; Smith, T. J., Identification of a Novel Activator of Mammalian Glutamate Dehydrogenase. *Biochemistry* **2016,** *55* (47), 6568-6576.
177. Shashidharan, P.; Plaitakis, A., The discovery of human of GLUD2 glutamate dehydrogenase and its implications for cell function in health and disease. *Neurochem Res* **2014,** *39* (3), 460-70.
178. Olson, J. A.; Anfinsen, C. B., The crystallization and characterization of L-glutamic acid dehydrogenase. *J Biol Chem* **1952,** *197* (1), 67-79.
179. Cassman, M.; Schachman, H. K., Sedimentation equilibrium studies on glutamic dehydrogenase. *Biochemistry* **1971,** *10* (6), 1015-24.
180. Frieden, C., Glutamic dehydrogenase. II. The effect of various nucleotides on the association-dissociation and kinetic properties. *J Biol Chem* **1959,** *234* (4), 815-20.
181. Fahien, L. A.; MacDonald, M. J.; Teller, J. K.; Fibich, B.; Fahien, C. M., Kinetic advantages of hetero-enzyme complexes with glutamate dehydrogenase and the alpha-ketoglutarate dehydrogenase complex. *J Biol Chem* **1989,** *264* (21), 12303-12.
182. Fahien, L. A.; Macdonald, M. J., The complex mechanism of glutamate dehydrogenase in insulin secretion. *Diabetes* **2011,** *60* (10), 2450-4.
183. Gylfe, E., Comparison of the effects of leucines, non-metabolizable leucine analogues and other insulin secretagogues on the activity of glutamate dehydrogenase. *Acta Diabetol Lat* **1976,** *13* (1-2), 20-4.
184. Fahien, L. A.; Teller, J. K.; Macdonald, M. J.; Fahien, C. M., Regulation of glutamate dehydrogenase by Mg2+ and magnification of leucine activation by Mg2+. *Mol Pharmacol* **1990,** *37* (6), 943-9.
185. DeLaBarre, B.; Gross, S.; Fang, C.; Gao, Y.; Jha, A.; Jiang, F.; Song, J. J.; Wei, W.; Hurov, J. B., Full-length human glutaminase in complex with an allosteric inhibitor. *Biochemistry* **2011,** *50* (50), 10764-70.
186. Szeliga, M.; Obara-Michlewska, M., Glutamine in neoplastic cells: focus on the expression and roles of glutaminases. *Neurochem Int* **2009,** *55* (1-3), 71-5.

187. DeBerardinis, R. J.; Cheng, T., Q's next: the diverse functions of glutamine in metabolism, cell biology and cancer. *Oncogene* **2010,** *29* (3), 313-24.
188. Mates, J. M.; Segura, J. A.; Campos-Sandoval, J. A.; Lobo, C.; Alonso, L.; Alonso, F. J.; Marquez, J., Glutamine homeostasis and mitochondrial dynamics. *Int J Biochem Cell Biol* **2009,** *41* (10), 2051-61.
189. Chaudhry, F. A.; Reimer, R. J.; Edwards, R. H., The glutamine commute: take the N line and transfer to the A. *The Journal of cell biology* **2002,** *157* (3), 349-55.
190. Hamberger, A.; Chiang, G. H.; Sandoval, E.; Cotman, C. W., Glutamate as a CNS transmitter. II. Regulation of synthesis in the releasable pool. *Brain Res* **1979,** *168* (3), 531-41.
191. Thanki, C. M.; Sugden, D.; Thomas, A. J.; Bradford, H. F., In vivo release from cerebral cortex of [14C]glutamate synthesized from [U-14C]glutamine. *J Neurochem* **1983,** *41* (3), 611-7.
192. Hoffman, E. M.; Schechter, R.; Miller, K. E., Fixative composition alters distributions of immunoreactivity for glutaminase and two markers of nociceptive neurons, Nav1.8 and TRPV1, in the rat dorsal root ganglion. *J Histochem Cytochem* **2010,** *58* (4), 329-44.
193. Erdmann, N.; Zhao, J.; Lopez, A. L.; Herek, S.; Curthoys, N.; Hexum, T. D.; Tsukamoto, T.; Ferraris, D.; Zheng, J., Glutamate production by HIV-1 infected human macrophage is blocked by the inhibition of glutaminase. *J Neurochem* **2007,** *102* (2), 539-49.
194. Erdmann, N.; Tian, C.; Huang, Y.; Zhao, J.; Herek, S.; Curthoys, N.; Zheng, J., In vitro glutaminase regulation and mechanisms of glutamate generation in HIV-1-infected macrophage. *J Neurochem* **2009,** *109* (2), 551-61.
195. Ferreira, A. P.; Cassago, A.; Goncalves Kde, A.; Dias, M. M.; Adamoski, D.; Ascencao, C. F.; Honorato, R. V.; de Oliveira, J. F.; Ferreira, I. M.; Fornezari, C.; Bettini, J.; Oliveira, P. S.; Paes Leme, A. F.; Portugal, R. V.; Ambrosio, A. L.; Dias, S. M., Active glutaminase C self-assembles into a supratetrameric oligomer that can be disrupted by an allosteric inhibitor. *J Biol Chem* **2013,** *288* (39), 28009-20.
196. Peitsaro, N.; Polianskyte, Z.; Tuimala, J.; Porn-Ares, I.; Liobikas, J.; Speer, O.; Lindholm, D.; Thompson, J.; Eriksson, O., Evolution of a family of metazoan active-site-serine enzymes from penicillin-binding proteins: a novel facet of the bacterial legacy. *BMC Evol Biol* **2008,** *8*, 26.
197. Smith, T. S.; Southan, C.; Ellington, K.; Campbell, D.; Tew, D. G.; Debouck, C., Identification, genomic organization, and mRNA expression of LACTB, encoding a serine beta-lactamase-like protein with an amino-terminal transmembrane domain. *Genomics* **2001,** *78* (1-2), 12-4.
198. Chen, Y.; Zhu, J.; Lum, P. Y.; Yang, X.; Pinto, S.; MacNeil, D. J.; Zhang, C.; Lamb, J.; Edwards, S.; Sieberts, S. K.; Leonardson, A.; Castellini, L. W.; Wang, S.; Champy, M. F.; Zhang, B.; Emilsson, V.; Doss, S.; Ghazalpour, A.; Horvath, S.; Drake, T. A.; Lusis, A. J.; Schadt, E. E., Variations in DNA elucidate molecular networks that cause disease. *Nature* **2008,** *452* (7186), 429-35.
199. Polianskyte, Z.; Peitsaro, N.; Dapkunas, A.; Liobikas, J.; Soliymani, R.; Lalowski, M.; Speer, O.; Seitsonen, J.; Butcher, S.; Cereghetti, G. M.; Linder, M. D.; Merckel, M.; Thompson, J.; Eriksson, O., LACTB is a filament-forming protein localized in mitochondria. *Proc Natl Acad Sci U S A* **2009,** *106* (45), 18960-5.
200. Noichri, Y.; Palais, G.; Ruby, V.; D'Autreaux, B.; Delaunay-Moisan, A.; Nystrom, T.; Molin, M.; Toledano, M. B., In vivo parameters influencing 2-Cys Prx oligomerization: The role of enzyme sulfinylation. *Redox Biol* **2015,** *6*, 326-333.
201. Puerto-Galan, L.; Perez-Ruiz, J. M.; Ferrandez, J.; Cano, B.; Naranjo, B.; Najera, V. A.; Gonzalez, M.; Lindahl, A. M.; Cejudo, F. J., Overoxidation of chloroplast 2-Cys peroxiredoxins: balancing toxic and signaling activities of hydrogen peroxide. *Front Plant Sci* **2013,** *4*, 310.
202. Angelucci, F.; Saccoccia, F.; Ardini, M.; Boumis, G.; Brunori, M.; Di Leandro, L.; Ippoliti, R.; Miele, A. E.; Natoli, G.; Scotti, S.; Bellelli, A., Switching between the alternative structures and functions of a 2-Cys peroxiredoxin, by site-directed mutagenesis. *J Mol Biol* **2013,** *425* (22), 4556-68.
203. Cimini, A.; Gentile, R.; Angelucci, F.; Benedetti, E.; Pitari, G.; Giordano, A.; Ippoliti, R., Neuroprotective effects of PrxI over-expression in an in vitro human Alzheimer's disease model. *J Cell Biochem* **2013,** *114* (3), 708-15.
204. Moon, J. C.; Hah, Y. S.; Kim, W. Y.; Jung, B. G.; Jang, H. H.; Lee, J. R.; Kim, S. Y.; Lee, Y. M.; Jeon, M. G.; Kim, C. W.; Cho, M. J.; Lee, S. Y., Oxidative stress-dependent structural and functional switching of a human 2-Cys peroxiredoxin isotype II that enhances HeLa cell resistance to H2O2-induced cell death. *J Biol Chem* **2005,** *280* (31), 28775-84.
205. Wood, Z. A.; Poole, L. B.; Hantgan, R. R.; Karplus, P. A., Dimers to doughnuts: redox-sensitive oligomerization of 2-cysteine peroxiredoxins. *Biochemistry* **2002,** *41* (17), 5493-504.
206. Teixeira, F.; Tse, E.; Castro, H.; Makepeace, K. A. T.; Meinen, B. A.; Borchers, C. H.; Poole, L. B.; Bardwell, J. C.; Tomas, A. M.; Southworth, D. R.; Jakob, U., Chaperone activation and client binding of a 2-cysteine peroxiredoxin. *Nat Commun* **2019,** *10* (1), 659.
207. Saccoccia, F.; Di Micco, P.; Boumis, G.; Brunori, M.; Koutris, I.; Miele, A. E.; Morea, V.; Sriratana, P.; Williams, D. L.; Bellelli, A.; Angelucci, F., Moonlighting by different stressors: crystal structure of the chaperone species of a 2-Cys peroxiredoxin. *Structure* **2012,** *20* (3), 429-39.
208. Kumsta, C.; Jakob, U., Redox-regulated chaperones. *Biochemistry* **2009,** *48* (22), 4666-76.
209. Ando, N.; Li, H.; Brignole, E. J.; Thompson, S.; McLaughlin, M. I.; Page, J. E.; Asturias, F. J.; Stubbe, J.; Drennan, C. L., Allosteric Inhibition of Human Ribonucleotide Reductase by dATP Entails the Stabilization of a Hexamer. *Biochemistry* **2016,** *55* (2), 373-81.
210. Thomas, W. C.; Brooks, F. P. I.; Burnim, A. A.; Bacik, J.-P.; Stubbe, J.; Kaelber, J. T.; Chen, J. Z.; Ando, N., Convergent Allostery in Ribonucleotide Reductase. *bioRxiv* **2018**.
211. Thomas, W. C.; Brooks, F. P., 3rd; Burnim, A. A.; Bacik, J. P.; Stubbe, J.; Kaelber, J. T.; Chen, J. Z.; Ando, N., Convergent allostery in ribonucleotide reductase. *Nat Commun* **2019,** *10* (1), 2653.
212. Lusetti, S. L.; Cox, M. M., The bacterial RecA protein and the recombinational DNA repair of stalled replication forks. *Annu Rev Biochem* **2002,** *71*, 71-100.
213. Radding, C. M., Recombination activities of E. coli recA protein. *Cell* **1981,** *25* (1), 3-4.
214. Chen, Z.; Yang, H.; Pavletich, N. P., Mechanism of homologous recombination from the RecA-ssDNA/dsDNA structures. *Nature* **2008,** *453* (7194), 489-4.
215. Katayama, T.; Ozaki, S.; Keyamura, K.; Fujimitsu, K., Regulation of the replication cycle: conserved and diverse regulatory systems for DnaA and oriC. *Nat Rev Microbiol* **2010,** *8* (3), 163-70.
216. Kaguni, J. M., DnaA: controlling the initiation of bacterial DNA replication and more. *Annu Rev Microbiol* **2006,** *60*, 351-75.
217. Leonard, A. C.; Grimwade, J. E., Regulating DnaA complex assembly: it is time to fill the gaps. *Current opinion in microbiology* **2010,** *13* (6), 766-72.
218. Duderstadt, K. E.; Chuang, K.; Berger, J. M., DNA stretching by bacterial initiators promotes replication origin opening. *Nature* **2011,** *478* (7368), 209-13.
219. Tautz, N.; Kaluza, K.; Frey, B.; Jarsch, M.; Schmitz, G. G.; Kessler, C., SgrAI, a novel class-II restriction endonuclease from Streptomyces griseus recognizing the octanucleotide sequence 5'-CR/CCGGYG-3' [corrected]. *Nucleic Acids Res* **1990,** *18* (10), 3087.
220. Pingoud, A.; Wilson, G. G.; Wende, W., Type II restriction endonucleases--a historical perspective and more. *Nucleic Acids Res* **2014,** *42* (12), 7489-527.
221. Kobayashi, I., DNA modification and restriction: Selfish behavior of an epigenetic system. In *Epigenetic Mechanisms of Gene Regulation*, Russo, V. E. A., Martienssen, R.A., and Riggs, A.D., Ed. Cold Spring Harbor Laboratory Press: Cold Spring Harbor, NY, 1996; pp 155-172.
222. Vasu, K.; Nagaraja, V., Diverse functions of restriction-modification systems in addition to cellular defense. *Microbiol Mol Biol Rev* **2013,** *77* (1), 53-72.
223. Bilcock, D. T.; Daniels, L. E.; Bath, A. J.; Halford, S. E., Reactions of type II restriction endonucleases with 8-base pair recognition sites. *J Biol Chem* **1999,** *274* (51), 36379-86.
224. Daniels, L. E.; Wood, K. M.; Scott, D. J.; Halford, S. E., Subunit assembly for DNA cleavage by restriction endonuclease SgrAI. *J Mol Biol* **2003,** *327* (3), 579-91.
225. Hingorani-Varma, K.; Bitinaite, J., Kinetic analysis of the coordinated interaction of SgrAI restriction endonuclease with different DNA targets. *J Biol Chem* **2003,** *278* (41), 40392-9.
226. Wood, K. M.; Daniels, L. E.; Halford, S. E., Long-range communications between DNA sites by the dimeric restriction endonuclease SgrAI. *J Mol Biol* **2005,** *350* (2), 240-53.
227. Dunten, P. W.; Little, E. J.; Gregory, M. T.; Manohar, V. M.; Dalton, M.; Hough, D.; Bitinaite, J.; Horton, N. C., The structure of SgrAI bound to DNA; recognition of an 8 base pair target. *Nucleic Acids Res* **2008,** *36*, 5405-5416.

228. Little, E. J.; Dunten, P. W.; Bitinaite, J.; Horton, N. C., New clues in the allosteric activation of DNA cleavage by SgrAI: structures of SgrAI bound to cleaved primary-site DNA and uncleaved secondary-site DNA. *Acta Crystallogr D Biol Crystallogr* **2011,** *67* (Pt 1), 67-74.
229. Wang, M.; Kaufman, R. J., Protein misfolding in the endoplasmic reticulum as a conduit to human disease. *Nature* **2016,** *529* (7586), 326-35.
230. Walter, P.; Ron, D., The unfolded protein response: from stress pathway to homeostatic regulation. *Science* **2011,** *334* (6059), 1081-6.
231. Gardner, B. M.; Walter, P., Unfolded proteins are Ire1-activating ligands that directly induce the unfolded protein response. *Science* **2011,** *333* (6051), 1891-4.
232. Karagoz, G. E.; Acosta-Alvear, D.; Nguyen, H. T.; Lee, C. P.; Chu, F.; Walter, P., An unfolded protein-induced conformational switch activates mammalian IRE1. *Elife* **2017,** *6*.
233. Credle, J. J.; Finer-Moore, J. S.; Papa, F. R.; Stroud, R. M.; Walter, P., On the mechanism of sensing unfolded protein in the endoplasmic reticulum. *Proc Natl Acad Sci U S A* **2005,** *102* (52), 18773-84.
234. Aragon, T.; van Anken, E.; Pincus, D.; Serafimova, I. M.; Korennykh, A. V.; Rubio, C. A.; Walter, P., Messenger RNA targeting to endoplasmic reticulum stress signalling sites. *Nature* **2009,** *457* (7230), 736-40.
235. Ishiwata-Kimata, Y.; Yamamoto, Y. H.; Takizawa, K.; Kohno, K.; Kimata, Y., F-actin and a type-II myosin are required for efficient clustering of the ER stress sensor Ire1. *Cell Struct Funct* **2013,** *38* (2), 135-43.
236. Li, H.; Korennykh, A. V.; Behrman, S. L.; Walter, P., Mammalian endoplasmic reticulum stress sensor IRE1 signals by dynamic clustering. *Proc Natl Acad Sci U S A* **2010,** *107* (37), 16113-8.
237. Chen, Y.; Brandizzi, F., IRE1: ER stress sensor and cell fate executor. *Trends Cell Biol* **2013,** *23* (11), 547-55.
238. Sidrauski, C.; Cox, J. S.; Walter, P., tRNA ligase is required for regulated mRNA splicing in the unfolded protein response. *Cell* **1996,** *87* (3), 405-13.
239. Lu, Y.; Liang, F. X.; Wang, X., A synthetic biology approach identifies the mammalian UPR RNA ligase RtcB. *Molecular cell* **2014,** *55* (5), 758-70.
240. Tam, A. B.; Koong, A. C.; Niwa, M., Ire1 has distinct catalytic mechanisms for XBP1/HAC1 splicing and RIDD. *Cell Rep* **2014,** *9* (3), 850-8.
241. Hollien, J.; Weissman, J. S., Decay of endoplasmic reticulum-localized mRNAs during the unfolded protein response. *Science* **2006,** *313* (5783), 104-7.
242. Hollien, J.; Lin, J. H.; Li, H.; Stevens, N.; Walter, P.; Weissman, J. S., Regulated Ire1-dependent decay of messenger RNAs in mammalian cells. *The Journal of cell biology* **2009,** *186* (3), 323-31.
243. Kimmig, P.; Diaz, M.; Zheng, J.; Williams, C. C.; Lang, A.; Aragon, T.; Li, H.; Walter, P., The unfolded protein response in fission yeast modulates stability of select mRNAs to maintain protein homeostasis. *Elife* **2012,** *1*, e00048.
244. Cox, J. S.; Shamu, C. E.; Walter, P., Transcriptional induction of genes encoding endoplasmic reticulum resident proteins requires a transmembrane protein kinase. *Cell* **1993,** *73* (6), 1197-206.
245. Mori, K.; Ma, W.; Gething, M. J.; Sambrook, J., A transmembrane protein with a cdc2+/CDC28-related kinase activity is required for signaling from the ER to the nucleus. *Cell* **1993,** *74* (4), 743-56.
246. Yoshida, H.; Matsui, T.; Yamamoto, A.; Okada, T.; Mori, K., XBP1 mRNA is induced by ATF6 and spliced by IRE1 in response to ER stress to produce a highly active transcription factor. *Cell* **2001,** *107* (7), 881-91.
247. Calfon, M.; Zeng, H.; Urano, F.; Till, J. H.; Hubbard, S. R.; Harding, H. P.; Clark, S. G.; Ron, D., IRE1 couples endoplasmic reticulum load to secretory capacity by processing the XBP-1 mRNA. *Nature* **2002,** *415* (6867), 92-6.
248. Naidoo, N.; Giang, W.; Galante, R. J.; Pack, A. I., Sleep deprivation induces the unfolded protein response in mouse cerebral cortex. *J Neurochem* **2005,** *92* (5), 1150-7.
249. Atkin, J. D.; Farg, M. A.; Walker, A. K.; McLean, C.; Tomas, D.; Horne, M. K., Endoplasmic reticulum stress and induction of the unfolded protein response in human sporadic amyotrophic lateral sclerosis. *Neurobiol Dis* **2008,** *30* (3), 400-7.
250. Li, W.; Okreglak, V.; Peschek, J.; Kimmig, P.; Zubradt, M.; Weissman, J. S.; Walter, P., Engineering ER-stress dependent non-conventional mRNA splicing. *Elife* **2018,** *7*.

251. Han, D.; Lerner, A. G.; Vande Walle, L.; Upton, J. P.; Xu, W.; Hagen, A.; Backes, B. J.; Oakes, S. A.; Papa, F. R., IRE1alpha kinase activation modes control alternate endoribonuclease outputs to determine divergent cell fates. *Cell* **2009,** *138* (3), 562-75.
252. Iwawaki, T.; Hosoda, A.; Okuda, T.; Kamigori, Y.; Nomura-Furuwatari, C.; Kimata, Y.; Tsuru, A.; Kohno, K., Translational control by the ER transmembrane kinase/ribonuclease IRE1 under ER stress. *Nat Cell Biol* **2001,** *3* (2), 158-64.
253. Mishiba, K.; Nagashima, Y.; Suzuki, E.; Hayashi, N.; Ogata, Y.; Shimada, Y.; Koizumi, N., Defects in IRE1 enhance cell death and fail to degrade mRNAs encoding secretory pathway proteins in the Arabidopsis unfolded protein response. *Proc Natl Acad Sci U S A* **2013,** *110* (14), 5713-8.
254. Maurel, M.; Chevet, E.; Tavernier, J.; Gerlo, S., Getting RIDD of RNA: IRE1 in cell fate regulation. *Trends in biochemical sciences* **2014,** *39* (5), 245-54.
255. Moore, K.; Hollien, J., Ire1-mediated decay in mammalian cells relies on mRNA sequence, structure, and translational status. *Molecular biology of the cell* **2015,** *26* (16), 2873-84.
256. Niwa, M.; Patil, C. K.; DeRisi, J.; Walter, P., Genome-scale approaches for discovering novel nonconventional splicing substrates of the Ire1 nuclease. *Genome Biol* **2005,** *6* (1), R3.
257. Nakamura, D.; Tsuru, A.; Ikegami, K.; Imagawa, Y.; Fujimoto, N.; Kohno, K., Mammalian ER stress sensor IRE1beta specifically down-regulates the synthesis of secretory pathway proteins. *FEBS Lett* **2011,** *585* (1), 133-8.
258. Imagawa, Y.; Hosoda, A.; Sasaka, S.; Tsuru, A.; Kohno, K., RNase domains determine the functional difference between IRE1alpha and IRE1beta. *FEBS Lett* **2008,** *582* (5), 656-60.
259. Upton, J. P.; Wang, L.; Han, D.; Wang, E. S.; Huskey, N. E.; Lim, L.; Truitt, M.; McManus, M. T.; Ruggero, D.; Goga, A.; Papa, F. R.; Oakes, S. A., IRE1alpha cleaves select microRNAs during ER stress to derepress translation of proapoptotic Caspase-2. *Science* **2012,** *338* (6108), 818-22.
260. Zhou, J.; Liu, C. Y.; Back, S. H.; Clark, R. L.; Peisach, D.; Xu, Z.; Kaufman, R. J., The crystal structure of human IRE1 luminal domain reveals a conserved dimerization interface required for activation of the unfolded protein response. *Proc Natl Acad Sci U S A* **2006,** *103* (39), 14343-8.
261. Lee, K. P.; Dey, M.; Neculai, D.; Cao, C.; Dever, T. E.; Sicheri, F., Structure of the dual enzyme Ire1 reveals the basis for catalysis and regulation in nonconventional RNA splicing. *Cell* **2008,** *132* (1), 89-100.
262. Ali, M. M.; Bagratuni, T.; Davenport, E. L.; Nowak, P. R.; Silva-Santisteban, M. C.; Hardcastle, A.; McAndrews, C.; Rowlands, M. G.; Morgan, G. J.; Aherne, W.; Collins, I.; Davies, F. E.; Pearl, L. H., Structure of the Ire1 autophosphorylation complex and implications for the unfolded protein response. *EMBO J* **2011,** *30* (5), 894-905.
263. Litchfield, D. W., Protein kinase CK2: structure, regulation and role in cellular decisions of life and death. *The Biochemical journal* **2003,** *369* (Pt 1), 1-15.
264. Meggio, F.; Pinna, L. A., One-thousand-and-one substrates of protein kinase CK2? *FASEB J* **2003,** *17* (3), 349-68.
265. Poole, A.; Poore, T.; Bandhakavi, S.; McCann, R. O.; Hanna, D. E.; Glover, C. V., A global view of CK2 function and regulation. *Mol Cell Biochem* **2005,** *274* (1-2), 163-70.
266. Valero, E.; De Bonis, S.; Filhol, O.; Wade, R. H.; Langowski, J.; Chambaz, E. M.; Cochet, C., Quaternary structure of casein kinase 2. Characterization of multiple oligomeric states and relation with its catalytic activity. *J Biol Chem* **1995,** *270* (14), 8345-52.
267. Glover, C. V., A filamentous form of Drosophila casein kinase II. *J Biol Chem* **1986,** *261* (30), 14349-54.
268. Mamrack, M. D., Stimulation of enzymatic activity in filament preparations of casein kinase II by polylysine, melittin, and spermine. *Mol Cell Biochem* **1989,** *85* (2), 147-57.
269. Seetoh, W. G.; Chan, D. S.; Matak-Vinkovic, D.; Abell, C., Mass Spectrometry Reveals Protein Kinase CK2 High-Order Oligomerization via the Circular and Linear Assembly. *ACS Chem Biol* **2016,** *11* (6), 1511-7.
270. Niefind, K.; Guerra, B.; Ermakowa, I.; Issinger, O. G., Crystal structure of human protein kinase CK2: insights into basic properties of the CK2 holoenzyme. *EMBO J* **2001,** *20* (19), 5320-31.
271. Lolli, G.; Ranchio, A.; Battistutta, R., Active form of the protein kinase CK2 alpha2beta2 holoenzyme is a strong complex with symmetric architecture. *ACS Chem Biol* **2014,** *9* (2), 366-71.

272. Schnitzler, A.; Olsen, B. B.; Issinger, O. G.; Niefind, K., The protein kinase CK2(Andante) holoenzyme structure supports proposed models of autoregulation and trans-autophosphorylation. *J Mol Biol* **2014,** *426* (9), 1871-82.
273. Hubner, G. M.; Larsen, J. N.; Guerra, B.; Niefind, K.; Vrecl, M.; Issinger, O. G., Evidence for aggregation of protein kinase CK2 in the cell: a novel strategy for studying CK2 holoenzyme interaction by BRET(2). *Mol Cell Biochem* **2014,** *397* (1-2), 285-93.
274. Wild, K.; Grafmuller, R.; Wagner, E.; Schulz, G. E., Structure, catalysis and supramolecular assembly of adenylate kinase from maize. *European journal of biochemistry / FEBS* **1997,** *250* (2), 326-31.
275. Moquin, D.; Chan, F. K., The molecular regulation of programmed necrotic cell injury. *Trends in biochemical sciences* **2010,** *35* (8), 434-41.
276. Cho, Y. S.; Challa, S.; Moquin, D.; Genga, R.; Ray, T. D.; Guildford, M.; Chan, F. K., Phosphorylation-driven assembly of the RIP1-RIP3 complex regulates programmed necrosis and virus-induced inflammation. *Cell* **2009,** *137* (6), 1112-23.
277. He, S.; Wang, L.; Miao, L.; Wang, T.; Du, F.; Zhao, L.; Wang, X., Receptor interacting protein kinase-3 determines cellular necrotic response to TNF-alpha. *Cell* **2009,** *137* (6), 1100-11.
278. Zhang, D. W.; Shao, J.; Lin, J.; Zhang, N.; Lu, B. J.; Lin, S. C.; Dong, M. Q.; Han, J., RIP3, an energy metabolism regulator that switches TNF-induced cell death from apoptosis to necrosis. *Science* **2009,** *325* (5938), 332-6.
279. Sun, L.; Wang, H.; Wang, Z.; He, S.; Chen, S.; Liao, D.; Wang, L.; Yan, J.; Liu, W.; Lei, X.; Wang, X., Mixed lineage kinase domain-like protein mediates necrosis signaling downstream of RIP3 kinase. *Cell* **2012,** *148* (1-2), 213-27.
280. Zhao, J.; Jitkaew, S.; Cai, Z.; Choksi, S.; Li, Q.; Luo, J.; Liu, Z. G., Mixed lineage kinase domain-like is a key receptor interacting protein 3 downstream component of TNF-induced necrosis. *Proc Natl Acad Sci U S A* **2012,** *109* (14), 5322-7.
281. Li, J.; McQuade, T.; Siemer, A. B.; Napetschnig, J.; Moriwaki, K.; Hsiao, Y. S.; Damko, E.; Moquin, D.; Walz, T.; McDermott, A.; Chan, F. K.; Wu, H., The RIP1/RIP3 necrosome forms a functional amyloid signaling complex required for programmed necrosis. *Cell* **2012,** *150* (2), 339-50.
282. Mompean, M.; Li, W.; Li, J.; Laage, S.; Siemer, A. B.; Bozkurt, G.; Wu, H.; McDermott, A. E., The Structure of the Necrosome RIPK1-RIPK3 Core, a Human Hetero-Amyloid Signaling Complex. *Cell* **2018,** *173* (5), 1244-1253 e10.
283. Chiti, F.; Dobson, C. M., Protein misfolding, functional amyloid, and human disease. *Annu Rev Biochem* **2006,** *75*, 333-66.
284. Uptain, S. M.; Lindquist, S., Prions as protein-based genetic elements. *Annu Rev Microbiol* **2002,** *56*, 703-41.
285. Ferrao, R.; Wu, H., Helical assembly in the death domain (DD) superfamily. *Curr Opin Struct Biol* **2012,** *22* (2), 241-7.
286. Yin, Q.; Fu, T. M.; Li, J.; Wu, H., Structural biology of innate immunity. *Annu Rev Immunol* **2015,** *33*, 393-416.
287. Zeng, W.; Sun, L.; Jiang, X.; Chen, X.; Hou, F.; Adhikari, A.; Xu, M.; Chen, Z. J., Reconstitution of the RIG-I pathway reveals a signaling role of unanchored polyubiquitin chains in innate immunity. *Cell* **2010,** *141* (2), 315-30.
288. Wu, B.; Peisley, A.; Tetrault, D.; Li, Z.; Egelman, E. H.; Magor, K. E.; Walz, T.; Penczek, P. A.; Hur, S., Molecular imprinting as a signal-activation mechanism of the viral RNA sensor RIG-I. *Molecular cell* **2014,** *55* (4), 511-23.
289. Gong, Q.; Long, Z.; Zhong, F. L.; Teo, D. E. T.; Jin, Y.; Yin, Z.; Boo, Z. Z.; Zhang, Y.; Zhang, J.; Yang, R.; Bhushan, S.; Reversade, B.; Li, Z.; Wu, B., Structural basis of RIP2 activation and signaling. *Nat Commun* **2018,** *9* (1), 4993.
290. Lin, S. C.; Lo, Y. C.; Wu, H., Helical assembly in the MyD88-IRAK4-IRAK2 complex in TLR/IL-1R signalling. *Nature* **2010,** *465* (7300), 885-90.
291. Diebolder, C. A.; Halff, E. F.; Koster, A. J.; Huizinga, E. G.; Koning, R. I., Cryoelectron Tomography of the NAIP5/NLRC4 Inflammasome: Implications for NLR Activation. *Structure* **2015,** *23* (12), 2349-2357.

292. Lu, A.; Li, Y.; Schmidt, F. I.; Yin, Q.; Chen, S.; Fu, T. M.; Tong, A. B.; Ploegh, H. L.; Mao, Y.; Wu, H., Molecular basis of caspase-1 polymerization and its inhibition by a new capping mechanism. *Nat Struct Mol Biol* **2016,** *23* (5), 416-25.
293. Lu, A.; Magupalli, V. G.; Ruan, J.; Yin, Q.; Atianand, M. K.; Vos, M. R.; Schroder, G. F.; Fitzgerald, K. A.; Wu, H.; Egelman, E. H., Unified polymerization mechanism for the assembly of ASC-dependent inflammasomes. *Cell* **2014,** *156* (6), 1193-1206.
294. Wang, F.; Eric Knabe, W.; Li, L.; Jo, I.; Mani, T.; Roehm, H.; Oh, K.; Li, J.; Khanna, M.; Meroueh, S. O., Design, synthesis, biochemical studies, cellular characterization, and structure-based computational studies of small molecules targeting the urokinase receptor. *Bioorg Med Chem* **2012,** *20* (15), 4760-73.
295. Lu, A.; Wu, H., Structural mechanisms of inflammasome assembly. *The FEBS journal* **2015,** *282* (3), 435-44.
296. Prouteau, M.; Desfosses, A.; Sieben, C.; Bourgoint, C.; Lydia Mozaffari, N.; Demurtas, D.; Mitra, A. K.; Guichard, P.; Manley, S.; Loewith, R., TORC1 organized in inhibited domains (TOROIDs) regulate TORC1 activity. *Nature* **2017,** *550* (7675), 265-269.
297. Zhang, S.; Ding, K.; Shen, Q. J.; Zhao, S.; Liu, J. L., Filamentation of asparagine synthetase in Saccharomyces cerevisiae. *PLoS Genet* **2018,** *14* (10), e1007737.
298. Campbell, S. G.; Hoyle, N. P.; Ashe, M. P., Dynamic cycling of eIF2 through a large eIF2B-containing cytoplasmic body: implications for translation control. *The Journal of cell biology* **2005,** *170* (6), 925-34.
299. Dever, T. E.; Yang, W.; Astrom, S.; Bystrom, A. S.; Hinnebusch, A. G., Modulation of tRNA(iMet), eIF-2, and eIF-2B expression shows that GCN4 translation is inversely coupled to the level of eIF-2.GTP.Met-tRNA(iMet) ternary complexes. *Mol Cell Biol* **1995,** *15* (11), 6351-63.
300. Nuske, E.; Marini, G.; Richter, D.; Leng, W.; Bogdanova, A.; Franszmann, T.; Pigino, G.; Alberti, S., Filament formation by the translation factor eIF2B regulates protein synthesis in starved cells. *bioRxiv* **2018**.
301. Angermuller, S.; Bruder, G.; Volkl, A.; Wesch, H.; Fahimi, H. D., Localization of xanthine oxidase in crystalline cores of peroxisomes. A cytochemical and biochemical study. *Eur J Cell Biol* **1987,** *45* (1), 137-44.
302. Saad, S.; Cereghetti, G.; Feng, Y.; Picotti, P.; Peter, M.; Dechant, R., Reversible protein aggregation is a protective mechanism to ensure cell cycle restart after stress. *Nat Cell Biol* **2017,** *19* (10), 1202-1213.
303. Miura, N.; Shinohara, M.; Tatsukami, Y.; Sato, Y.; Morisaka, H.; Kuroda, K.; Ueda, M., Spatial reorganization of Saccharomyces cerevisiae enolase to alter carbon metabolism under hypoxia. *Eukaryot Cell* **2013,** *12* (8), 1106-19.
304. Jain, S.; Wheeler, J. R.; Walters, R. W.; Agrawal, A.; Barsic, A.; Parker, R., ATPase-Modulated Stress Granules Contain a Diverse Proteome and Substructure. *Cell* **2016,** *164* (3), 487-98.
305. Gilliland, W. D.; Vietti, D. L.; Schweppe, N. M.; Guo, F.; Johnson, T. J.; Hawley, R. S., Hypoxia transiently sequesters mps1 and polo to collagenase-sensitive filaments in Drosophila prometaphase oocytes. *PLoS One* **2009,** *4* (10), e7544.
306. Pandey, R.; Heeger, S.; Lehner, C. F., Rapid effects of acute anoxia on spindle kinetochore interactions activate the mitotic spindle checkpoint. *J Cell Sci* **2007,** *120* (Pt 16), 2807-18.
307. Pollegioni, L.; Piubelli, L.; Sacchi, S.; Pilone, M. S.; Molla, G., Physiological functions of D-amino acid oxidases: from yeast to humans. *Cell Mol Life Sci* **2007,** *64* (11), 1373-94.
308. Antonini, E.; Brunori, M.; Bruzzesi, R.; Chiancone, E.; Massey, V., Association-dissociation phenomena of D-amino acid oxidase. *J Biol Chem* **1966,** *241* (10), 2358-66.
309. Tojo, H.; Horiike, K.; Shiga, K.; Nishina, Y.; Watari, H.; Yamano, T., Self-association mode of a flavoenzyme D-amino acid oxidase from hog kidney. II. Stoichiometry of holoenzyme association and energetics of subunit association. *J Biol Chem* **1985,** *260* (23), 12615-21.
310. Tojo, H.; Horiike, K.; Shiga, K.; Nishina, Y.; Watari, H.; Yamano, T., Self-association mode of a flavoenzyme D-amino acid oxidase from hog kidney. I. Analysis of apparent weight-average molecular weight data for the apoenzyme in terms of models. *J Biol Chem* **1985,** *260* (23), 12607-14.

311. Jacobson, T.; Navarrete, C.; Sharma, S. K.; Sideri, T. C.; Ibstedt, S.; Priya, S.; Grant, C. M.; Christen, P.; Goloubinoff, P.; Tamas, M. J., Arsenite interferes with protein folding and triggers formation of protein aggregates in yeast. *J Cell Sci* **2012,** *125* (Pt 21), 5073-83.
312. Alberti, S.; Halfmann, R.; King, O.; Kapila, A.; Lindquist, S., A systematic survey identifies prions and illuminates sequence features of prionogenic proteins. *Cell* **2009,** *137* (1), 146-58.
313. Saarikangas, J.; Barral, Y., Protein aggregation as a mechanism of adaptive cellular responses. *Curr Genet* **2016,** *62* (4), 711-724.
314. Aulas, A.; Vande Velde, C., Alterations in stress granule dynamics driven by TDP-43 and FUS: a link to pathological inclusions in ALS? *Frontiers in Cellular Neuroscience* **2015,** *9*.
315. Woodward, J. D.; Trompetter, I.; Sewell, B. T.; Piotrowski, M., Substrate specificity of plant nitrilase complexes is affected by their helical twist. *Commun Biol* **2018,** *1*, 186.
316. Breinig, S.; Kervinen, J.; Stith, L.; Wasson, A. S.; Fairman, R.; Wlodawer, A.; Zdanov, A.; Jaffe, E. K., Control of tetrapyrrole biosynthesis by alternate quaternary forms of porphobilinogen synthase. *Nat Struct Biol* **2003,** *10* (9), 757-63.
317. Selwood, T.; Tang, L.; Lawrence, S. H.; Anokhina, Y.; Jaffe, E. K., Kinetics and thermodynamics of the interchange of the morpheein forms of human porphobilinogen synthase. *Biochemistry* **2008,** *47* (10), 3245-57.
318. Tang, L.; Breinig, S.; Stith, L.; Mischel, A.; Tannir, J.; Kokona, B.; Fairman, R.; Jaffe, E. K., Single amino acid mutations alter the distribution of human porphobilinogen synthase quaternary structure isoforms (morpheeins). *J Biol Chem* **2006,** *281* (10), 6682-90.
319. Tang, L.; Stith, L.; Jaffe, E. K., Substrate-induced interconversion of protein quaternary structure isoforms. *J Biol Chem* **2005,** *280* (16), 15786-93.
320. Goodsell, D. S.; Olson, A. J., Structural symmetry and protein function. *Annual review of biophysics and biomolecular structure* **2000,** *29*, 105-53.
321. Thayer, N. H.; Leverich, C. K.; Fitzgibbon, M. P.; Nelson, Z. W.; Henderson, K. A.; Gafken, P. R.; Hsu, J. J.; Gottschling, D. E., Identification of long-lived proteins retained in cells undergoing repeated asymmetric divisions. *Proc Natl Acad Sci U S A* **2014,** *111* (39), 14019-26.
322. Jang, S. R.; Xuan, Z.; Lagoy, R. C.; Jawerth, L. M.; Gonzalez, I.; Singh, M.; Prashad, S.; Kim, H. S.; Patel, A. R.; Albrecht, D. R.; Hyman, A. A.; Colon-Ramos, D. A., The Glycolytic Protein Phosphofructokinase Dynamically Relocalizes into
Subcellular Compartments with Liquid-like Properties in vivo. *BioRxiv* **2019**.
323. Chan, K. Y.; Gumbart, J.; McGreevy, R.; Watermeyer, J. M.; Sewell, B. T.; Schulten, K., Symmetry-restrained flexible fitting for symmetric EM maps. *Structure* **2011,** *19* (9), 1211-8.
324. Phalen, T. J.; Weirather, K.; Deming, P. B.; Anathy, V.; Howe, A. K.; van der Vliet, A.; Jonsson, T. J.; Poole, L. B.; Heintz, N. H., Oxidation state governs structural transitions in peroxiredoxin II that correlate with cell cycle arrest and recovery. *The Journal of cell biology* **2006,** *175* (5), 779-89.
325. Gourlay, L. J.; Bhella, D.; Kelly, S. M.; Price, N. C.; Lindsay, J. G., Structure-function analysis of recombinant substrate protein 22 kDa (SP-22). A mitochondrial 2-CYS peroxiredoxin organized as a decameric toroid. *J Biol Chem* **2003,** *278* (35), 32631-7.
326. Yu, X.; Jacobs, S. A.; West, S. C.; Ogawa, T.; Egelman, E. H., Domain structure and dynamics in the helical filaments formed by RecA and Rad51 on DNA. *Proc Natl Acad Sci U S A* **2001,** *98* (15), 8419-24.
327. Bleichert, F.; Botchan, M. R.; Berger, J. M., Mechanisms for initiating cellular DNA replication. *Science* **2017,** *355* (6327).
328. Poothong, J.; Tirasophon, W.; Kaufman, R. J., Functional analysis of the mammalian RNA ligase for IRE1 in the unfolded protein response. *Biosci Rep* **2017,** *37* (2).
329. Ghosh, R.; Wang, L.; Wang, E. S.; Perera, B. G.; Igbaria, A.; Morita, S.; Prado, K.; Thamsen, M.; Caswell, D.; Macias, H.; Weiberth, K. F.; Gliedt, M. J.; Alavi, M. V.; Hari, S. B.; Mitra, A. K.; Bhhatarai, B.; Schurer, S. C.; Snapp, E. L.; Gould, D. B.; German, M. S.; Backes, B. J.; Maly, D. J.; Oakes, S. A.; Papa, F. R., Allosteric inhibition of the IRE1alpha RNase preserves cell viability and function during endoplasmic reticulum stress. *Cell* **2014,** *158* (3), 534-48.
330. Telford, J. N.; Lad, P. M.; Hammes, G. G., Electron microscope study of native and crosslinked rabbit muscle phosphofructokinase. *Proc Natl Acad Sci U S A* **1975,** *72* (8), 3054-6.

331. Schwock, J.; Kirchberger, J.; Edelmann, A.; Kriegel, T. M.; Kopperschlager, G., Interaction of 6-phosphofructokinase with cytosolic proteins of Saccharomyces cerevisiae. *Yeast* **2004,** *21* (6), 483-94.
332. Ranjit, S.; Dvornikov, A.; Holland, D. A.; Reinhart, G. D.; Jameson, D. M.; Gratton, E., Application of three-photon excitation FCS to the study of protein oligomerization. *J Phys Chem B* **2014,** *118* (50), 14627-31.
333. Nagai, M.; Natsumeda, Y.; Konno, Y.; Hoffman, R.; Irino, S.; Weber, G., Selective up-regulation of type II inosine 5'-monophosphate dehydrogenase messenger RNA expression in human leukemias. *Cancer Res* **1991,** *51* (15), 3886-90.
334. Zimmermann, A. G.; Gu, J. J.; Laliberte, J.; Mitchell, B. S., Inosine-5'-monophosphate dehydrogenase: regulation of expression and role in cellular proliferation and T lymphocyte activation. *Progress in nucleic acid research and molecular biology* **1998,** *61*, 181-209.
335. Nisius, A., The stromacentre inAvena plastids: an aggregation ofbeta-glucosidase responsible for the activation of oat-leaf saponins. *Planta* **1988,** *173* (4), 474-81.
336. Kessler, D.; Leibrecht, I.; Knappe, J., Pyruvate-formate-lyase-deactivase and acetyl-CoA reductase activities of Escherichia coli reside on a polymeric protein particle encoded by adhE. *FEBS Lett* **1991,** *281* (1-2), 59-63.
337. Barahona, C. J.; Basantes, L. E.; Tompkins, K. J.; Heitman, D. M.; Chukwu, B. I.; Sanchez, J.; Sanchez, J. L.; Ghadirian, N.; Park, C. K.; Horton, N. C., The Need for Speed: Run-On Oligomer Filament Formation Provides Maximum Speed with Maximum Sequestration of Activity. *Submitted for publication* **2018**.
338. Erzberger, J. P.; Mott, M. L.; Berger, J. M., Structural basis for ATP-dependent DnaA assembly and replication-origin remodeling. *Nat Struct Mol Biol* **2006,** *13* (8), 676-83.